\documentclass[12pt]{article}
\usepackage{jheppub}
\usepackage{breqn}

\pdfoutput=1

\usepackage[table]{xcolor}
\usepackage{amsmath,bbm,array,amsfonts,graphicx,wrapfig,lscape,float,mathtools,multirow,longtable}
\usepackage{stackrel}
\usepackage[all]{xy}
\usepackage{subcaption}

\newcommand{\be}{\begin{equation}}
\newcommand{\ee}{\end{equation}}
\newcommand{\beq}{\begin{equation}}
\newcommand{\beql}[1]{\begin{equation}\label{#1}}
\newcommand{\eeq}{\end{equation}}
\newcommand{\ba}{\begin{array}}
\newcommand{\ea}{\end{array}}
\newcommand{\bea}{\begin{eqnarray}}
\newcommand{\beal}[1]{\begin{eqnarray}\label{#1}}
\newcommand{\eea}{\end{eqnarray}}
\newcommand{\ben}{\begin{enumerate}}
\newcommand{\een}{\end{enumerate}}
\newcommand{\bean}{\begin{eqnarray*}}
\newcommand{\eean}{\end{eqnarray*}}

\newcommand{\sref}[1]{\S\ref{#1}}

\newcommand{\fref}[1]{Figure \ref{#1}}
\newcommand{\btab}[1]{\begin{tabular}{#1}}
\newcommand{\etab}{\end{tabular}}

\newcommand{\comment}[1]{}

\newcommand{\qed}{\nobreak \ifvmode \relax \else
      \ifdim\lastskip<1.5em \hskip-\lastskip
      \hskip1.5em plus0em minus0.5em \fi \nobreak
      \vrule height0.75em width0.5em depth0.25em\fi}

\definecolor{darkspringgreen}{rgb}{0.09, 0.45, 0.27}
\definecolor{forestgreen}{rgb}{0.13, 0.55, 0.13}

\usepackage{array}

\newcolumntype{C}[1]{>{\centering\let\newline\\\arraybackslash\hspace{0pt}}m{#1}}

\title{Crystal Melting, Triality and Partition Functions for Toric Calabi-Yau Fourfolds} 
\author[a,b]{Mario Carcamo}
\author[a,b,c]{and Sebasti\'an Franco}

\affiliation[a]{Physics Department, The City College of the CUNY\\
	160 Convent Avenue, New York, NY 10031, USA}
	
\affiliation[b]{Physics Program and \textsuperscript{$c$}Initiative for the Theoretical Sciences\\
	The Graduate School and University Center, The City University of New York\\
	365 Fifth Avenue, New York NY 10016, USA}

\emailAdd{mcarcamo@ccny.cuny.edu,sfranco@ccny.cuny.edu}

\abstract{We extend the study of the recently introduced crystal melting models associated to toric Calabi-Yau 4-folds in several directions. In particular, we investigate in greater detail the structure of these models for general toric CY 4-folds and flavor configurations, using the explicit example of $Q^{1,1,1}$ to illustrate our ideas. To this end, we develop an efficient algorithm for constructing crystals based on periodic quivers. A central goal of this work is to understand the behavior of crystals and their partition functions under triality. We analyze the evolution of crystals along periodic triality cascades and generate detailed data for these systems, including Hasse diagrams, partition functions, and the multiplicities of melting configurations. We introduce the notion of stable variables and show that they lead to the stabilization of the partition functions along cascades. Finally, we define the profile of the crystal partition function and observe that, when expressed in terms of stable variables, it displays interesting behavior. A further motivation for this work is to generate empirical data that may guide the search for a physically motivated generalization of cluster algebras associated with $2d$ (0,2) quiver theories and their triality transformations.
}

\begin{document}

\maketitle

%=================================================================
\section{Introduction}
%=================================================================

D-branes probing Calabi-Yau (CY) singularities provide a powerful framework for realizing quantum field theories in various dimensions in string theory \cite{Klebanov:1998hh,Morrison:1998cs,Klebanov:2000hb,Aldazabal:2000sa,Verlinde:2005jr}. Such setups often make the interplay between the dynamics of the quantum field theories and the underlying geometry explicit. 

The $4d$ $\mathcal{N} = 1$ gauge theories on the worldvolume of D3-branes probing toric CY 3-folds are one of the best understood classes of theories of this type. These gauge theories are fully encoded by brane tilings, a class of brane configurations T-dual to branes at singularities \cite{Hanany:2005ve,Franco:2005rj,Franco:2005sm}. Moreover, brane tilings streamline the connection between gauge theory and the underlying geometry.

The BPS spectrum of D-branes on a toric CY 3-folds is captured by a statistical model of crystal melting \cite{Okounkov:2003sp,Iqbal:2003ds,Ooguri:2009ijd,Ooguri:2009ri} (see also \cite{Szendroi:2007nu,Mozgovoy:2008fd} for related ideas). Interestingly, the underlying structure of such crystals is dictated by the corresponding brane tilings or, equivalently, the associated periodic quivers.

In recent years, Nekrasov initiated a parallel line of investigation for CY 4-folds with the introduction of the {\it Magnificent Four} \cite{Nekrasov:2017cih,Nekrasov:2018xsb,Nekrasov:2023nai}. This statistical model, whose random variables are solid partitions, computes the refined index of a system of D0-branes in the presence of a D8-$\overline{\rm{D8}}$ configuration in $\mathbb{C}^4$ with a background $B$-field.

Interestingly, there exists a class of brane constructions, {\it brane brick models}, which play a role for CY 4-folds analogous to that of brane tilings for CY 3-folds \cite{Franco:2015tya}. They are related to the D1-branes at the singularities via T-duality, neatly encode the gauge theory and simplify the connection to geometry (we refer the interested reader to \cite{Franco:2015tna,Franco:2016nwv,Franco:2016qxh,Franco:2016fxm,Franco:2017cjj,Franco:2018qsc,Franco:2019bmx,Franco:2020avj,Franco:2022iap,Franco:2022gvl,Franco:2022isw,Franco:2023tyf,Carcamo:2025shw} for various results regarding branes on toric CY 4-folds and brane brick models). 

Motivated by these recent developments, a new class of crystal melting models capturing the BPS bound states of D-branes on toric CY 4-folds was introduced in \cite{Franco:2023tly}. These crystals are 4-dimensional and their structure is determined by the associated brane brick models. These models were further studied in \cite{Bao:2024ygr}

This paper extends the study of crystal melting models for toric CY$_4$-folds in several directions. First, we explicitly analyze crystals for general toric CY 4-folds and flavor configurations. To this end, we develop efficient methods for their construction and introduce new tools for their systematic investigation. A central theme of this work is the evolution of these crystals under triality.

Our results substantially deepen the understanding of CY$_4$ crystals and are intrinsically interesting in their own right. Having said that, in addition, our study is motivated by a broader question. Since the discovery of triality for $2d$ (0,2) theories, it has been natural to ask whether a generalization of cluster algebras based on these theories and their mutations exists. This perspective has been further strengthened by the advent of brane brick models, which endow an infinite class of $2d$ (0,2) theories with a combinatorial structure. We propose that the crystals associated with toric CY 4-folds, together with their transformations under triality, provide a natural class of systems that may be captured by such hypothetical mathematical structures. From this viewpoint, the results presented here can be regarded as empirical data that help guide the search for this generalization.

This paper is organized as follows. In Section \sref{section_CY4_crystals}, we review the crystal melting models introduced in \cite{Franco:2023tly} for general toric CY 4-folds and flavor configurations. Section \sref{section_generalization_cluster_algebras} reviews ordinary cluster algebras and discusses the search for a generalization of these structures based on $2d$ (0,2) quivers as an additional motivation our work. Section \sref{section_Q111} reviews one of the toric phases for D1-branes over $Q^{1,1,1}$, which is the main example used throughout the paper to illustrate our ideas. Section \sref{section_periodic_cascade_Q111} discusses a periodic cascade for this theory and Section \sref{section_flavored_cascade} investigates its flavored version. In Section \sref{section_crystal_algorithm_periodic_quivers}, we introduce an efficient algorithm for generating crystals based on periodic quivers. In Section \sref{section_crystals_Q111} we construct the crystals for various steps in the $Q^{1,1,1}$ cascade and present detailed information about them, including Hasse diagrams, fully refined partition functions and multiplicities of melting configurations. We also investigate the unrefined partition functions and introduce their profile as an interesting object to study. Finally, in Section \sref{section_stable_variables} we introduce a new set of variables that lead to an interesting convergence of partition functions. Moreover, we observe that, when expressed in terms of these variables, the profiles of the partition functions seem to display an interesting behavior at large steps in the cascade. We present our conclusions in Section \sref{section_conclusions}.

%=================================================================
\section{Toric CY$_4$ Crystals}
%=================================================================

\label{section_CY4_crystals}

In this section we review the statistical model of crystal melting for toric CY 4-folds introduced in \cite{Franco:2023tly}. This model is defined by a flavored quiver. The flavor configuration is determined by the specific system of branes in the CY$_4$ under consideration.\footnote{Our discussion applies to general flavor configurations, regardless of whether they are realized by a brane setup. This more general combinatorial problem is interesting in its own right. When engineered in terms of branes, flavors come from higher-dimensional branes wrapping non-compact cycles inside the CY$_4$.} The quiver theories we consider can be interpreted either as $2d$ $(0,2)$ gauge theories on D1-branes or as $\mathcal{N}=2$ quantum mechanics on D0-branes, extended by higher-dimensional flavor branes. We will freely switch between these two interpretations. The crystal can be defined both in terms of the periodic quiver and the brane brick model. We now discuss the periodic quiver construction and refer readers to \cite{Franco:2023tly} for more details (see also \cite{Franco:2015tya} for background on brane brick models). This construction applies to general toric CY 4-folds and flavor configurations.

%=================================================================
\subsection{The Unmolten Crystal}
%=================================================================

The starting point in the construction is the universal cover of the periodic quiver for the CY$_4$ under consideration, which we denote $\tilde{Q}$.\footnote{Generically, there can be multiple periodic quivers, or equivalently brane brick modes, for a toric CY$_4$. They correspond to the so-called {\it toric phases} and are related by triality \cite{Gadde:2013lxa}.} The crystal has one type of atom for every gauge node in the original quiver. We label each type of atom with an index $i$, with $i=1,\ldots, G$, where $G$ is the number of gauge nodes. We denote the $3$-dimensional space in which $\tilde{Q}$ lives as {\it quiver space} and assign coordinates $(x,y,z)$ to it.

In general, the configuration of flavor fields consists of $N_q$ incoming chirals $q_{i}$, $N_{\tilde{q}}$ outgoing chirals $\tilde{q}_{j}$ and $N_{\Psi}$ Fermis $\Psi_{k}$, where $i$, $j$ and $k$ indicate the nodes in the quiver to which the flavors are connected and $N_q,N_{\tilde{q}},N_{\Psi} \geq 0$.\footnote{We denote as flavor a field that transforms in the (anti)fundamental representation of the gauge nodes in the quiver, as opposed to bifundamental or adjoint fields. We can think about flavors as bifundamental fields connecting a gauge node and a global symmetry node.} Flavors can participate in $J$- and $E$-terms, represented by gauge invariant terms of the general forms:
\beq
q_{i} \mathcal{O}_{i,j} \Psi_{j}  \ \ , \ \ \overline{\Psi}_{i} \mathcal{O}_{i,j} \tilde{q}_j \ \ , \ \ q_{i} \Phi_{i,j} \tilde{q}_j \ \ , \ \  q_{i} \overline
{\Phi}_{i,j} \tilde{q}_j
\eeq
where $\mathcal{O}_{i,j}$ and $\Phi_{i,j}$ are operators made of D0-D0 fields.\footnote{More generally, to include general configurations that do not necessarily come from branes, we should understand the term D0-D0 fields as a synonym of fields in the unflavored quiver.} The $\mathcal{O}_{i,j}$ operators contain only chiral fields, the $\Phi_{i,j}$ operators contain one Fermi and possibly chirals, and the $\overline{\Phi}_{i,j}$ operators consist of one conjugate Fermi and possibly chirals. These interactions should be added to the original $J$- and $E$-terms that involve exclusively D0-D0 fields, i.e. those encoded in the periodic quiver/brane brick model.

We define the crystal such that every atom in it is in one-to-one correspondence with an oriented path of chiral fields in $\tilde{Q}$ starting from an incoming chiral flavor $q_i$ modulo $J$- and $E$-term relations of both the D0-D0 Fermi fields and the Fermi flavors $\Psi_j$. Vanishing $J$- and $E$-term relations lead to equivalences between paths in the periodic quiver with the same endpoints. In other words, atoms can generically be reached in multiple, equivalent ways.

The crystal is built out of atoms stacked on top of the nodes of $\tilde{Q}$. The crystal has a fourth dimension, that we denote the {\it depth}. Chiral fields determine the relative depth of the atoms connected by them. If there is a chiral arrow from atom $i$ to atom $j$, atom $j$ is at a higher depth than $i$. We can think about two such atoms as partially overlapping. The $(x,y,z)$ position of an atom is the one of the corresponding node in $\tilde{Q}$, while the depth is proportional to the $R$-charge of the corresponding chiral operator.\footnote{Any valid $R$-charge assignment works. The purpose of the the fourth dimension is to distinguish operators that differ by closed loops.} Modulo $J$- and $E$-term constraints, every oriented path $\gamma_{i_0,j}$ from a top atom $i_0$ defining an atom $j$ of the crystal can be expressed as 
\beq
\gamma_{i_0,j}=v_{i_0,j} \omega^n
\label{general_form_path_atom}
\eeq
where $v_{i_0,j}$ is a shortest path connecting $i_0$ to $j$, $\omega$ is the closed loop associated to a chiral cycle, and $n\geq 0$. Chiral cycles were introduced in \cite{Franco:2019bmx} and they are defined as follows. Consider the $J$- and $E$-terms associated to a Fermi field $\Lambda_a$
\beq
\begin{array}{cl}
J_a & = J_a^+ - J_a^- \\[.75mm]
E_a & = E_a^+ - E_a^-
\end{array}
\eeq
where $J_a^\pm$ and $E_a^\pm$ are monomials in chiral fields and we have made the toric structure of the  $J$- and $E$-terms \cite{Franco:2015tna}, consisting of two monomials with opposite signs, explicit. Their product takes the form
\beq
J_a E_a=J_a^+E_a^+-J_a^+E_a^- -J_a^-E_a^+ + J_a^-E_a^- \, .
\eeq
We refer to each of the terms on the right as a chiral cycle. From the perspective of the periodic quiver, chiral
cycles are ``minimal" closed oriented loops of chiral fields. As explained in \cite{Franco:2023tly}, all chiral cycles are equivalent modulo vanishing $J$- and $E$-terms, and therefore can be identified with a single variable $\omega$, as done in \eqref{general_form_path_atom}. We can interpret $v_{i_0,j}$ as defining an atom at the top layer of the crystal. An atom with an additional factor of $\omega^n$, is located directly below, $n$ levels down.

The resulting crystals can be regarding as discretized version of the underlying toric CY 4-fold. General flavor configurations can correspond to ``resolutions", in which certain cycles grow to finite size. CY$_3$ examples showing analogous behavior can be found in \cite{Chuang:2009crq,Eager:2011ns}.

%=================================================================
\subsection{Finite and Infinite Crystals}
%=================================================================

\label{section_finite_and_infinite_crystals}

Depending on the relation between $N_\Psi$ and $N_q$, the resulting crystals can be finite or infinite. For example, if $N_\Psi > N_q$, the number of relations coming from the $J$- and $E$-terms for $\Psi_j$ fields exceeds the number of $q_i$ fields. As a result, the chiral operators associated to atoms are truncated, resulted in a finite crystal. The explicit examples in Sections \sref{section_flavored_cascade} and \sref{section_crystals_Q111} will provide additional details on this process. Conversely, if $N_\Psi < N_q$, we obtain infinite crystals. Finally, the case $N_\Psi= N_q$ typically results in infinite, but effectively lower dimensional, crystals. This means that the crystal extends in less than four dimensions. One way in which this arises is whenever the full quiver theory, including the flavors, is obtained from a 4- or higher-dimensional theory via dimensional reduction. For similar phenomena for CY 3-folds, which are related to different stability chambers, see \cite{Chuang:2009crq,Eager:2011ns}.

%======================================================
\subsection{Melting Configurations}
%======================================================

\label{section_melting_configurations}

We now consider consider {\it molten crystals}, i.e. configurations that are obtained by removing atoms from the unmolten crystal. We will denote any crystal configuration (i.e. molten or not) as $\mathcal{I}_\mu$. We define the corresponding complement $\Omega_\mu$ as the difference between the unmolten crystal and $\mathcal{I}_\mu$, i.e. $\Omega_\mu$ is the set of removed atoms. We will refer to the $\Omega_\mu$ as {\it melting configurations}.

Let us momentarily focus on the unflavored quiver $Q$. Let us denote $Q_0$ and $Q_X$ the sets of nodes and chiral arrows in $Q$, respectively. The set of all open oriented chiral paths in $Q$ gives rise to an algebra $\mathbb{C}[Q_0,Q_X]$, that we will call the {\it chiral path algebra}. Given the ideal of relations coming from vanishing $J$- and $E$-terms
\beq
\mathcal{I}_{J, E} = \langle J_{a}^{+} - J_{a}^{-} = 0, E_{a}^{+} - E_{a}^{-} = 0 \rangle \, ,
\eeq
where $a$ runs over all Fermis, it is natural to define the factor algebra $A=\mathbb{C}[Q_0,Q_X]/ \mathcal{I}_{J, E}$. $A$ consists of the open chiral paths in the unflavored quiver modulo vanishing $J$- and $E$-terms.

Melting configurations are constructed according to the following {\it melting rule}.

\medskip

%=================================================================
\begin{center}
\begin{tabular}{| m{0.85\textwidth}|  }
\hline 
{\bf Melting rule:}  If $\gamma_{i_0,i} \alpha_{i,j}$ is in $\Omega_\mu$ for some $\alpha_{i,j} \in A$, then $\gamma_{i_0,i}$ should also be in $\Omega_\mu$. 
\\ \hline
\end{tabular}
\end{center}
%=================================================================

\medskip

\noindent Heuristically, this means that if an atom is removed in a given melting configuration, then all atoms on top of it must be removed too. More precisely, starting from atom $j$, we can go up the crystal by following the path $\alpha_{i,j}$ in the reverse direction, encountering atom $i$, which should also be removed. 

It is straightforward to see that every molten crystal $\mathcal{I}_\mu$ defines an ideal of $A$. To show this, we consider the contraposition of the melting rule, which implies that for any $\gamma \in \mathcal{I}_\mu$ and any $\alpha \in A$, then $\gamma \alpha$ is also in $\mathcal{I}_\mu$. In simple words, starting from any atom in a molten crystal and moving from it along a path $\alpha \in A$, results in another atom in the molten crystal, i.e. an atom that has not been removed.

Explicit constructions of the unmolten crystal and melting configurations will be presented in Section \sref{section_crystals_Q111}.

%=================================================================
\subsection{Partition Functions}
%=================================================================

\label{section_partition_functions}

Given an unmolten crystal $\mathcal{I}_0$, it is useful to define a partition function encoding its melting configurations. The partition function has a variable $y_i$, $i=1,\ldots,G$, for each type of atom in the crystal, i.e. for every gauge node in the quiver and takes the form
\beq
Z=\sum_{\Omega_\mu} \prod_i y_i^{n^{(\mu)}_i} \, ,
\label{partition_function}
\eeq
where the sum runs over melting configurations $\Omega_\mu$ and $n^{(\mu)}_i$ is the number of atoms of type $i$ in $\Omega_\mu$.

In Section \sref{section_partition_functions}, we will present explicit examples of these partition functions.

%=================================================================
\section{Further Motivation: Towards a Generalization of Cluster Algebras}
%=================================================================

\label{section_generalization_cluster_algebras} 

Cluster algebras \cite{Fomin:2001mwn,Fomin:2016caz} play a central role in modern mathematics and have recently found important applications in physics (see, for example, \cite{Arkani-Hamed:2012zlh,Elvang:2015rqa,Goncharov:2010jf,Golden:2013xva,Golden:2014xqa, Golden:2014pua,Drummond:2014ffa,Goncharov:2011hp,Eager:2011dp,Nagao:2011aa,Terashima:2013fg}). Arguably, the interest in cluster algebras is fueled by their universality. They capture structures that appear in a wide range of contexts, including algebraic geometry (e.g. Grassmannians), triangulations of surfaces, scattering amplitudes, integrable systems, SUSY gauge theories, and toric CY$_3$ crystals, to name a few.

As explained below, the essential ingredients of cluster algebras are ordinary quivers and their mutations. In physics, these correspond to \(4d\) \(\mathcal{N}=1\) gauge theories and Seiberg duality \cite{Seiberg:1994pq}, respectively. Moreover, this class of theories can be engineered using D3-branes probing CY 3-folds. When the CY 3-folds are toric, the associated gauge theories are elegantly encoded by brane tilings \cite{Franco:2005rj,Franco:2005sm}, which underlie the integrable systems and crystal models discussed in the previous paragraph.

It is natural to ask whether $2d$ (0,2) gauge theories (which can be interpreted as graded quivers \cite{Franco:2017lpa}) together with triality give rise to an interesting and physically well-motivated generalization of cluster algebras. Further motivation comes from the fact that the $2d$ (0,2) theories on D1-branes probing toric CY 4-folds are elegantly encoded by brane brick models, which play a role analogous to brane tilings.

This raises two questions: how can we guide or inspire the discovery of such a generalization, and, to do so, can we identify a class of physical systems that is intrinsically described by these new mathematical structures? 

We propose that the partition functions of toric CY$_4$ crystals related by triality provide natural candidates for exploring these questions. Accordingly, we will study sequences of crystals connected by triality. Although our ideas apply to both infinite and finite crystals, we will focus on the latter, as they yield finite, albeit often huge, expressions that can be compared explicitly.

%=================================================================
\subsection{Cluster Algebras}
%================================================================= 

Let us briefly review some of the basic ingredients of cluster algebras. We refer the reader to \cite{Fomin:2001mwn,Fomin:2016caz} for in depth introductions to the field. A {\it labeled seed} is a triple $({\bf Z},{\bf y},B)$ consisting of: \footnote{Our notation for the cluster variables $({\bf Z},{\bf y},B)$ differs from the more standard one $({\bf x},{\bf y},B)$.}
\begin{itemize}
\item ${\bf Z}=(Z_1,\ldots,Z_n)$ a cluster.
\item ${\bf y}=(y_1,\ldots,y_n)$ an $n$-tuple of coefficients.
\item $B = (b_{ij})$ an $n \times n$ matrix. We say that $B$ is \emph{skew-symmetrizable} if there exist positive integers $d_1, \ldots, d_n$ such that $d_i\, b_{ij} = -\, d_j\, b_{ji}$, with no sum over repeated indices. This information can alternatively be encoded, namely it is equivalent to, a quiver. For any two nodes  $i \neq j$, there are $[b_{ij}]_+$ arrows from $i$ to $j$ in the quiver. Cluster algebras have been extended to include superpotentials \cite{Derksen:2010}. Skew-symmetrizability requires all diagonal entries in $B$ to be equal to zero. Such entries would corresponds to arrows starting and ending on the same node of the quiver, namely adjoint fields. This condition can be relaxed in physics, since it is well-known how to deal with quivers with adjoint fields. Notice, however, that mutations on nodes with adjoints are not permitted.
\end{itemize}

%=================================================================
\paragraph{Mutation.} 
%=================================================================
Given a labeled seed, we define the mutation of the seed on node $k$ as follows.

\begin{itemize}
\item {\bf Quiver Mutation.} The matrix $B$ transforms as follows
\beq
b'_{ij} = 
\begin{cases}
-\, b_{ij}, & \text{if } i = k \text{ or } j = k, \\[6pt]
b_{ij} + \text{sgn}(b_{ik})\, [b_{ik} b_{kj}], & \text{otherwise}.
\end{cases}
\label{quiver_mutation_ordinary_cluster}
\eeq
If the quiver is interpreted as defining a $4d$ $\mathcal{N}=1$ gauge theory, this implements the well-known transformation associated to Seiberg duality \cite{Seiberg:1994pq}. Moreover, the superpotential also transforms according to the rules of Seiberg duality.

\item {\bf Coefficients.} The coefficients transform as follows\footnote{The transformation is in fact slightly more general: arrows incoming into $k$ are replaced by outgoing arrows when the exponent of $y^k$ is negative. It is straightforward to incorporate this case, but we omit it here to keep the discussion simple. Furthermore, this situation does not arise in the specific examples that we are interested in (see e.g. \cite{Eager:2011ns}).}
\beq
y_j'=\left\{
\begin{array}{ll}
y_k^{-1}, & \text{if } j = k,\\[7pt]
y_j \displaystyle\prod_{j\to k} y_k, \ \ \ & \text{if } j \neq k.`
\end{array}
\right.
\label{cluster_algebra_transformation_coefficients}
\eeq

\item {\bf Cluster transformation.}
Finally, the cluster variables transform as follows
\beq
Z'_k={\prod_{j\to k} Z_j + y_k \prod_{k\to j} Z_j \over Z_k}
\label{cluster_transformation}
\eeq
\end{itemize}

%=================================================================
\section{$Q^{1,1,1}$ and its Quiver Gauge Theory}
%=================================================================

\label{section_Q111}

Throughout the rest of the paper, we will focus on D-branes probing the real cone over the $7d$ Sasaki-Einstein manifold $Q^{1,1,1}$, since this example provides an ideal playground to illustrate our ideas. The concepts we develop extend in obvious ways to general toric CY 4-folds. 

\begin{figure}[H]
	\centering
	\includegraphics[width=2.8cm]{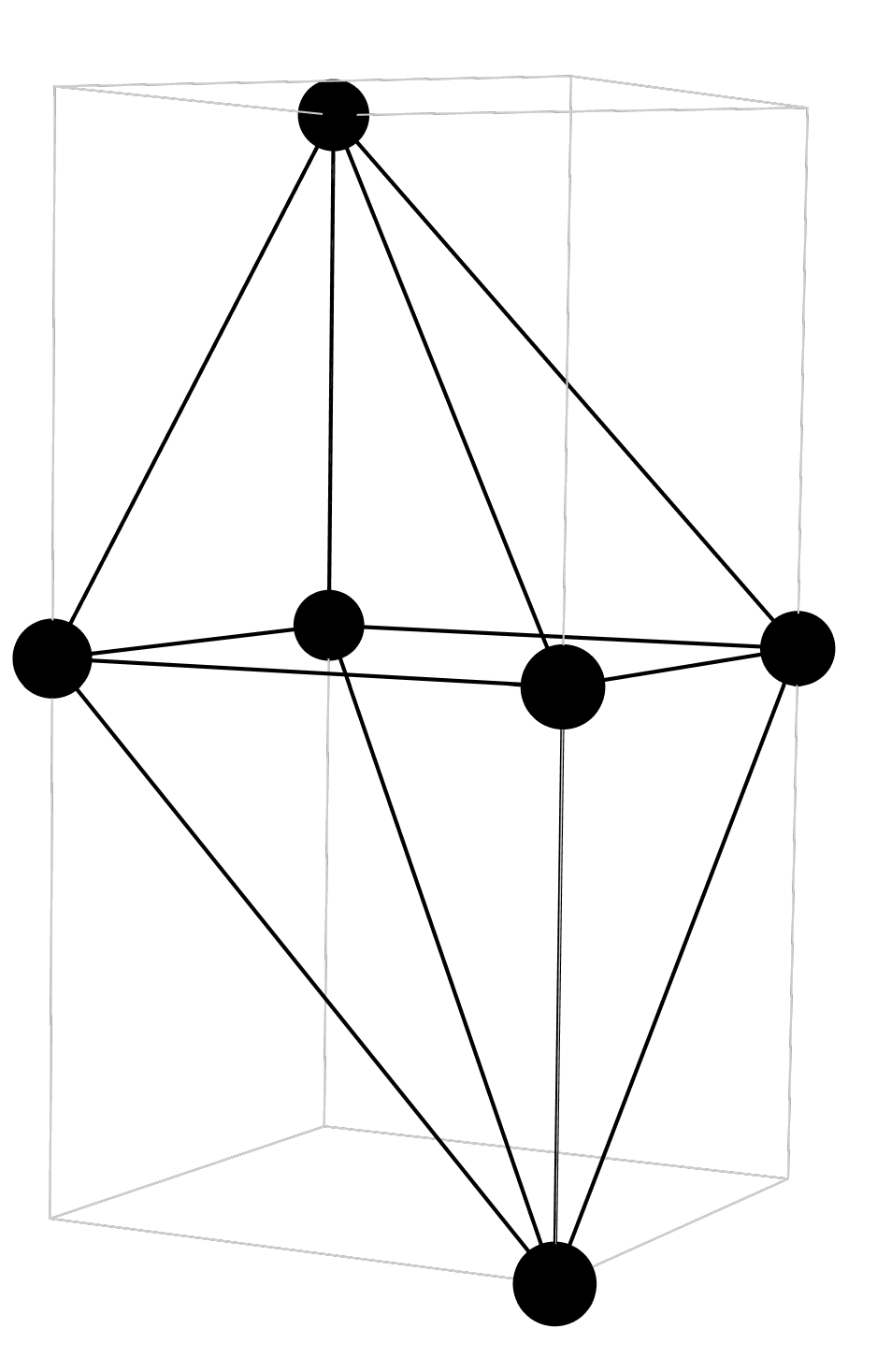}
\caption{Toric diagram for $Q^{1,1,1}$.}
	\label{toric_diagram_Q111}
\end{figure}

$Q^{1,1,1}$ is the homogeneous coset space
\beq
SU(2) \times SU(2) \times SU(2) \over U(1) \times U(1)
\eeq
and has a $U(1)_R \times SU(2)^3$ isometry \cite{D'Auria:1983vy,Nilsson:1984bj,Sorokin:1984ca,Sorokin:1985ap}. It can be expressed as a $U(1)$ fibration over $S^2 \times S^2 \times S^2$. For brevity, we will often refer to the full cone geometry simply as $Q^{1,1,1}$. \fref{toric_diagram_Q111} shows the corresponding toric diagram.

%===================================================================
%===================================================================

\begin{figure}[h]
	\centering
	\includegraphics[width=3.8cm]{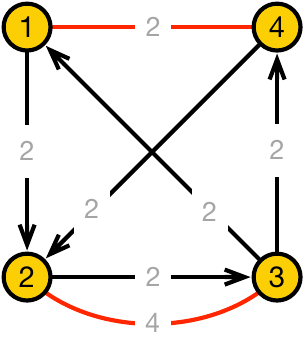}
\caption{Quiver for phase A of $Q^{1,1,1}$.}
	\label{quiver_Q111}
\end{figure}

The $2d$ $(0,2)$ gauge theory on D1-branes probing $Q^{1,1,1}$ has been extensively studied (see .e.g \cite{Franco:2015tna,Franco:2015tya,Franco:2016nwv}). The theory has multiple toric phases. We will consider the phase that was called phase A in \cite{Franco:2016nwv}, whose quiver is shown in \fref{quiver_Q111}.

%===================================================================
%===================================================================

The $J$- and $E$-terms for this theory are
{\small
\beq
\begin{array}{lcccc}
& &  J & &  E
\\
   \Lambda_{41}^{+} & :\ \ \ & X^{+}_{12} \cdot X^{-}_{23} \cdot X^{-}_{34} - X^{-}_{12} \cdot X^{-}_{23} \cdot X^{+}_{34}    & \ \ \ \ &  X^{+}_{42} \cdot X^{+}_{23} \cdot X^{-}_{31} -  X^{-}_{42} \cdot X^{+}_{23} \cdot X^{+}_{31}  
\\
   \Lambda_{41}^{-} & :\ \ \ & X^{-}_{12} \cdot X^{+}_{23} \cdot X^{+}_{34}  - X^{+}_{12} \cdot X^{+}_{23} \cdot X^{-}_{34}   & \ \ \ \ & X^{+}_{42} \cdot X^{-}_{23} \cdot X^{-}_{31} - X^{-}_{42} \cdot X^{-}_{23} \cdot X^{+}_{31}  
\\
   \Lambda_{32}^{++} & :\ \ \ & X^{+}_{23} \cdot X^{-}_{34} \cdot X^{-}_{42} \cdot X^{-}_{23} - X^{-}_{23} \cdot X^{-}_{31} \cdot X^{-}_{12} \cdot X^{+}_{23}   & \ \ \ \ & X^{+}_{34} \cdot X^{®†+}_{42} -  X^{+}_{31} \cdot X^{+}_{12}  
\\
   \Lambda_{32}^{--} & :\ \ \ & X^{+}_{23} \cdot X^{+}_{31} \cdot X^{+}_{12} \cdot X^{-}_{23} - X^{-}_{23} \cdot X^{+}_{34} \cdot X^{+}_{42} \cdot X^{+}_{23}  & \ \ \ \ &  X^{-}_{34} \cdot X^{-}_{42} - X^{-}_{31} \cdot X^{-}_{12}  
\\
    \Lambda_{32}^{+-} & :\ \ \ & X^{-}_{23} \cdot X^{+}_{31} \cdot X^{-}_{12} \cdot X^{+}_{23}  - X^{+}_{23} \cdot X^{-}_{34} \cdot X^{+}_{42} \cdot X^{-}_{23}   & \ \ \ \ &  X^{+}_{34} \cdot X^{-}_{42} - X^{-}_{31} \cdot X^{+}_{12}   
\\
   \Lambda_{32}^{-+} & :\ \ \ & X^{-}_{23} \cdot X^{+}_{34} \cdot X^{-}_{42} \cdot X^{+}_{23} - X^{+}_{23} \cdot X^{-}_{31} \cdot X^{+}_{12} \cdot X^{-}_{23}   & \ \ \ \ &   X^{-}_{34} \cdot X^{+}_{42} - X^{+}_{31} \cdot X^{-}_{12}  
\\ 
   \end{array}
\label{je-q111a}
\eeq
}

\begin{figure}[ht!]
	\centering
	\includegraphics[width=9cm]{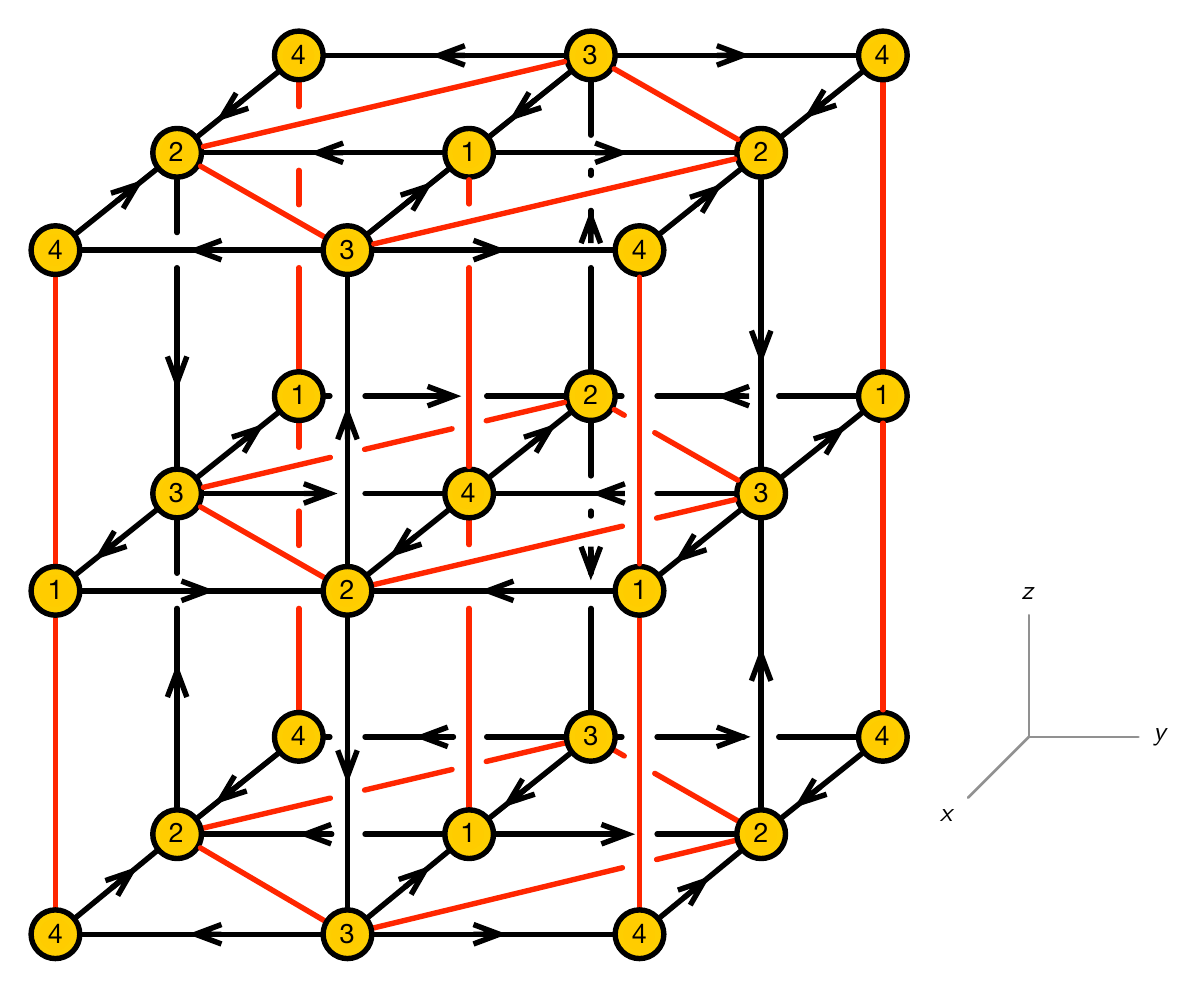}
\caption{Periodic quiver for phase A of $Q^{1,1,1}$.}
	\label{periodic_quiver_Q111}
\end{figure}

The periodic quiver for this theory was first presented in \cite{Franco:2015tya} and is shown in \fref{periodic_quiver_Q111}.

%===================================================================
%===================================================================

%=================================================================
\section{A Periodic Cascade for $Q^{1,1,1}$}
%=================================================================

\label{section_periodic_cascade_Q111}

\begin{figure}[H]
	\centering
	\includegraphics[width=\textwidth]{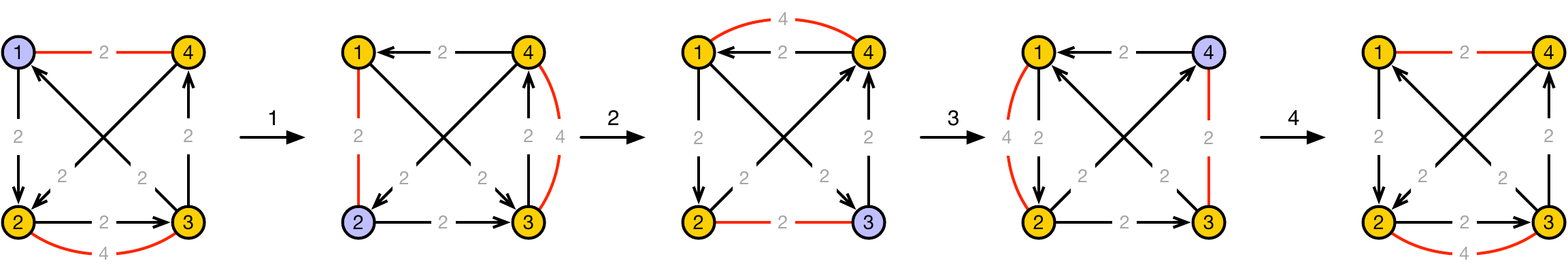}
\caption{A period in a periodic triality cascade for phase A of $Q^{1,1,1}$.}
	\label{cascade_Q111}
\end{figure}

An interesting feature of the theory under consideration is that it admits a periodic triality cascade in which, moreover, every step corresponds to the same toric phase. This cascade is shown in \fref{cascade_Q111}, where the node on which triality acts at each step is indicated in blue. Remarkably, after each triality, we recover the same quiver up to a 90$^\circ$ counterclockwise rotation While, for brevity, we do not write down the $J$- and $E$-terms at each step, they are simply given by the appropriate rotation of \eqref{je-q111a}. After four consecutive trialities, one on each node of the quiver, the cascade returns to the initial theory and keeps repeating itself. Further studies of sequences of trialities and triality cascades have appeared in \cite{Closset:2018axq,Franco:2020ijt,Carcamo:2025het}.

%===================================================================
%===================================================================

%=================================================================
\section{Adding Flavor to the Cascade}
%=================================================================

\label{section_flavored_cascade}

\begin{figure}[ht]
	\centering
	\includegraphics[width=3.8cm]{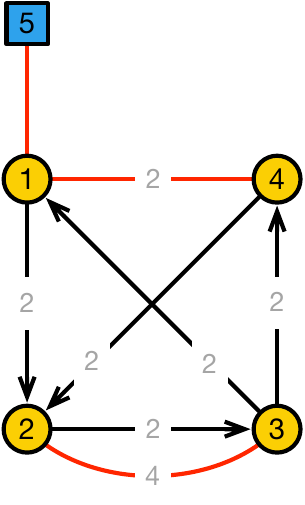}
\caption{Starting point for the flavored cascade of $Q^{1,1,1}$.}
	\label{quiver_Q111_1_Fermi}
\end{figure}

Since we are interested in constructing crystals, the next step is to add flavors to this theory. There is an infinite number of possible flavor configurations, both in terms of flavor fields and the corresponding $J$- and $E$-terms. Since, as stated in Section \sref{section_generalization_cluster_algebras}, we are particularly interested in finite crystals, the general arguments of Section \sref{section_finite_and_infinite_crystals} indicate that we should include more Fermi flavors than chiral flavors. A minimal way to achieve this, and thus a natural starting point, is to add a single Fermi field, which we choose to connect to node 1, as shown in \fref{quiver_Q111_1_Fermi}. We have singled out this node, since it is the one that triality acts on to take us to the next step in the cascade. Node 4 plays an analogous role if we cascade using inverse triality. The uniqueness of this configuration is further enhanced by the fact that it does not admit any $J$- or $E$-terms, leaving no freedom in their choice.

We can think about flavors as bifundamental fields connected to a global symmetry node, shown in blue in \fref{quiver_Q111_1_Fermi}, although this is not necessary.

%===================================================================
%===================================================================

Next, we study the cascade in \fref{cascade_Q111} once flavors are added. Triality has two interesting effects. First of all, it modifies the flavor configuration as a result of the following three processes: changing the types of the existing fields charged under the dualized gauge group (incoming chiral $\to$ outgoing chiral, outgoing chiral $\to$ Fermi, and Fermi $\to$ incoming chiral), introducing new flavors corresponding to mesons, and integrating out massive flavors. Moreover, these new flavors come with their $J$- and $E$-terms, which are dictated by the rules of triality (for a detailed discussion, see e.g. \cite{Franco:2017lpa}). We have worked out the theories at arbitrary steps in the cascade and present the results below. To simplify their discussion, it is convenient to discuss even and odd steps separately.

%=================================================================
\subsection{Even Steps}
%=================================================================

\begin{figure}[H]
	\centering
	\includegraphics[width=4.3cm]{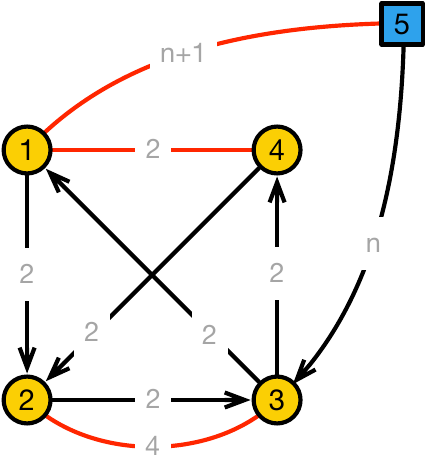}
\caption{Quiver at step $2n$ in the cascade, with $n=0,1,\ldots$.}
	\label{quiver_Q111_flavors_even_steps}
\end{figure}

\fref{quiver_Q111_flavors_even_steps} shows the quiver at an even step $2n$ in the cascade, with $n=0,1,\ldots$. To unify the discussion for arbitrary $n$, we have undone the 90$^\circ$ rotation of the quiver at every step, which is shown in \fref{cascade_Q111}. It is straightforward to return to the rotated quivers. The theory has $n$ incoming chiral flavors $q_{53}^{(i)}$, $i=1,\ldots,n$. Therefore, the number of top atoms in the resulting crystal is $N_{top}=n$, as we will see in Section \sref{section_crystals_Q111}. In addition, there are $n+1$ Fermi flavors, $\Lambda_{15}^{(j)}$, $j=1,\ldots,n$. The difference between the number of Fermi and chiral flavors remains equal to 1 throughout the entire cascade.

%===================================================================
%===================================================================

To fully specify the flavor configuration, we need to determine the $J$- and $E$-terms. In addition to the original terms in \eqref{je-q111a}, we have the following flavor couplings
\beq
\begin{array}{rl}
W_{flav}= - & \Lambda_{15}^{(1)} \cdot q_{53}^{(1)}  \cdot X_{31}^- \\[1.5 mm]
+ & \sum_{i=1}^n \Lambda_{15}^{(i)} \cdot \left(q_{53}^{(i-1)}  \cdot X_{31}^+ - q_{53}^{(i)}  \cdot X_{31}^-\right) \\[1.5 mm]
+ & \Lambda_{15}^{(n+1)} \cdot q_{53}^{(n)}  \cdot X_{31}^+
\end{array}
\label{W_flavor_even_steps}
\eeq
where, for convenience, we have used a plaquette notation, in which the $J$- and $E$-terms are combined into a single ``superpotential.'' This is achieved by completing each term into a gauge invariant by appropriately including the corresponding Fermi or its conjugate. The explicit form of \eqref{W_flavor_even_steps} depends on, or equivalently fixes, our choice of what is meant by $q_{53}^{(1)}$ and $q_{53}^{(n)}$. In the convention we use for the orientation of the Fermi flavors, all their couplings are $J$-terms.

Vanishing of the $J$-terms for the Fermi flavors $\Lambda_{15}^{(j)}$, $j=1,\ldots,n+1$, leads to the following equations
\beq
\begin{array}{ccc}
q_{53}^{(1)}  \cdot X_{31}^- = 0 & & \\[1.5 mm]
q_{53}^{(i-1)}  \cdot X_{31}^+ = q_{53}^{(i)}  \cdot X_{31}^-  & \ \ \ \ & i=2,\ldots,n \\[1.5 mm]
q_{53}^{(n)}  \cdot X_{31}^+=0
\end{array}
\label{relations_flavor_even_steps}
\eeq
The relations in the middle state the equivalence of certain atoms that can be reached by paths that start at consecutive chiral flavors. As we consider atoms at higher depth ($R$-charge), atoms can be reached by paths originating from more distant flavors as a result of the path relations.

\paragraph{A comment on anomalies.} Looking at the quiver in \fref{quiver_Q111_flavors_even_steps}, we observe that gauge nodes are generically anomalous if we assume that they all have the same rank. This is not an issue for the problem we are focusing on, namely for building crystals from these quivers. Moreover, if we assume that the ranks of all gauge groups remain equal along the cascade, the anomaly of node 5 remains constant, but the anomalies of other nodes obviously change. This seems to be in tension with the expectation that triality conserves anomalies (for both dualized and spectator nodes). However, for anomalous quivers, anomaly need not be preserved even if we transform the ranks of nodes appropriately. Take for example the first step of this cascade, with starting point given by \fref{quiver_Q111_1_Fermi}. The new rank of node 1 depends only on the incoming chirals and its original rank. If we set the ranks of all gauge nodes equal to begin with, which would make the original node 1 anomalous, its anomaly clearly changes, since the contribution of flavors goes from one Fermi to one chiral (see the quiver for odd step for $n=1$ and apply the obvious renaming of nodes). An analogous issue with anomalies is present in the CY$_3$ case (see e.g. \cite{Chuang:2009crq,Eager:2011ns}). In these cases, we think about triality/Seiberg duality more as a change of basis, as in the discussion in \cite{Chuang:2009crq}.

%=================================================================
\subsection{Odd Steps}
%=================================================================

\begin{figure}[h]
	\centering
	\includegraphics[width=4.3cm]{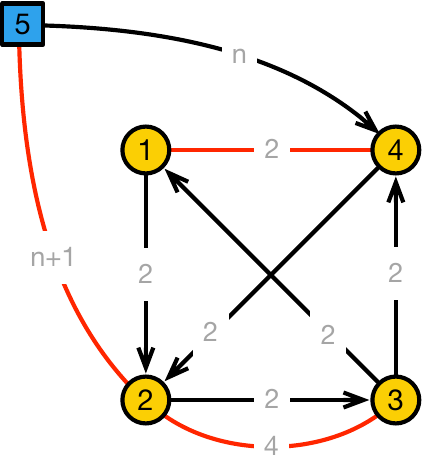}
\caption{Quiver at step $2n-1$  in the cascade, with $n=1,2,\ldots$.}
	\label{quiver_Q111_flavors_odd_steps}
\end{figure}

\fref{quiver_Q111_flavors_odd_steps} shows the quiver at an odd step $2n-1$ in the cascade, with $n=1,2,\ldots$. Once again, we have undone the 90$^\circ$ rotation of the quiver at every step. The theory has $n$ incoming chiral flavors $q_{54}^{(i)}$, $i=1,\ldots,n$. Therefore, the number of top atoms in the crystal is $N_{top}=n$. In addition, there are $n+1$ Fermi flavors, $\Lambda_{25}^{(j)}$, $j=1,\ldots,n+1$.

%===================================================================
%===================================================================

In addition to the original $J$- and $E$-terms in \eqref{je-q111a}, we have the following flavor couplings
\beq
\begin{array}{rl}
W_{flav}= - & \Lambda_{25}^{(1)} \cdot q_{54}^{(1)}  \cdot X_{42}^- \\[1.5 mm]
+ & \sum_{i=1}^n \Lambda_{15}^{(i)} \cdot \left(q_{54}^{(i-1)}  \cdot X_{42}^+ - q_{54}^{(i)}  \cdot X_{42}^-\right) \\[1.5 mm]
+ & \Lambda_{25}^{(n+1)} \cdot q_{54}^{(n)}  \cdot X_{42}^+
\end{array}
\label{W_flavor_odd_steps}
\eeq

Vanishing of the $J$-terms for the Fermi flavors $\Lambda_{25}^{(j)}$, $j=1,\ldots,n+1$, leads to the following equations
\beq
\begin{array}{ccc}
q_{54}^{(1)}  \cdot X_{42}^- = 0 & & \\[1.5 mm]
q_{54}^{(i-1)}  \cdot X_{42}^+ = q_{54}^{(i)}  \cdot X_{42}^-  & \ \ \ \ & i=1,\ldots,n \\[1.5 mm]
q_{54}^{(n)}  \cdot X_{42}^+ =0
\label{relations_flavor_odd_steps}
\end{array}
\eeq

%=================================================================
\section{Efficient Generation of Crystals with Periodic Quivers}
%=================================================================

\label{section_crystal_algorithm_periodic_quivers}

While the definition of a crystal is rather straightforward, its practical implementation may be challenging. In particular, for a long chiral operator in the quiver emanating from a framing flavor, determining whether it corresponds to an atom in the crystal or vanishes may require repeatedly deforming the path using $J$- and $E$-term equivalence relations until the fundamental vanishing conditions, such as the ones in the first and last lines of \eqref{relations_flavor_even_steps} and \eqref{relations_flavor_odd_steps}, become manifest. Even with computer assistance, identifying the appropriate strategy for deforming the path in a useful way can be highly nontrivial. In this process, it is also important to keep track of the order in which fields are concatenated.

Remarkably, the periodic quiver, or more precisely the $4d$ crystal space, trivializes this problem, since $J$- and $E$-term equivalences are built in. Two operators are simply equivalent if they correspond to the same $4d$ vector! This observation leads to an efficient algorithm for constructing crystals, in which fields are translated into vectors in crystal space, making the method amenable to computer implementation, as described below.

%=================================================================
\subsection{An Algorithm to Build Crystals from Vectors} 
%=================================================================

\label{section_crystals_from_vectors}

Below we summarize the steps of the algorithm. To simplify our discussion, we will classify fields as either {\it flavors} or {\it quiver fields}, with the latter denoting fields stretching between pairs of gauged nodes.

\begin{enumerate}

\item Assign a $4d$ vector $\mathcal{X}_i$ to every chiral quiver field $X_i$ according to the prescription outlined in Section \sref{section_CY4_crystals}. This is done for both flavor and quiver chiral fields.

\item Assign a $4d$ vector to each of the framing flavors $q_{a_i}^{(i)}$, $i=1,\ldots,N_{top}$. By convention, we set their $R$-charge to zero. The relative positions of these vectors in quiver space are determined by the flavor $J$- and $E$-terms and the vectors in step 1. For simplicity, in this paper we restrict to examples in which all flavors are charged under the same node $a$ of the quiver, so that $a_i=a$ for all $i$. We will adopt this assumption throughout, noting that our discussion extends straightforwardly to general $a_i$. This step corresponds to fixing the positions of the top atoms in the crystal. 

\item Assign a $4d$ vector to each of the top vanishing atoms. These atoms correspond to the minimal vanishing paths, which in the example under consideration can be found in the top and last lines of \eqref{relations_flavor_even_steps} and \eqref{relations_flavor_odd_steps}. The top atoms form what was called the {\it spine} of the crystal in \cite{Eager:2011ns}. The atoms at this step are at the endpoints of the crystal's spine.

\item Generate all possible oriented paths consisting of concatenations of $n_\chi$ chiral quiver fields, for $n=1,\ldots,n_{max}$.\footnote{Of course, the number of fields in an operator is not a fundamental quantity. Some operators can correspond to equivalent combinations involving different numbers of chiral fields.} Here, $n_{max}$ is the maximum number of chiral fields in the operators we consider and, as we explain below, will be set high enough so the construction converges. The generation of the paths can be efficiently implemented as follows. The paths starting from node $a$ and ending on node $b$, are simply given by $(A^{n_\chi})_{ab}$, with $A$ the adjacency matrix. In our examples, $a$ will be the node type for the top atoms and $b$ will run over the different nodes in the quiver, which in turn correspond to the different types of atoms. This step generates the operators as products of chiral fields. We then translate these operators into vectors. Schematically, this is done as follows
\beq
\prod_i X_i \to \sum_i \mathcal{X}_i \, .
\eeq

In principle, since the chiral fields in an operator correspond to matrices, they should be multiplied in an order determined by their gauge indices. However, this relative ordering becomes unimportant once we replace each of them by the corresponding vectors, a key advantage of this approach. This ordering also becomes unimportant once we think about each node in cover of the periodic quiver as Abelian. Even with this simplification, mapping fields to vectors has the advantage of automatically taking into account relations.

\end{enumerate}

\noindent We will next construct three sets of atoms, which we refer to as {\it primary}, {\it vanishing} and {\it final}. They are defined as follows:

\begin{enumerate}
\setcounter{enumi}{4}

\item {\bf Primary Atoms:} Generate the vectors associated to primary atoms by adding the paths in Step 4, to each of the vectors for the top atoms in Step 2. These are the atoms associated to all possible paths starting from the framing flavors.

\item {\bf Vanishing Atoms:} Generate the vectors associated to vanishing atoms by adding the paths in Step 4, to each of the vectors for the top vanishing atoms in Step 3. These are the atoms associated to to paths that vanish due to the vanishing relations coming from the flavors.

\item {\bf Final Atoms:} The final atoms are those that remain after removing those atoms that are both primary and vanishing.
Being in both sets would mean that the path can be deformed to contain the fundamental vanishing path.
\end{enumerate}
The crystals studied in this paper have a finite number of atoms. The process above is repeated for increasing values of the cutoff $n_{max}$ until we can confirm convergence. In the next section we illustrate these ideas for $Q^{1,1,1}$.

%=================================================================
\section{Crystals for $Q^{1,1,1}$}
%=================================================================

\label{section_crystals_Q111}

In this section, we apply the algorithm introduced in the previous section to construct the crystals for $Q^{1,1,1}$ corresponding to the quivers and triality cascade in Section \sref{section_periodic_cascade_Q111}.

%=================================================================
\subsection*{\underline{Quiver Fields}} 
%=================================================================

From the periodic quiver in \fref{periodic_quiver_Q111}, we determine the following $4d$ vectors for the quiver fields
\beq
\begin{array}{ccc}
X_{12}^+ & \ \ & (0, 1, 0, 1/2) \\
X_{12}^- & & (0, -1, 0, 1/2) \\[1mm]
X_{42}^+ & & (1, 0, 0, 1/2) \\[1mm]
X_{42}^- & & (-1, 0, 0, 1/2) \\[1mm]
X_{34}^+ & & (0, 1, 0, 1/2) \\[1mm]
X_{34}^- & & (0, -1, 0, 1/2) \\[1mm]
X_{31}^+ & & (1, 0, 0, 1/2) \\[1mm]
X_{31}^- & & (-1, 0, 0, 1/2) \\[1mm]
X_{23}^+ & & (0, 0, 1, 0) \\[1mm]
X_{23}^- & & (0, 0, -1, 0) 
\end{array}
\label{vectors_quiver_fields_Q111}
\eeq
where the fourth coordinate is a possible $R$-charge assignment.\footnote{To keep notation simple, from now on we drop the calligraphic font, and use the same one to denote fields and the corresponding vectors. Which of them we are referring to should be clear from the context.}

\smallskip

%=================================================================
\subsection*{\underline{Top atoms}} 
%=================================================================

In what follows, we set set $(0,0,0)$ in quiver space to be a node of type 3. We use this for both the even and odd step crystals. Following the discussion in Section \sref{section_flavored_cascade}, we need to distinguish the even and odd steps in the cascade.

%=================================================================
\paragraph*{Even steps.} 
%=================================================================

To determine the relative positions of the framing flavors, we consider the flavor superpotential \eqref{W_flavor_even_steps} and the resulting relations \eqref{relations_flavor_even_steps}.
\beq
q_{53}^{(i-1)}  \cdot X_{31}^+ = q_{53}^{(i)}  \cdot X_{31}^- \, ,
\eeq
for $i=2,\ldots,n$. This leads into the following relation between the corresponding vectors
\beq
q_{53}^{(i)} - q_{53}^{(i-1)}  = X_{31}^+ - X_{31}^- = (2,0,0,0) \, .
\eeq
Then, we can set
\beq
\begin{array}{ccc}
q_{53}^{(i)} & \ \ & ((i-1) 2,0,0,0) \\[1mm]
\end{array}
\eeq
$i=1,\ldots,N_{top}$, where we have used the fact that $N_{top}=n$.

%=================================================================
\paragraph*{Odd steps.} 
%=================================================================

We can proceed similarly for the odd steps in the cascade. From the flavor superpotential \eqref{W_flavor_odd_steps} and the corresponding relations \eqref{relations_flavor_odd_steps}, we have
\beq
q_{54}^{(i-1)}  \cdot X_{42}^+=q_{54}^{(i)}  \cdot X_{42}^- \, ,
\eeq
for $i=1,\ldots,n$. This translates into the following relation between the corresponding vectors
\beq
q_{54}^{(i)} - q_{54}^{(i-1)} = X_{42}^+ - X_{42}^- = (2,0,0,0) \, ,
\eeq
where, in the last step, we have used the vectors in \eqref{vectors_quiver_fields_Q111}. Then, we can set
\beq
\begin{array}{ccc}
q_{54}^{(i)} & \ \ & ((i-1) 2,1,0,0)) \\[1mm]
\end{array}
\eeq
$i=1,\ldots,N_{top}$. The $+1$ in the second coordinate indicates the relative position of some node 4 with respect to the node 3 that we have set as the origin in quiver space.

\bigskip

%=================================================================
\subsection*{\underline{Top vanishing atoms}} 
%=================================================================

The next step is to determine the top vanishing atoms. Once again, we separate the even and odd steps in the cascade.

%=================================================================
\paragraph*{Even steps.} 
%=================================================================
From \eqref{relations_flavor_even_steps}, the top vanishing atoms are given by
\beq
\begin{array}{ccl}
q_{53}^{(1)}  \cdot X_{31}^- & \ \ \ \to \ \ \ & (0,0,0,0) + (-1,0,0,1/2) = (-1,0,0,1/2) \\[1.5 mm]
q_{53}^{(n)}  \cdot X_{31}^+=0 & \ \ \ \to \ \ \ & (2n-2,0,0,0) + (1,0,0,1/2) = (2n-1,0,0,1/2)
\end{array}
\label{relations_top_vanishing_even_steps}
\eeq

%=================================================================
\paragraph*{Odd steps.} 
%=================================================================
From \eqref{relations_flavor_odd_steps}, the top vanishing atoms are given by
\beq
\begin{array}{ccl}
q_{54}^{(1)}  \cdot X_{42}^- & \ \ \ \to \ \ \ & (0,1,0,0) + (-1,0,0,1/2) = (-1,1,0,1/2) \\[1.5 mm]
q_{54}^{(n)}  \cdot X_{42}^+ & \ \ \ \to \ \ \ & (2n-2,1,0,0) + (1,0,0,1/2) = (2n-1,1,0,1/2)
\label{relations_top_vanishing_odd_steps}
\end{array}
\eeq

\bigskip

%=================================================================
\subsection*{\underline{Paths from flavors} }
%=================================================================

All paths from flavors containing a given number $n_\chi$ of chiral fields are generated by taking the $n_\chi$-th power of the adjacency matrix and proceeding as explained in Step 4 of the algorithm in Section \sref{section_crystals_from_vectors}. For this theory, the adjacency matrix is
\beq
A=\left(
\begin{array}{cccc}
 0 & 0 & 0 & X_{14}^- + X_{14}^+ \\
 0 & 0 & 0 & X_{24}^- + X_{24}^+ \\
X_{31}^- + X_{31}^+ & X_{32}^- + X_{32}^+ & 0 & 0 \\
 0 & 0 & X_{43}^- + X_{43}^+ & 0 \\
\end{array}
\right) \, ,
\eeq
It is worth noting that operators with the same number of chiral fields can have different $R$-charges. Moreover, as previously mentioned, operators with different numbers of chirals may be equivalent.

%=================================================================
\subsection{Building the Crystals}
%=================================================================

We now illustrate the construction with the explicit example of step 6 in the cascade, which has $N_{top}=3$ top atoms. We will call
\beq
\begin{array}{ccl}
\mathcal{S}_{prim}(R) & : & \mbox{set of primary atoms with $R$-charge $R$.} \\[1.5mm]
\mathcal{S}_{van}(R) & : & \mbox{set of vanishing atoms with $R$-charge $R$.}
\end{array}
\eeq

For every $R$, the final atoms $\mathcal{S}_{fin}(R)$ are obtained by keeping those primary atoms that are not vanishing, namely
\beq
\mathcal{S}_{fin}(R) = \mathcal{S}_{prim}(R) \backslash \mathcal{S}_{van}(R) \, .
\eeq

\begin{figure}[H]
	\centering
	\includegraphics[width=.9\textwidth]{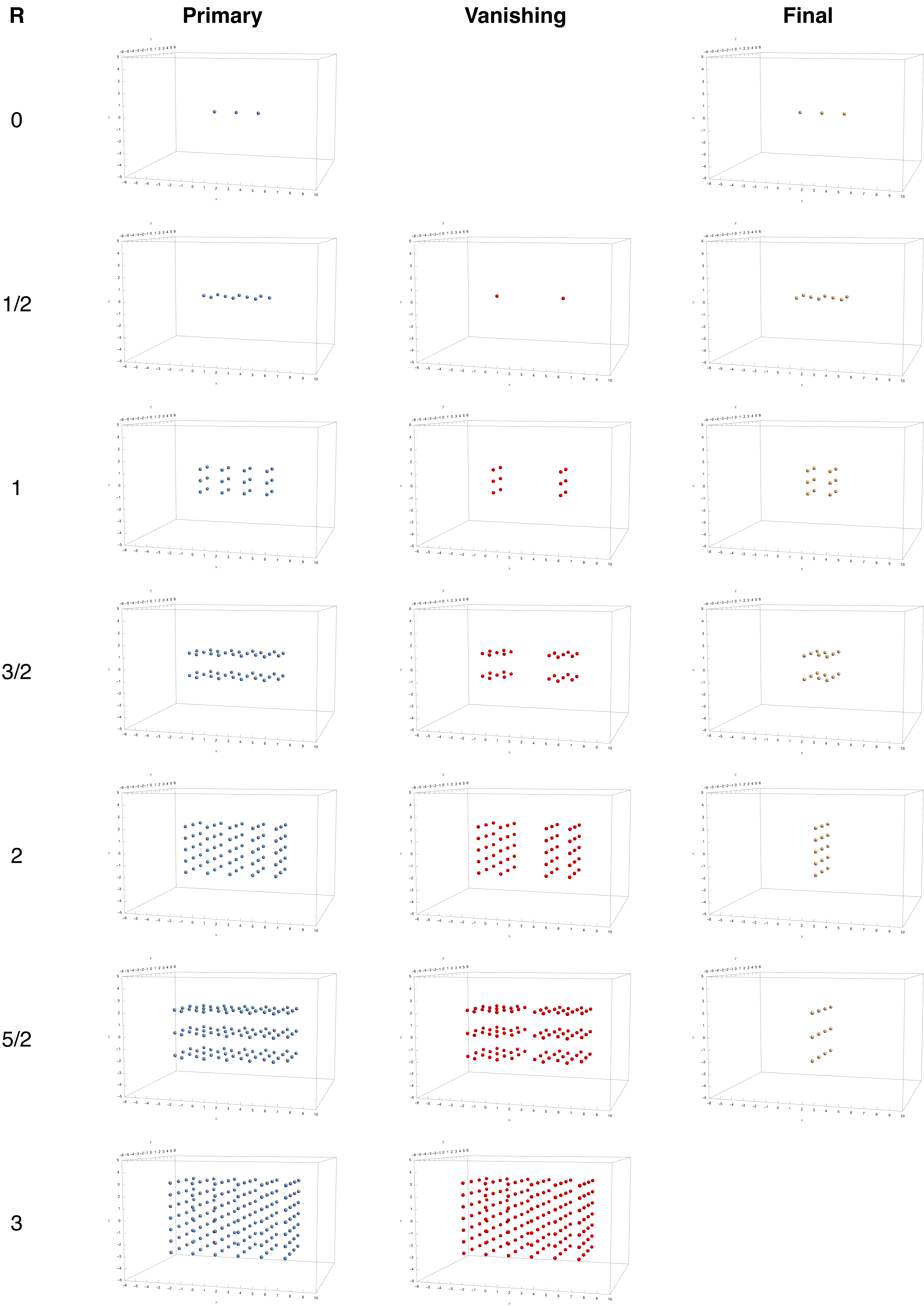}
\caption{Crystal for $Q^{1,1,1}$ at step 6 in the cascade. We show primary, vanishing and final atoms for $R=0,\ldots,3$.}
	\label{Building_crystal_Q111}
\end{figure}

\fref{Building_crystal_Q111} shows the atoms in each of these sets for increasing values of the $R$-charge and provides a detailed perspective on how the crystal is generated. We observe the following behavior:

\paragraph{Primary atoms.} The primary atoms emanate from the framing flavors and expand around them as the depth in the crystal, namely the $R$-charge, increases.

\newpage

\paragraph{Vanishing atoms.} Vanishing atoms emanate from the flavors at the two endpoints of the spine. Each of these two regions grows with depth until they collide and merge. At some depth, the entire sets of primary and vanishing atoms coincide. In this example, this occurs at $R=3$. For depths greater or equal than the merger, the primary and vanishing atoms coincide and no new final atoms are generated. In practice, generating operators for a few values of the $R$-charge beyond the merger is sufficient for convincing ourselves that all final atoms have been found.

\bigskip

From \fref{Building_crystal_Q111}, the distribution of atoms of each type across the different R-charges is as follows:
\beq
\begin{array}{|c|c|c|c|}
\hline
\ \ R \ \ \ & \ \ \mbox{Primary} \ \ \ & \ \ \mbox{Vanishing} \ \ \ & \ \ \mbox{Final} \ \ \ \\ \hline
0 & 3 & 0 & 3 \\ \hline
1/2 & 10 & 2 & 8 \\ \hline
1 & 24 & 12 & 12 \\ \hline
3/2 & 44 & 28 & 16 \\ \hline
2 & 75 &  60 & 15 \\ \hline
5/2 & 114 & 102 & 12 \\ \hline
3 & 168 & 168 & 0 \\  \hline
\end{array}
\eeq

\bigskip

%===================================================================
\paragraph{Overlapping atoms.} 
%===================================================================
There are 66 atoms in this crystal, with a projection onto quiver space consisting of 63 atoms. This is due to three pairs of overlapping atoms, which share the $(x,y,z)$ coordinates, but whose $R$-charges differ by 2. They are:
\beq
\begin{array}{ccc}
(2,0,0,0) & \ \ \ \ \ & (2,0,0,2) \\
(2,-1,0,1/2) & \ \ \ \ \ & (2,-1,0,5/2) \\
(2,1,0,1/2) & \ \ \ \ \ & (2,1,0,5/2)
\end{array}
\eeq

%===================================================================
%===================================================================

%=================================================================
\subsection{Crystalizing by Triality: Hasse Diagrams} 
%=================================================================

\begin{figure}[H]
	\centering
	\includegraphics[width=\textwidth]{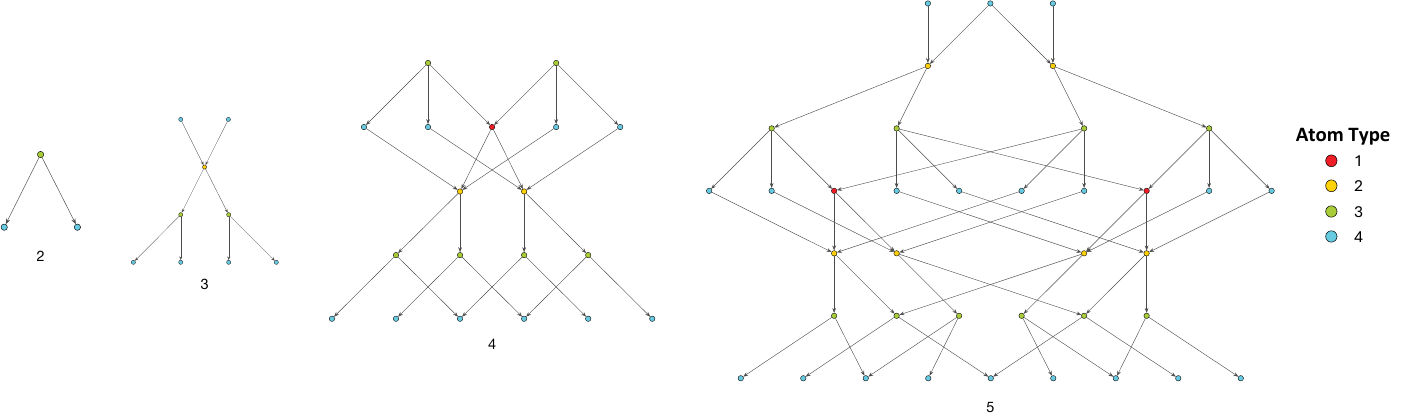}
\caption{Hasse diagrams of the crystals at steps 2 to 5 in the cascade. Each type of node in the quiver is indicated with a different color.}
	\label{Hasse_Q111_2_to_5}
\end{figure}

\begin{figure}[H]
	\centering
	\includegraphics[width=\textwidth]{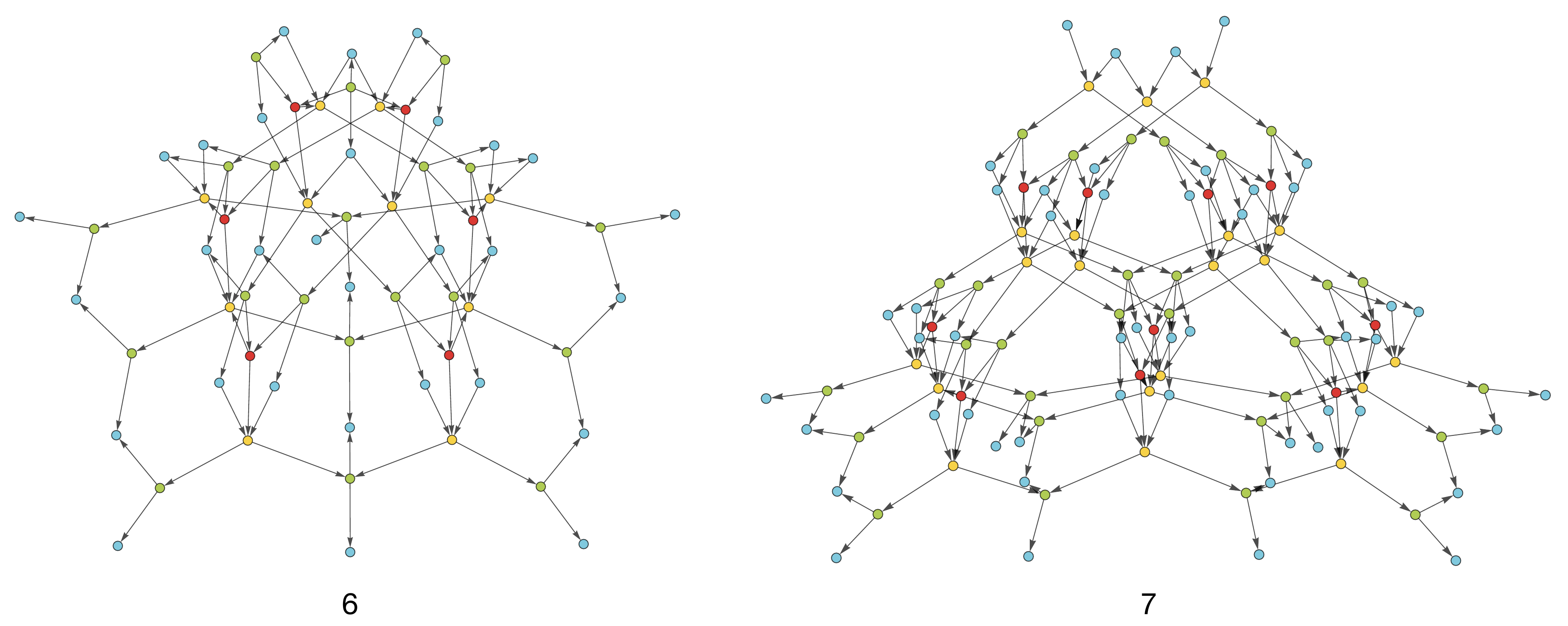}
\caption{Hasse diagrams of the crystals at steps 6 and 7 in the cascade.}
	\label{Hasse_Q111_6_to_7}
\end{figure}

\begin{figure}[H]
	\centering
	\includegraphics[width=\textwidth]{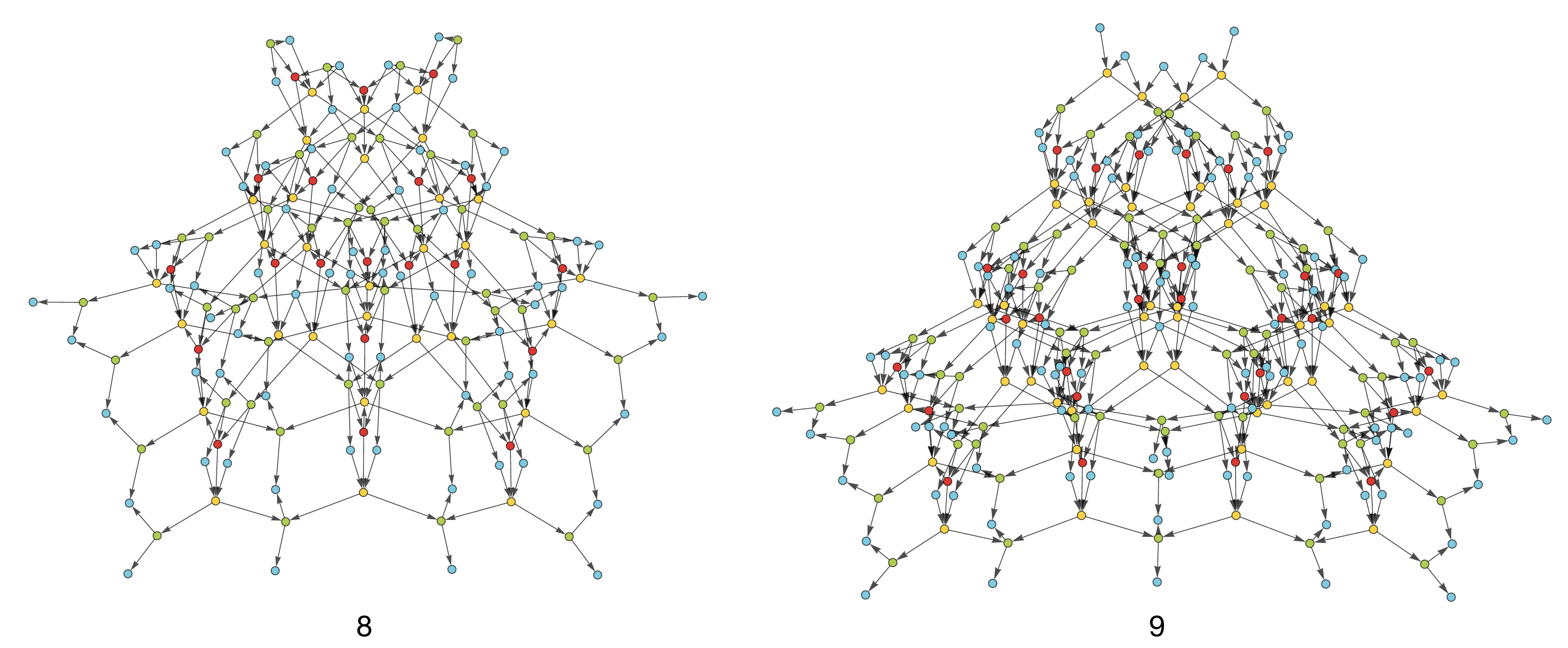}
\caption{Hasse diagrams of the crystals at steps 8 and 9 in the cascade.}
	\label{Hasse_Q111_8_to_9}
\end{figure}

\begin{figure}[H]
	\centering
	\includegraphics[width=.9\textwidth]{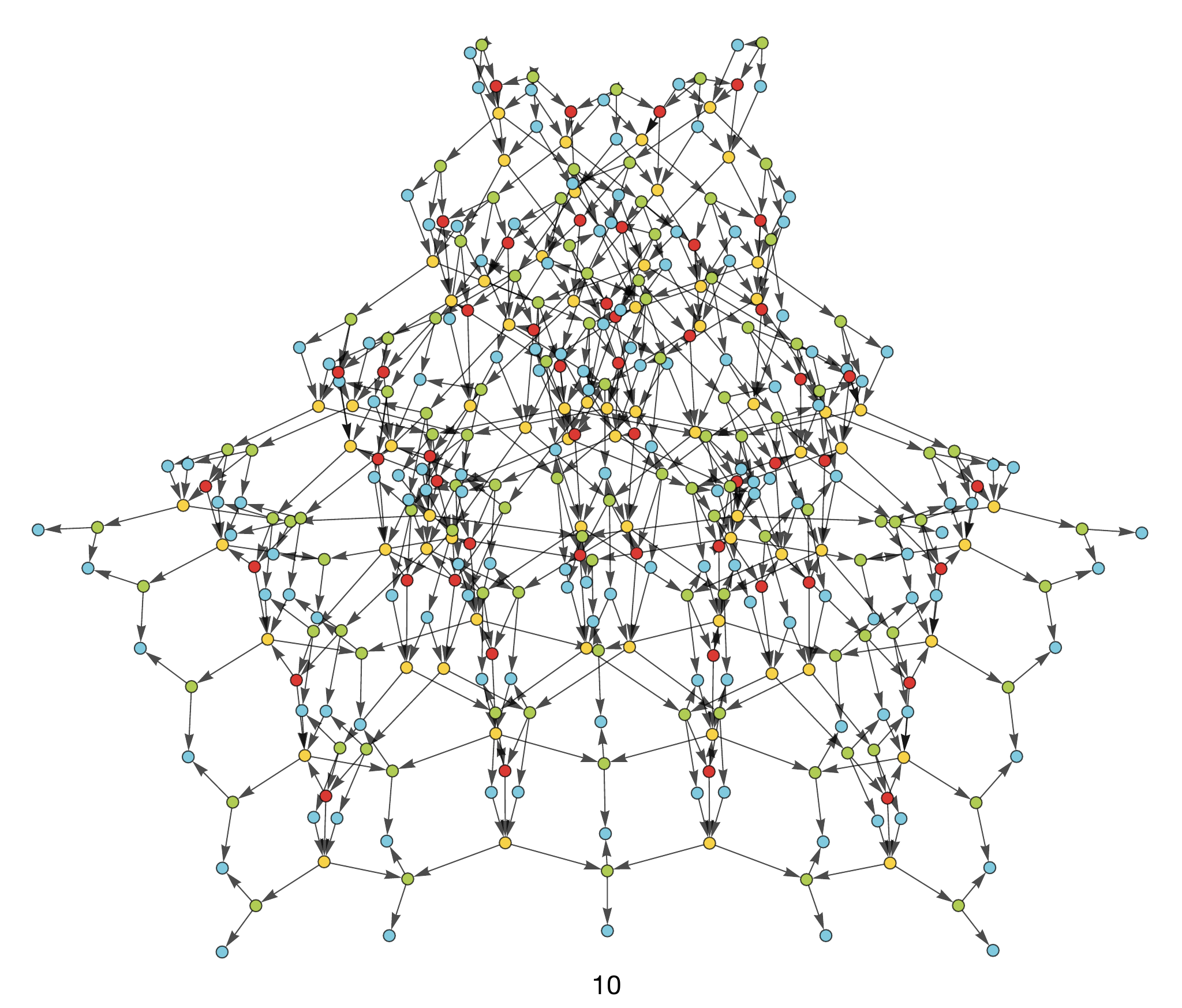}
\caption{Hasse diagram of the crystal at step 10 in the cascade}
	\label{Hasse_Q111_10}
\end{figure}

We have explicitly constructed the crystals for $Q^{1,1,1}$ with this particular flavoring, up to step 10 in the triality cascade. Their beautiful structure can be elegantly encoded in Hasse diagrams, shown in Figures \ref{Hasse_Q111_2_to_5} to \ref{Hasse_Q111_10}. At the first step of the cascade, the crystal consists of a single atom of type 3, so we do not show a diagram for it. Every node in a Hasse diagram represents an atom in the crystal, while an arrow from atom $a$ to atom $b$ corresponds to a chiral field and indicates that $a$ lies above $b$. The top atoms are those without incoming arrows. For further details on Hasse diagrams and how they can be used to determine melting configurations, see \cite{Franco:2023tly}.

%===================================================================
%===================================================================

%===================================================================
%===================================================================

%===================================================================
%===================================================================

%===================================================================
%===================================================================

Below we summarize key information about these crystals: the number of top atoms, the total number of atoms, and the breakdown by atom type.
\beq
\begin{array}{|cccc|}
\hline
\ \mbox{Step} \ & N_{top} & N_{atoms} & (n_1,n_2,n_3,n_4) \\ \hline
1 & 1 & 1 & (0,0,1,0) \\
2 & 1 & 3 & (0,0,1,2) \\
3 & 2 & 9 & (0,1,2,6) \\
4 & 2 & 19 & (1,2,6,10) \\
5 & 3 & 38 & (20,2,6,10)  \\
6 & 3 & 66 & (6,10,20,30) \\  
7 & 4 & 110 & (50,10,20,30) \\
8 & 4 & 170 & (20,30,50,70) \\
9 & 5 & 255 & (105,30,50,70) \\
10 & 5 & 365 & (140, 50, 70, 105) \\ \hline
\end{array}
\eeq
We note some interesting patterns, like the three lowest multiplicities at a given step coinciding with the three highest ones at the previous step.\footnote{If this pattern is correct, it represents an interesting, albeit minor, confirmation of our calculations. This is a useful cross-check, since we calculate the Hasse diagrams for odd and even steps independently, using the corresponding quivers in Section \sref{section_flavored_cascade}. The multiplicities at step $n$ can be inferred from those at steps $n+1$ and $n-1$. For example, from the odd-step multiplicities at steps 7 and 9, which are (10,20,30,50) and (30,50,70,105), respectively (here and in the following we sort them in increasing order), we can infer that the multiplicities at the even step 8 are (20,30,50,70), in agreement with the independent calculation. In this way, the multiplicities at odd steps determine those at even steps, and vice versa.} Interestingly, thes sequence $(0,1,2,6,10,20,30,50,70,105,140,\ldots)$ coincides with the partial sums of the duplicated tetrahedral numbers (see sequence A096338 in \cite{OEIS}, where additional information and a recursive definition are given).\footnote{We are indebted to Gregg Musiker for bringing this observation to our attention.} It would be interesting to understand more deeply why this sequence arises in the present problem.

%=================================================================
\subsection{Partition Functions}
%=================================================================

\label{section_partition_functions}

We will now present the partition functions, defined as in \eqref{partition_function}, for the first six steps of the triality cascade. It is worth emphasizing that these partition functions were explicitly constructed following the general construction outlined in Section \sref{section_melting_configurations}. Specifically, for each crystal, this involved: identifying the atoms in the crystal using the periodic quiver, constructing the associated Hasse diagram, and computing the partition function through a direct application of the melting rule. This approach contrasts, for example, with the recursive computation of partition functions for large crystals associated to toric CY 3-folds, as developed in \cite{Eager:2011ns}, which relies on cluster transformations. Indeed, as stated in Section \sref{section_generalization_cluster_algebras}, one of the goals of this paper is precisely to collect data and establish the foundation for developing analogous recursive techniques for computing partitions for toric CY 4-folds.

%=================================================================
\subsubsection*{Fully Refined Partition Functions}
%=================================================================

The partition functions $Z_j$ at step $j$ in the cascade, fully refined with a variable $y_i$, $i=1,\ldots,4$, for each node of the quiver are presented below.

\beq
\resizebox{\textwidth}{!}{$
\begin{array}{cl}
Z_0 & = 1 \\[.3cm]
%=================================================================
Z_1 & = 1 + y_4 \\[.3cm]
%=================================================================
Z_2 & = 1 + y_3 + 2 y_3 y_4+ y_3 y_4^2 \\[.3cm]
%=================================================================
Z_3 & = 1+ 2 y_4+ y_4^2+ y_2 y_4^2+ 2 y_2 y_3 y_4^2+ y_2 y_3^2 y_4^2+ 4 y_2 y_3 y_4^3+ 4 
y_2 y_3^2 y_4^3+ 2 y_2 y_3 y_4^4+ 6 y_2 y_3^2 y_4^4+ 4 y_2 y_3^2 y_4^5+ y_2 y_3^2 
y_4^6 \\[.3cm]
%=================================================================
Z_4 & = 1+ 2 y_3+ 4 y_3 y_4+ 2 y_3 y_4^2+ y_3^2+ y_1 y_3^2+ 4 y_3^2 y_4+ 4 y_1 y_3^2 y_4+ 
6 y_3^2 y_4^2+ 6 y_1 y_3^2 y_4^2+ 4 y_3^2 y_4^3+ 2 y_1 y_2 y_3^2 y_4^2+ 4 y_1 
y_3^2 y_4^3+ y_3^2 y_4^4 + 4 y_1 y_2 y_3^2 y_4^3
\\[.05cm] &+ y_1 y_3^2 y_4^4+ 2 y_1 y_2 y_3^2 
y_4^4+ y_1 y_2^2 y_3^2 y_4^4+ 4 y_1 y_2 y_3^3 y_4^2+ 12 y_1 y_2 y_3^3 y_4^3+ 12 y_1 
y_2 y_3^3 y_4^4+ 4 y_1 y_2^2 y_3^3 y_4^4+ 4 y_1 y_2 y_3^3 y_4^5+ 4 y_1 y_2^2 y_3^3 
y_4^5+ 2 y_1 y_2 y_3^4 y_4^2
\\[.05cm] & + 8 y_1 y_2 y_3^4 y_4^3+ 12 y_1 y_2 y_3^4 y_4^4+ 6 y_1 
y_2^2 y_3^4 y_4^4+ 8 y_1 y_2 y_3^4 y_4^5+ 14 y_1 y_2^2 y_3^4 y_4^5+ 2 y_1 y_2 y_3^4 
y_4^6+ 10 y_1 y_2^2 y_3^4 y_4^6+ 2 y_1 y_2^2 y_3^4 y_4^7+ 4 y_1 y_2^2 y_3^5 y_4^4+ 
16 y_1 y_2^2 y_3^5 y_4^5
\\[.05cm] & + 24 y_1 y_2^2 y_3^5 y_4^6 + 16 y_1 y_2^2 y_3^5 y_4^7+ 4 
y_1 y_2^2 y_3^5 y_4^8+ y_1 y_2^2 y_3^6 y_4^4+ 6 y_1 y_2^2 y_3^6 y_4^5+ 15 y_1 y_2^2 
y_3^6 y_4^6+ 20 y_1 y_2^2 y_3^6 y_4^7+ 15 y_1 y_2^2 y_3^6 y_4^8+ 6 y_1 y_2^2 y_3^6 
y_4^9+ y_1 y_2^2 y_3^6 y_4^{10} \\[.3cm]
 %=================================================================
 Z_5 & = 1+ 3 y_4+ 3 y_4^2+ 2 y_2 y_4^2+ 4 y_2 y_3 y_4^2+ 2 y_2 y_3^2 y_4^2+ y_4^3+ 2 y_2 
y_4^3+ 12 y_2 y_3 y_4^3+ y_2^2 y_4^3+ 10 y_2 y_3^2 y_4^3+ 4 y_2^2 y_3 y_4^3+ 6 
y_2^2 y_3^2 y_4^3+ 2 y_1 y_2^2 y_3^2 y_4^3+ 4 y_2^2 y_3^3 y_4^3 \\[.05cm] & + 4 y_1 y_2^2 y_3^3 
y_4^3+ y_2^2 y_3^4 y_4^3+ 2 y_1 y_2^2 y_3^4 y_4^3+ y_1^2 y_2^2 y_3^4 y_4^3+ 12 y_2 
y_3 y_4^4+ 20 y_2 y_3^2 y_4^4+ 8 y_2^2 y_3 y_4^4+ 24 y_2^2 y_3^2 y_4^4+ 8 y_1 
y_2^2 y_3^2 y_4^4+ 24 y_2^2 y_3^3 y_4^4+ 24 y_1 y_2^2 y_3^3 y_4^4 \\[.05cm] & + 8 y_2^2 y_3^4 
y_4^4+ 16 y_1 y_2^2 y_3^4 y_4^4+ 8 y_1^2 y_2^2 y_3^4 y_4^4+ 4 y_2 y_3 y_4^5+ 20 
y_2 y_3^2 y_4^5+ 4 y_2^2 y_3 y_4^5+ 36 y_2^2 y_3^2 y_4^5+ 12 y_1 y_2^2 y_3^2 
y_4^5+ 60 y_2^2 y_3^3 y_4^5+ 60 y_1 y_2^2 y_3^3 y_4^5 + 28 y_2^2 y_3^4 y_4^5 \\[.05cm] & + 4 
y_1 y_2^3 y_3^2 y_4^5+ 56 y_1 y_2^2 y_3^4 y_4^5+ 12 y_1 y_2^3 y_3^3 y_4^5+ 28 
y_1^2 y_2^2 y_3^4 y_4^5+ 12 y_1 y_2^3 y_3^4 y_4^5+ 4 y_1^2 y_2^3 y_3^4 y_4^5+ 4 
y_1 y_2^3 y_3^5 y_4^5+ 4 y_1^2 y_2^3 y_3^5 y_4^5+ 10 y_2 y_3^2 y_4^6+ 24 y_2^2 
y_3^2 y_4^6 \\[.05cm] & + 8 y_1 y_2^2 y_3^2 y_4^6+ 80 y_2^2 y_3^3 y_4^6+ 80 y_1 y_2^2 y_3^3 
y_4^6+ 56 y_2^2 y_3^4 y_4^6+ 8 y_1 y_2^3 y_3^2 y_4^6+ 112 y_1 y_2^2 y_3^4 y_4^6+ 
44 y_1 y_2^3 y_3^3 y_4^6+ 56 y_1^2 y_2^2 y_3^4 y_4^6+ 64 y_1 y_2^3 y_3^4 y_4^6+ 
24 y_1^2 y_2^3 y_3^4 y_4^6 \\[.05cm] & + 28 y_1 y_2^3 y_3^5 y_4^6 + 28 y_1^2 y_2^3 y_3^5 y_4^6+ 
2 y_2 y_3^2 y_4^7+ 6 y_2^2 y_3^2 y_4^7+ 2 y_1 y_2^2 y_3^2 y_4^7+ 60 y_2^2 y_3^3 
y_4^7+ 60 y_1 y_2^2 y_3^3 y_4^7+ 70 y_2^2 y_3^4 y_4^7+ 4 y_1 y_2^3 y_3^2 y_4^7+ 
140 y_1 y_2^2 y_3^4 y_4^7 \\[.05cm] & + 60 y_1 y_2^3 y_3^3 y_4^7+ 2 y_1 y_2^4 y_3^2 y_4^7+ 70 
y_1^2 y_2^2 y_3^4 y_4^7+ 140 y_1 y_2^3 y_3^4 y_4^7+ 8 y_1 y_2^4 y_3^3 y_4^7+ 60 
y_1^2 y_2^3 y_3^4 y_4^7+ 84 y_1 y_2^3 y_3^5 y_4^7+ 12 y_1 y_2^4 y_3^4 y_4^7+ 84 
y_1^2 y_2^3 y_3^5 y_4^7 \\[.05cm] & + 6 y_1^2 y_2^4 y_3^4 y_4^7+ 8 y_1 y_2^4 y_3^5 y_4^7+ 14 
y_1^2 y_2^4 y_3^5 y_4^7+ 2 y_1 y_2^4 y_3^6 y_4^7+ 10 y_1^2 y_2^4 y_3^6 y_4^7+ 2 
y_1^2 y_2^4 y_3^7 y_4^7+ 24 y_2^2 y_3^3 y_4^8+ 24 y_1 y_2^2 y_3^3 y_4^8+ 56 y_2^2 
y_3^4 y_4^8 + 112 y_1 y_2^2 y_3^4 y_4^8 \\[.05cm] & + 36 y_1 y_2^3 y_3^3 y_4^8+ 56 y_1^2 y_2^2 
y_3^4 y_4^8+ 160 y_1 y_2^3 y_3^4 y_4^8+ 12 y_1 y_2^4 y_3^3 y_4^8+ 80 y_1^2 y_2^3 
y_3^4 y_4^8+ 140 y_1 y_2^3 y_3^5 y_4^8+ 38 y_1 y_2^4 y_3^4 y_4^8+ 140 y_1^2 y_2^3 
y_3^5 y_4^8+ 24 y_1^2 y_2^4 y_3^4 y_4^8 \\[.05cm] & + 40 y_1 y_2^4 y_3^5 y_4^8+ 70 y_1^2 y_2^4 
y_3^5 y_4^8+ 14 y_1 y_2^4 y_3^6 y_4^8+ 62 y_1^2 y_2^4 y_3^6 y_4^8+ 14 y_1^2 y_2^4 
y_3^7 y_4^8+ 4 y_2^2 y_3^3 y_4^9+ 4 y_1 y_2^2 y_3^3 y_4^9+ 28 y_2^2 y_3^4 y_4^9+ 
56 y_1 y_2^2 y_3^4 y_4^9+ 8 y_1 y_2^3 y_3^3 y_4^9 \\[.05cm] & + 28 y_1^2 y_2^2 y_3^4 y_4^9+ 
100 y_1 y_2^3 y_3^4 y_4^9+ 4 y_1 y_2^4 y_3^3 y_4^9+ 60 y_1^2 y_2^3 y_3^4 y_4^9+ 
140 y_1 y_2^3 y_3^5 y_4^9+ 42 y_1 y_2^4 y_3^4 y_4^9+ 140 y_1^2 y_2^3 y_3^5 y_4^9+ 
36 y_1^2 y_2^4 y_3^4 y_4^9+ 80 y_1 y_2^4 y_3^5 y_4^9 \\[.05cm] & + 140 y_1^2 y_2^4 y_3^5 
y_4^9+ 42 y_1 y_2^4 y_3^6 y_4^9+ 4 y_1^2 y_2^5 y_3^4 y_4^9+ 162 y_1^2 y_2^4 y_3^6 
y_4^9+ 16 y_1^2 y_2^5 y_3^5 y_4^9+ 42 y_1^2 y_2^4 y_3^7 y_4^9+ 24 y_1^2 y_2^5 
y_3^6 y_4^9+ 16 y_1^2 y_2^5 y_3^7 y_4^9+ 4 y_1^2 y_2^5 y_3^8 y_4^9 \\[.05cm] & + 8 y_2^2 y_3^4 
y_4^{10}+ 16 y_1 y_2^2 y_3^4 y_4^{10}+ 8 y_1^2 y_2^2 y_3^4 y_4^{10}+ 32 y_1 y_2^3 y_3^4 
y_4^{10}+ 24 y_1^2 y_2^3 y_3^4 y_4^{10}+ 84 y_1 y_2^3 y_3^5 y_4^{10}+ 18 y_1 y_2^4 
y_3^4 y_4^{10}+ 84 y_1^2 y_2^3 y_3^5 y_4^{10}+ 24 y_1^2 y_2^4 y_3^4 y_4^{10} \\[.05cm] & + 80 y_1 
y_2^4 y_3^5 y_4^{10}+ 140 y_1^2 y_2^4 y_3^5 y_4^{10}+ 70 y_1 y_2^4 y_3^6 y_4^{10}+ 8 
y_1^2 y_2^5 y_3^4 y_4^{10}+ 230 y_1^2 y_2^4 y_3^6 y_4^{10}+ 48 y_1^2 y_2^5 y_3^5 
y_4^{10}+ 70 y_1^2 y_2^4 y_3^7 y_4^{10}+ 100 y_1^2 y_2^5 y_3^6 y_4^{10} \\[.05cm] & + 88 y_1^2 
y_2^5 y_3^7 y_4^{10}+ 28 y_1^2 y_2^5 y_3^8 y_4^{10}+ y_2^2 y_3^4 y_4^{11}+ 2 y_1 y_2^2 
y_3^4 y_4^{11}+ y_1^2 y_2^2 y_3^4 y_4^{11}+ 4 y_1 y_2^3 y_3^4 y_4^{11}+ 4 y_1^2 y_2^3 
y_3^4 y_4^{11}+ 28 y_1 y_2^3 y_3^5 y_4^{11}+ 2 y_1 y_2^4 y_3^4 y_4^{11} \\[.05cm] & + 28 y_1^2 y_2^3 
y_3^5 y_4^{11}+ 6 y_1^2 y_2^4 y_3^4 y_4^{11}+ 40 y_1 y_2^4 y_3^5 y_4^{11}+ 70 y_1^2 
y_2^4 y_3^5 y_4^{11}+ 70 y_1 y_2^4 y_3^6 y_4^{11}+ 4 y_1^2 y_2^5 y_3^4 y_4^{11}+ 190 
y_1^2 y_2^4 y_3^6 y_4^{11}+ 48 y_1^2 y_2^5 y_3^5 y_4^{11} \\[.05cm] & + y_1^2 y_2^6 y_3^4 y_4^{11}+ 
70 y_1^2 y_2^4 y_3^7 y_4^{11}+ 160 y_1^2 y_2^5 y_3^6 y_4^{11}+ 6 y_1^2 y_2^6 y_3^5 
y_4^{11}+ 200 y_1^2 y_2^5 y_3^7 y_4^{11}+ 15 y_1^2 y_2^6 y_3^6 y_4^{11}+ 84 y_1^2 
y_2^5 y_3^8 y_4^{11}+ 20 y_1^2 y_2^6 y_3^7 y_4^{11} \\[.05cm] & + 15 y_1^2 y_2^6 y_3^8 y_4^{11}+ 6 
y_1^2 y_2^6 y_3^9 y_4^{11}+ y_1^2 y_2^6 y_3^{10} y_4^{11}+ 4 y_1 y_2^3 y_3^5 y_4^{12}+ 4 
y_1^2 y_2^3 y_3^5 y_4^{12}+ 8 y_1 y_2^4 y_3^5 y_4^{12}+ 14 y_1^2 y_2^4 y_3^5 y_4^{12}+ 
42 y_1 y_2^4 y_3^6 y_4^{12}+ 90 y_1^2 y_2^4 y_3^6 y_4^{12} \\[.05cm] & + 16 y_1^2 y_2^5 y_3^5 
y_4^{12}+ 42 y_1^2 y_2^4 y_3^7 y_4^{12}+ 120 y_1^2 y_2^5 y_3^6 y_4^{12}+ 6 y_1^2 y_2^6 
y_3^5 y_4^{12}+ 240 y_1^2 y_2^5 y_3^7 y_4^{12}+ 33 y_1^2 y_2^6 y_3^6 y_4^{12}+ 140 
y_1^2 y_2^5 y_3^8 y_4^{12}+ 72 y_1^2 y_2^6 y_3^7 y_4^{12} \\[.05cm] & + 78 y_1^2 y_2^6 y_3^8 
y_4^{12}+ 42 y_1^2 y_2^6 y_3^9 y_4^{12}+ 9 y_1^2 y_2^6 y_3^{10} y_4^{12}+ 14 y_1 y_2^4 
y_3^6 y_4^{13}+ 22 y_1^2 y_2^4 y_3^6 y_4^{13}+ 14 y_1^2 y_2^4 y_3^7 y_4^{13}+ 40 y_1^2 
y_2^5 y_3^6 y_4^{13}+ 160 y_1^2 y_2^5 y_3^7 y_4^{13} \\[.05cm] & + 21 y_1^2 y_2^6 y_3^6 y_4^{13}+ 
140 y_1^2 y_2^5 y_3^8 y_4^{13}+ 96 y_1^2 y_2^6 y_3^7 y_4^{13}+ 165 y_1^2 y_2^6 y_3^8 
y_4^{13}+ 126 y_1^2 y_2^6 y_3^9 y_4^{13}+ 36 y_1^2 y_2^6 y_3^{10} y_4^{13}+ 2 y_1 y_2^4 
y_3^6 y_4^{14}+ 2 y_1^2 y_2^4 y_3^6 y_4^{14} \\[.05cm] & + 2 y_1^2 y_2^4 y_3^7 y_4^{14}+ 4 y_1^2 
y_2^5 y_3^6 y_4^{14}+ 56 y_1^2 y_2^5 y_3^7 y_4^{14}+ 3 y_1^2 y_2^6 y_3^6 y_4^{14}+ 84 
y_1^2 y_2^5 y_3^8 y_4^{14}+ 56 y_1^2 y_2^6 y_3^7 y_4^{14}+ 180 y_1^2 y_2^6 y_3^8 
y_4^{14}+ 210 y_1^2 y_2^6 y_3^9 y_4^{14} \\[.05cm] & + 84 y_1^2 y_2^6 y_3^{10} y_4^{14}+ 8 y_1^2 
y_2^5 y_3^7 y_4^{15}+ 28 y_1^2 y_2^5 y_3^8 y_4^{15}+ 12 y_1^2 y_2^6 y_3^7 y_4^{15}+ 
105 y_1^2 y_2^6 y_3^8 y_4^{15}+ 210 y_1^2 y_2^6 y_3^9 y_4^{15}+ 126 y_1^2 y_2^6 
y_3^{10} y_4^{15}+ 4 y_1^2 y_2^5 y_3^8 y_4^{16} \\[.05cm] & + 30 y_1^2 y_2^6 y_3^8 y_4^{16}+ 126 
y_1^2 y_2^6 y_3^9 y_4^{16}+ 126 y_1^2 y_2^6 y_3^{10} y_4^{16}+ 3 y_1^2 y_2^6 y_3^8 
y_4^{17}+ 42 y_1^2 y_2^6 y_3^9 y_4^{17}+ 84 y_1^2 y_2^6 y_3^{10} y_4^{17}+ 6 y_1^2 y_2^6 
y_3^9 y_4^{18}+ 36 y_1^2 y_2^6 y_3^{10} y_4^{18} \\[.05cm] & + 9 y_1^2 y_2^6 y_3^{10} y_4^{19}+ y_1^2 
y_2^6 y_3^{10} y_4^{20} 
\end{array}
$}
\label{Z1_to_Z5}
\eeq

\beq
\resizebox{.92\textwidth}{!}{$
\begin{array}{cl}
%=================================================================
 Z_6 & = 1+ 3 y_3+ 6 y_3 y_4+ 3 y_3 y_4^2+ 3 y_3^2+ 2 y_1 y_3^2+ 12 y_3^2 y_4+ 8 y_1 y_3^2 
y_4+ 18 y_3^2 y_4^2+ 12 y_1 y_3^2 y_4^2+ 12 y_3^2 y_4^3+ 4 y_1 y_2 y_3^2 y_4^2+ 8 
y_1 y_3^2 y_4^3+ 3 y_3^2 y_4^4+ 8 y_1 y_2 y_3^2 y_4^3 \\[.05cm] & + 2 y_1 y_3^2 y_4^4+ 4 y_1 y_2 
y_3^2 y_4^4+ 2 y_1 y_2^2 y_3^2 y_4^4+ y_3^3+ 2 y_1 y_3^3+ 6 y_3^3 y_4+ y_1^2 
y_3^3+ 12 y_1 y_3^3 y_4+ 15 y_3^3 y_4^2+ 6 y_1^2 y_3^3 y_4+ 30 y_1 y_3^3 y_4^2+ 
20 y_3^3 y_4^3+ 12 y_1 y_2 y_3^3 y_4^2 \\[.05cm] & + 15 y_1^2 y_3^3 y_4^2+ 40 y_1 y_3^3 y_4^3 + 
15 y_3^3 y_4^4+ 4 y_1^2 y_2 y_3^3 y_4^2+ 40 y_1 y_2 y_3^3 y_4^3+ 20 y_1^2 y_3^3 
y_4^3+ 30 y_1 y_3^3 y_4^4+ 6 y_3^3 y_4^5+ 16 y_1^2 y_2 y_3^3 y_4^3+ 48 y_1 y_2 
y_3^3 y_4^4+ 15 y_1^2 y_3^3 y_4^4 \\[.05cm] & + 12 y_1 y_3^3 y_4^5+ y_3^3 y_4^6+ 2 y_1^2 y_2^2 
y_3^3 y_4^3+ 10 y_1 y_2^2 y_3^3 y_4^4+ 24 y_1^2 y_2 y_3^3 y_4^4+ 24 y_1 y_2 y_3^3 
y_4^5+ 6 y_1^2 y_3^3 y_4^5+ 2 y_1 y_3^3 y_4^6+ 10 y_1^2 y_2^2 y_3^3 y_4^4+ 12 y_1 
y_2^2 y_3^3 y_4^5+ 16 y_1^2 y_2 y_3^3 y_4^5 \\[.05cm] & + 4 y_1 y_2 y_3^3 y_4^6+ y_1^2 y_3^3 
y_4^6+ 14 y_1^2 y_2^2 y_3^3 y_4^5+ 2 y_1 y_2^2 y_3^3 y_4^6+ 4 y_1^2 y_2 y_3^3 
y_4^6+ 4 y_1^2 y_2^3 y_3^3 y_4^5+ 6 y_1^2 y_2^2 y_3^3 y_4^6+ 4 y_1^2 y_2^3 y_3^3 
y_4^6+ y_1^2 y_2^4 y_3^3 y_4^6+ 12 y_1 y_2 y_3^4 y_4^2+ 8 y_1^2 y_2 y_3^4 y_4^2 \\[.05cm] & + 
56 y_1 y_2 y_3^4 y_4^3+ 40 y_1^2 y_2 y_3^4 y_4^3+ 104 y_1 y_2 y_3^4 y_4^4+ 8 y_1^2 
y_2^2 y_3^4 y_4^3+ 20 y_1 y_2^2 y_3^4 y_4^4+ 80 y_1^2 y_2 y_3^4 y_4^4+ 96 y_1 y_2 
y_3^4 y_4^5+ 48 y_1^2 y_2^2 y_3^4 y_4^4+ 52 y_1 y_2^2 y_3^4 y_4^5+ 80 y_1^2 y_2 
y_3^4 y_4^5 \\[.05cm] & + 44 y_1 y_2 y_3^4 y_4^6+ 96 y_1^2 y_2^2 y_3^4 y_4^5+ 44 y_1 y_2^2 
y_3^4 y_4^6+ 40 y_1^2 y_2 y_3^4 y_4^6+ 8 y_1 y_2 y_3^4 y_4^7+ 24 y_1^2 y_2^3 y_3^4 
y_4^5+ 80 y_1^2 y_2^2 y_3^4 y_4^6+ 12 y_1 y_2^2 y_3^4 y_4^7+ 8 y_1^2 y_2 y_3^4 
y_4^7+ 48 y_1^2 y_2^3 y_3^4 y_4^6 \\[.05cm] & + 24 y_1^2 y_2^2 y_3^4 y_4^7+ 8 y_1^2 y_2^4 
y_3^4 y_4^6+ 24 y_1^2 y_2^3 y_3^4 y_4^7+ 8 y_1^2 y_2^4 y_3^4 y_4^7+ 4 y_1 y_2 
y_3^5 y_4^2+ 4 y_1^2 y_2 y_3^5 y_4^2+ 24 y_1 y_2 y_3^5 y_4^3+ 24 y_1^2 y_2 y_3^5 
y_4^3+ 60 y_1 y_2 y_3^5 y_4^4+ 12 y_1^2 y_2^2 y_3^5 y_4^3 \\[.05cm] & + 20 y_1 y_2^2 y_3^5 
y_4^4+ 60 y_1^2 y_2 y_3^5 y_4^4+ 80 y_1 y_2 y_3^5 y_4^5+ 4 y_1^3 y_2^2 y_3^5 
y_4^3+ 84 y_1^2 y_2^2 y_3^5 y_4^4+ 84 y_1 y_2^2 y_3^5 y_4^5+ 80 y_1^2 y_2 y_3^5 
y_4^5+ 60 y_1 y_2 y_3^5 y_4^6+ 20 y_1^3 y_2^2 y_3^5 y_4^4+ 220 y_1^2 y_2^2 y_3^5 
y_4^5 \\[.05cm] & + 136 y_1 y_2^2 y_3^5 y_4^6+ 60 y_1^2 y_2 y_3^5 y_4^6+ 24 y_1 y_2 y_3^5 
y_4^7+ 60 y_1^2 y_2^3 y_3^5 y_4^5+ 40 y_1^3 y_2^2 y_3^5 y_4^5+ 280 y_1^2 y_2^2 
y_3^5 y_4^6+ 104 y_1 y_2^2 y_3^5 y_4^7+ 24 y_1^2 y_2 y_3^5 y_4^7+ 4 y_1 y_2 y_3^5 
y_4^8+ 12 y_1^3 y_2^3 y_3^5 y_4^5 \\[.05cm] & + 188 y_1^2 y_2^3 y_3^5 y_4^6+ 40 y_1^3 y_2^2 
y_3^5 y_4^6+ 180 y_1^2 y_2^2 y_3^5 y_4^7+ 36 y_1 y_2^2 y_3^5 y_4^8+ 4 y_1^2 y_2 
y_3^5 y_4^8+ 28 y_1^2 y_2^4 y_3^5 y_4^6+ 36 y_1^3 y_2^3 y_3^5 y_4^6+ 204 y_1^2 
y_2^3 y_3^5 y_4^7+ 20 y_1^3 y_2^2 y_3^5 y_4^7+ 52 y_1^2 y_2^2 y_3^5 y_4^8 \\[.05cm] & + 4 y_1 
y_2^2 y_3^5 y_4^9+ 4 y_1^3 y_2^4 y_3^5 y_4^6+ 60 y_1^2 y_2^4 y_3^5 y_4^7+ 36 
y_1^3 y_2^3 y_3^5 y_4^7+ 84 y_1^2 y_2^3 y_3^5 y_4^8+ 4 y_1^3 y_2^2 y_3^5 y_4^8+ 4 
y_1^2 y_2^2 y_3^5 y_4^9+ 16 y_1^3 y_2^4 y_3^5 y_4^7+ 36 y_1^2 y_2^4 y_3^5 y_4^8+ 
12 y_1^3 y_2^3 y_3^5 y_4^8 \\[.05cm] & + 8 y_1^2 y_2^3 y_3^5 y_4^9+ 12 y_1^3 y_2^4 y_3^5 
y_4^8+ 4 y_1^2 y_2^4 y_3^5 y_4^9+ 4 y_1^3 y_2^5 y_3^5 y_4^8+ 8 y_1^2 y_2^2 y_3^6 
y_4^3+ 10 y_1 y_2^2 y_3^6 y_4^4+ 8 y_1^3 y_2^2 y_3^6 y_4^3+ 64 y_1^2 y_2^2 y_3^6 
y_4^4+ 60 y_1 y_2^2 y_3^6 y_4^5+ 48 y_1^3 y_2^2 y_3^6 y_4^4 \\[.05cm] & + 208 y_1^2 y_2^2 
y_3^6 y_4^5+ 150 y_1 y_2^2 y_3^6 y_4^6+ 80 y_1^2 y_2^3 y_3^6 y_4^5+ 120 y_1^3 
y_2^2 y_3^6 y_4^5+ 360 y_1^2 y_2^2 y_3^6 y_4^6+ 200 y_1 y_2^2 y_3^6 y_4^7+ 44 
y_1^3 y_2^3 y_3^6 y_4^5+ 352 y_1^2 y_2^3 y_3^6 y_4^6+ 160 y_1^3 y_2^2 y_3^6 
y_4^6 \\[.05cm] & + 360 y_1^2 y_2^2 y_3^6 y_4^7+ 150 y_1 y_2^2 y_3^6 y_4^8+ 56 y_1^2 y_2^4 
y_3^6 y_4^6+ 184 y_1^3 y_2^3 y_3^6 y_4^6+ 608 y_1^2 y_2^3 y_3^6 y_4^7+ 120 y_1^3 
y_2^2 y_3^6 y_4^7+ 208 y_1^2 y_2^2 y_3^6 y_4^8+ 60 y_1 y_2^2 y_3^6 y_4^9+ 24 
y_1^3 y_2^4 y_3^6 y_4^6 \\[.05cm] & + 192 y_1^2 y_2^4 y_3^6 y_4^7+ 296 y_1^3 y_2^3 y_3^6 
y_4^7+ 512 y_1^2 y_2^3 y_3^6 y_4^8+ 48 y_1^3 y_2^2 y_3^6 y_4^8+ 64 y_1^2 y_2^2 
y_3^6 y_4^9+ 10 y_1 y_2^2 y_3^6 y_4^{10}+ 120 y_1^3 y_2^4 y_3^6 y_4^7+ 240 y_1^2 
y_2^4 y_3^6 y_4^8+ 224 y_1^3 y_2^3 y_3^6 y_4^8 \\[.05cm] & + 208 y_1^2 y_2^3 y_3^6 y_4^9+ 8 
y_1^3 y_2^2 y_3^6 y_4^9+ 8 y_1^2 y_2^2 y_3^6 y_4^{10}+ 184 y_1^3 y_2^4 y_3^6 y_4^8+ 
128 y_1^2 y_2^4 y_3^6 y_4^9+ 76 y_1^3 y_2^3 y_3^6 y_4^9+ 32 y_1^2 y_2^3 y_3^6 
y_4^{10}+ 28 y_1^3 y_2^5 y_3^6 y_4^8+ 104 y_1^3 y_2^4 y_3^6 y_4^9 \\[.05cm] & + 24 y_1^2 y_2^4 
y_3^6 y_4^{10}+ 8 y_1^3 y_2^3 y_3^6 y_4^{10}+ 36 y_1^3 y_2^5 y_3^6 y_4^9+ 16 y_1^3 
y_2^4 y_3^6 y_4^{10}+ 8 y_1^3 y_2^5 y_3^6 y_4^{10}+ 2 y_1^2 y_2^2 y_3^7 y_4^3+ 2 y_1 
y_2^2 y_3^7 y_4^4+ 4 y_1^3 y_2^2 y_3^7 y_4^3+ 18 y_1^2 y_2^2 y_3^7 y_4^4+ 16 y_1 
y_2^2 y_3^7 y_4^5 \\[.05cm] & + 2 y_1^4 y_2^2 y_3^7 y_4^3+ 28 y_1^3 y_2^2 y_3^7 y_4^4+ 70 
y_1^2 y_2^2 y_3^7 y_4^5+ 56 y_1 y_2^2 y_3^7 y_4^6+ 14 y_1^4 y_2^2 y_3^7 y_4^4+ 60 
y_1^2 y_2^3 y_3^7 y_4^5+ 84 y_1^3 y_2^2 y_3^7 y_4^5+ 156 y_1^2 y_2^2 y_3^7 y_4^6+ 
112 y_1 y_2^2 y_3^7 y_4^7+ 60 y_1^3 y_2^3 y_3^7 y_4^5 \\[.05cm] & + 42 y_1^4 y_2^2 y_3^7 
y_4^5+ 348 y_1^2 y_2^3 y_3^7 y_4^6+ 140 y_1^3 y_2^2 y_3^7 y_4^6+ 222 y_1^2 y_2^2 
y_3^7 y_4^7+ 140 y_1 y_2^2 y_3^7 y_4^8+ 8 y_1^4 y_2^3 y_3^7 y_4^5+ 70 y_1^2 y_2^4 
y_3^7 y_4^6+ 332 y_1^3 y_2^3 y_3^7 y_4^6+ 70 y_1^4 y_2^2 y_3^7 y_4^6 \\[.05cm] & + 844 y_1^2 
y_2^3 y_3^7 y_4^7+ 140 y_1^3 y_2^2 y_3^7 y_4^7+ 212 y_1^2 y_2^2 y_3^7 y_4^8+ 112 
y_1 y_2^2 y_3^7 y_4^9+ 60 y_1^3 y_2^4 y_3^7 y_4^6+ 40 y_1^4 y_2^3 y_3^7 y_4^6+ 
340 y_1^2 y_2^4 y_3^7 y_4^7+ 760 y_1^3 y_2^3 y_3^7 y_4^7+ 70 y_1^4 y_2^2 y_3^7 
y_4^7 \\[.05cm] & + 1100 y_1^2 y_2^3 y_3^7 y_4^8+ 84 y_1^3 y_2^2 y_3^7 y_4^8+ 138 y_1^2 y_2^2 
y_3^7 y_4^9+ 56 y_1 y_2^2 y_3^7 y_4^{10}+ 6 y_1^4 y_2^4 y_3^7 y_4^6+ 368 y_1^3 
y_2^4 y_3^7 y_4^7+ 80 y_1^4 y_2^3 y_3^7 y_4^7+ 666 y_1^2 y_2^4 y_3^7 y_4^8+ 920 
y_1^3 y_2^3 y_3^7 y_4^8 \\[.05cm] & + 42 y_1^4 y_2^2 y_3^7 y_4^8+ 820 y_1^2 y_2^3 y_3^7 y_4^9+ 
28 y_1^3 y_2^2 y_3^7 y_4^9+ 60 y_1^2 y_2^2 y_3^7 y_4^{10}+ 16 y_1 y_2^2 y_3^7 
y_4^{11}+ 38 y_1^4 y_2^4 y_3^7 y_4^7+ 836 y_1^3 y_2^4 y_3^7 y_4^8+ 80 y_1^4 y_2^3 
y_3^7 y_4^8+ 664 y_1^2 y_2^4 y_3^7 y_4^9 \\[.05cm] & + 620 y_1^3 y_2^3 y_3^7 y_4^9+ 14 y_1^4 
y_2^2 y_3^7 y_4^9+ 340 y_1^2 y_2^3 y_3^7 y_4^{10}+ 4 y_1^3 y_2^2 y_3^7 y_4^{10}+ 16 
y_1^2 y_2^2 y_3^7 y_4^{11}+ 2 y_1 y_2^2 y_3^7 y_4^{12}+ 84 y_1^3 y_2^5 y_3^7 y_4^8+ 
84 y_1^4 y_2^4 y_3^7 y_4^8+ 904 y_1^3 y_2^4 y_3^7 y_4^9 \\[.05cm] & + 40 y_1^4 y_2^3 y_3^7 
y_4^9+ 346 y_1^2 y_2^4 y_3^7 y_4^{10}+ 220 y_1^3 y_2^3 y_3^7 y_4^{10}+ 2 y_1^4 y_2^2 
y_3^7 y_4^{10}+ 68 y_1^2 y_2^3 y_3^7 y_4^{11}+ 2 y_1^2 y_2^2 y_3^7 y_4^{12}+ 14 y_1^4 
y_2^5 y_3^7 y_4^8+ 224 y_1^3 y_2^5 y_3^7 y_4^9+ 86 y_1^4 y_2^4 y_3^7 y_4^9 \\[.05cm] & + 476 
y_1^3 y_2^4 y_3^7 y_4^{10}+ 8 y_1^4 y_2^3 y_3^7 y_4^{10}+ 84 y_1^2 y_2^4 y_3^7 
y_4^{11}+ 32 y_1^3 y_2^3 y_3^7 y_4^{11}+ 4 y_1^2 y_2^3 y_3^7 y_4^{12}+ 44 y_1^4 y_2^5 
y_3^7 y_4^9+ 200 y_1^3 y_2^5 y_3^7 y_4^{10}+ 44 y_1^4 y_2^4 y_3^7 y_4^{10}+ 104 
y_1^3 y_2^4 y_3^7 y_4^{11} \\[.05cm] & + 6 y_1^2 y_2^4 y_3^7 y_4^{12}+ 48 y_1^4 y_2^5 y_3^7 
y_4^{10}+ 64 y_1^3 y_2^5 y_3^7 y_4^{11}+ 12 y_1^4 y_2^4 y_3^7 y_4^{11}+ 4 y_1^3 y_2^4 
y_3^7 y_4^{12}+ 10 y_1^4 y_2^6 y_3^7 y_4^{10}+ 24 y_1^4 y_2^5 y_3^7 y_4^{11}+ 4 y_1^3 
y_2^5 y_3^7 y_4^{12}+ 2 y_1^4 y_2^4 y_3^7 y_4^{12} \\[.05cm] & + 12 y_1^4 y_2^6 y_3^7 y_4^{11}+ 6 
y_1^4 y_2^5 y_3^7 y_4^{12}+ 6 y_1^4 y_2^6 y_3^7 y_4^{12}+ 2 y_1^4 y_2^7 y_3^7 y_4^{12}+ 
24 y_1^2 y_2^3 y_3^8 y_4^5+ 36 y_1^3 y_2^3 y_3^8 y_4^5+ 176 y_1^2 y_2^3 y_3^8 
y_4^6+ 12 y_1^4 y_2^3 y_3^8 y_4^5+ 56 y_1^2 y_2^4 y_3^8 y_4^6 \\[.05cm] & + 256 y_1^3 y_2^3 
y_3^8 y_4^6+ 560 y_1^2 y_2^3 y_3^8 y_4^7+ 80 y_1^3 y_2^4 y_3^8 y_4^6+ 80 y_1^4 
y_2^3 y_3^8 y_4^6+ 360 y_1^2 y_2^4 y_3^8 y_4^7+ 788 y_1^3 y_2^3 y_3^8 y_4^7+ 
1008 y_1^2 y_2^3 y_3^8 y_4^8+ 24 y_1^4 y_2^4 y_3^8 y_4^6+ 592 y_1^3 y_2^4 y_3^8 
y_4^7 \\[.05cm] & + 228 y_1^4 y_2^3 y_3^8 y_4^7+ 984 y_1^2 y_2^4 y_3^8 y_4^8+ 1368 y_1^3 
y_2^3 y_3^8 y_4^8+ 1120 y_1^2 y_2^3 y_3^8 y_4^9+ 182 y_1^4 y_2^4 y_3^8 y_4^7+ 
1784 y_1^3 y_2^4 y_3^8 y_4^8+ 360 y_1^4 y_2^3 y_3^8 y_4^8+ 1480 y_1^2 y_2^4 
y_3^8 y_4^9+ 1460 y_1^3 y_2^3 y_3^8 y_4^9 \\[.05cm] & + 784 y_1^2 y_2^3 y_3^8 y_4^{10}+ 140 
y_1^3 y_2^5 y_3^8 y_4^8+ 544 y_1^4 y_2^4 y_3^8 y_4^8+ 2856 y_1^3 y_2^4 y_3^8 
y_4^9+ 340 y_1^4 y_2^3 y_3^8 y_4^9+ 1320 y_1^2 y_2^4 y_3^8 y_4^{10}+ 976 y_1^3 
y_2^3 y_3^8 y_4^{10}+ 336 y_1^2 y_2^3 y_3^8 y_4^{11}+ 70 y_1^4 y_2^5 y_3^8 y_4^8 \\[.05cm] & + 
580 y_1^3 y_2^5 y_3^8 y_4^9+ 844 y_1^4 y_2^4 y_3^8 y_4^9+ 2624 y_1^3 y_2^4 y_3^8 
y_4^{10}+ 192 y_1^4 y_2^3 y_3^8 y_4^{10}+ 696 y_1^2 y_2^4 y_3^8 y_4^{11}+ 396 y_1^3 
y_2^3 y_3^8 y_4^{11}+ 80 y_1^2 y_2^3 y_3^8 y_4^{12}+ 320 y_1^4 y_2^5 y_3^8 y_4^9+ 
936 y_1^3 y_2^5 y_3^8 y_4^{10} \\[.05cm] & + 736 y_1^4 y_2^4 y_3^8 y_4^{10}+ 1376 y_1^3 y_2^4 
y_3^8 y_4^{11}+ 60 y_1^4 y_2^3 y_3^8 y_4^{11}+ 200 y_1^2 y_2^4 y_3^8 y_4^{12}+ 88 
y_1^3 y_2^3 y_3^8 y_4^{12}+ 8 y_1^2 y_2^3 y_3^8 y_4^{13}+ 556 y_1^4 y_2^5 y_3^8 
y_4^{10}+ 728 y_1^3 y_2^5 y_3^8 y_4^{11}+ 358 y_1^4 y_2^4 y_3^8 y_4^{11} \\[.05cm] & + 376 y_1^3 
y_2^4 y_3^8 y_4^{12}+ 8 y_1^4 y_2^3 y_3^8 y_4^{12}+ 24 y_1^2 y_2^4 y_3^8 y_4^{13}+ 8 
y_1^3 y_2^3 y_3^8 y_4^{13}+ 62 y_1^4 y_2^6 y_3^8 y_4^{10}+ 460 y_1^4 y_2^5 y_3^8 
y_4^{11}+ 268 y_1^3 y_2^5 y_3^8 y_4^{12}+ 88 y_1^4 y_2^4 y_3^8 y_4^{12}+ 40 y_1^3 
y_2^4 y_3^8 y_4^{13} \\[.05cm] & + 146 y_1^4 y_2^6 y_3^8 y_4^{11}+ 182 y_1^4 y_2^5 y_3^8 y_4^{12}+ 
36 y_1^3 y_2^5 y_3^8 y_4^{13}+ 8 y_1^4 y_2^4 y_3^8 y_4^{13}+ 116 y_1^4 y_2^6 y_3^8 
y_4^{12}+ 28 y_1^4 y_2^5 y_3^8 y_4^{13}+ 14 y_1^4 y_2^7 y_3^8 y_4^{12}+ 32 y_1^4 y_2^6 
y_3^8 y_4^{13}+ 12 y_1^4 y_2^7 y_3^8 y_4^{13} \\[.05cm] & + 4 y_1^2 y_2^3 y_3^9 y_4^5+ 8 y_1^3 
y_2^3 y_3^9 y_4^5+ 36 y_1^2 y_2^3 y_3^9 y_4^6+ 4 y_1^4 y_2^3 y_3^9 y_4^5+ 28 
y_1^2 y_2^4 y_3^9 y_4^6+ 72 y_1^3 y_2^3 y_3^9 y_4^6+ 144 y_1^2 y_2^3 y_3^9 y_4^7+ 
60 y_1^3 y_2^4 y_3^9 y_4^6+ 36 y_1^4 y_2^3 y_3^9 y_4^6+ 228 y_1^2 y_2^4 y_3^9 
y_4^7 \\[.05cm] & + 288 y_1^3 y_2^3 y_3^9 y_4^7+ 336 y_1^2 y_2^3 y_3^9 y_4^8+ 36 y_1^4 y_2^4 
y_3^9 y_4^6+ 528 y_1^3 y_2^4 y_3^9 y_4^7+ 144 y_1^4 y_2^3 y_3^9 y_4^7+ 816 y_1^2 
y_2^4 y_3^9 y_4^8+ 672 y_1^3 y_2^3 y_3^9 y_4^8+ 504 y_1^2 y_2^3 y_3^9 y_4^9+ 4 
y_1^5 y_2^4 y_3^9 y_4^6 \\[.05cm] & + 322 y_1^4 y_2^4 y_3^9 y_4^7+ 2000 y_1^3 y_2^4 y_3^9 
y_4^8+ 336 y_1^4 y_2^3 y_3^9 y_4^8+ 1680 y_1^2 y_2^4 y_3^9 y_4^9+ 1008 y_1^3 
y_2^3 y_3^9 y_4^9+ 504 y_1^2 y_2^3 y_3^9 y_4^{10}+ 32 y_1^5 y_2^4 y_3^9 y_4^7+ 140 
y_1^3 y_2^5 y_3^9 y_4^8+ 1222 y_1^4 y_2^4 y_3^9 y_4^8 \\[.05cm] & + 4280 y_1^3 y_2^4 y_3^9 
y_4^9+ 504 y_1^4 y_2^3 y_3^9 y_4^9+ 2184 y_1^2 y_2^4 y_3^9 y_4^{10}+ 1008 y_1^3 
y_2^3 y_3^9 y_4^{10}+ 336 y_1^2 y_2^3 y_3^9 y_4^{11}+ 140 y_1^4 y_2^5 y_3^9 y_4^8+ 
112 y_1^5 y_2^4 y_3^9 y_4^8+ 800 y_1^3 y_2^5 y_3^9 y_4^9+ 2596 y_1^4 y_2^4 y_3^9 
y_4^9 \\[.05cm] & + 5688 y_1^3 y_2^4 y_3^9 y_4^{10}+ 504 y_1^4 y_2^3 y_3^9 y_4^{10}+ 1848 y_1^2 
y_2^4 y_3^9 y_4^{11}+ 672 y_1^3 y_2^3 y_3^9 y_4^{11}+ 144 y_1^2 y_2^3 y_3^9 y_4^{12}+ 
16 y_1^5 y_2^5 y_3^9 y_4^8+ 852 y_1^4 y_2^5 y_3^9 y_4^9+ 224 y_1^5 y_2^4 y_3^9 
y_4^9+ 1908 y_1^3 y_2^5 y_3^9 y_4^{10} \\[.05cm] & + 3404 y_1^4 y_2^4 y_3^9 y_4^{10}+ 4840 y_1^3 
y_2^4 y_3^9 y_4^{11}+ 336 y_1^4 y_2^3 y_3^9 y_4^{11}+ 1008 y_1^2 y_2^4 y_3^9 y_4^{12}+ 
288 y_1^3 y_2^3 y_3^9 y_4^{12}+ 36 y_1^2 y_2^3 y_3^9 y_4^{13}+ 96 y_1^5 y_2^5 y_3^9 
y_4^9+ 2124 y_1^4 y_2^5 y_3^9 y_4^{10}+ 280 y_1^5 y_2^4 y_3^9 y_4^{10} \\[.05cm] & + 2440 y_1^3 
y_2^5 y_3^9 y_4^{11}+ 2846 y_1^4 y_2^4 y_3^9 y_4^{11}+ 2608 y_1^3 y_2^4 y_3^9 
y_4^{12}+ 144 y_1^4 y_2^3 y_3^9 y_4^{12}+ 336 y_1^2 y_2^4 y_3^9 y_4^{13}+ 72 y_1^3 
y_2^3 y_3^9 y_4^{13}+ 4 y_1^2 y_2^3 y_3^9 y_4^{14}+ 162 y_1^4 y_2^6 y_3^9 y_4^{10}+ 
240 y_1^5 y_2^5 y_3^9 y_4^{10} \\[.05cm] & + 2800 y_1^4 y_2^5 y_3^9 y_4^{11}+ 224 y_1^5 y_2^4 
y_3^9 y_4^{11}+ 1780 y_1^3 y_2^5 y_3^9 y_4^{12}+ 1502 y_1^4 y_2^4 y_3^9 y_4^{12}+ 840 
y_1^3 y_2^4 y_3^9 y_4^{13}+ 36 y_1^4 y_2^3 y_3^9 y_4^{13}+ 60 y_1^2 y_2^4 y_3^9 
y_4^{14}+ 8 y_1^3 y_2^3 y_3^9 y_4^{14}+ 24 y_1^5 y_2^6 y_3^9 y_4^{10} \\[.05cm] & + 588 y_1^4 y_2^6 
y_3^9 y_4^{11}+ 320 y_1^5 y_2^5 y_3^9 y_4^{11}+ 2088 y_1^4 y_2^5 y_3^9 y_4^{12}+ 112 
y_1^5 y_2^4 y_3^9 y_4^{12}+ 720 y_1^3 y_2^5 y_3^9 y_4^{13}+ 472 y_1^4 y_2^4 y_3^9 
y_4^{13}+ 140 y_1^3 y_2^4 y_3^9 y_4^{14}+ 4 y_1^4 y_2^3 y_3^9 y_4^{14}+ 4 y_1^2 y_2^4 
y_3^9 y_4^{15} \\[.05cm] & + 96 y_1^5 y_2^6 y_3^9 y_4^{11}+ 812 y_1^4 y_2^6 y_3^9 y_4^{12}+ 240 
y_1^5 y_2^5 y_3^9 y_4^{12}+ 864 y_1^4 y_2^5 y_3^9 y_4^{13}+ 32 y_1^5 y_2^4 y_3^9 
y_4^{13}+ 140 y_1^3 y_2^5 y_3^9 y_4^{14}+ 76 y_1^4 y_2^4 y_3^9 y_4^{14}+ 8 y_1^3 y_2^4 
y_3^9 y_4^{15}+ 42 y_1^4 y_2^7 y_3^9 y_4^{12} \\[.05cm] & + 144 y_1^5 y_2^6 y_3^9 y_4^{12}+ 520 
y_1^4 y_2^6 y_3^9 y_4^{13}+ 96 y_1^5 y_2^5 y_3^9 y_4^{13}+ 176 y_1^4 y_2^5 y_3^9 
y_4^{14}+ 4 y_1^5 y_2^4 y_3^9 y_4^{14}+ 8 y_1^3 y_2^5 y_3^9 y_4^{15}+ 4 y_1^4 y_2^4 
y_3^9 y_4^{15}+ 16 y_1^5 y_2^7 y_3^9 y_4^{12}+ 80 y_1^4 y_2^7 y_3^9 y_4^{13} \\[.05cm] & + 96 y_1^5 
y_2^6 y_3^9 y_4^{13}+ 146 y_1^4 y_2^6 y_3^9 y_4^{14}+ 16 y_1^5 y_2^5 y_3^9 y_4^{14}+ 
12 y_1^4 y_2^5 y_3^9 y_4^{15}+ 32 y_1^5 y_2^7 y_3^9 y_4^{13}+ 42 y_1^4 y_2^7 y_3^9 
y_4^{14}+ 24 y_1^5 y_2^6 y_3^9 y_4^{14}+ 12 y_1^4 y_2^6 y_3^9 y_4^{15}+ 16 y_1^5 y_2^7 
y_3^9 y_4^{14} \\[.05cm] & + 4 y_1^4 y_2^7 y_3^9 y_4^{15}+ 4 y_1^5 y_2^8 y_3^9 y_4^{14}+ 8 y_1^2 
y_2^4 y_3^{10} y_4^6+ 24 y_1^3 y_2^4 y_3^{10} y_4^6+ 80 y_1^2 y_2^4 y_3^{10} y_4^7+ 24 
y_1^4 y_2^4 y_3^{10} y_4^6+ 248 y_1^3 y_2^4 y_3^{10} y_4^7+ 360 y_1^2 y_2^4 y_3^{10} 
y_4^8+ 8 y_1^5 y_2^4 y_3^{10} y_4^6 \\[.05cm] & + 250 y_1^4 y_2^4 y_3^{10} y_4^7+ 1144 y_1^3 
y_2^4 y_3^{10} y_4^8+ 960 y_1^2 y_2^4 y_3^{10} y_4^9+ 80 y_1^5 y_2^4 y_3^{10} y_4^7+ 
84 y_1^3 y_2^5 y_3^{10} y_4^8+ 1156 y_1^4 y_2^4 y_3^{10} y_4^8+ 3104 y_1^3 y_2^4 
y_3^{10} y_4^9+ 1680 y_1^2 y_2^4 y_3^{10} y_4^{10} \\[.05cm] & + 140 y_1^4 y_2^5 y_3^{10} y_4^8+ 360 
y_1^5 y_2^4 y_3^{10} y_4^8+ 620 y_1^3 y_2^5 y_3^{10} y_4^9+ 3134 y_1^4 y_2^4 y_3^{10} 
y_4^9+ 5488 y_1^3 y_2^4 y_3^{10} y_4^{10}+ 2016 y_1^2 y_2^4 y_3^{10} y_4^{11}+ 48 y_1^5 
y_2^5 y_3^{10} y_4^8+ 1076 y_1^4 y_2^5 y_3^{10} y_4^9 \\[.05cm] & + 960 y_1^5 y_2^4 y_3^{10} y_4^9+ 
1988 y_1^3 y_2^5 y_3^{10} y_4^{10}+ 5528 y_1^4 y_2^4 y_3^{10} y_4^{10}+ 6608 y_1^3 
y_2^4 y_3^{10} y_4^{11}+ 1680 y_1^2 y_2^4 y_3^{10} y_4^{12}+ 364 y_1^5 y_2^5 y_3^{10} 
y_4^9+ 3548 y_1^4 y_2^5 y_3^{10} y_4^{10}+ 1680 y_1^5 y_2^4 y_3^{10} y_4^{10} \\[.05cm] & + 3612 
y_1^3 y_2^5 y_3^{10} y_4^{11}+ 6638 y_1^4 y_2^4 y_3^{10} y_4^{11}+ 5488 y_1^3 y_2^4 
y_3^{10} y_4^{12}+ 960 y_1^2 y_2^4 y_3^{10} y_4^{13}+ 230 y_1^4 y_2^6 y_3^{10} y_4^{10}+ 
1208 y_1^5 y_2^5 y_3^{10} y_4^{10}+ 6568 y_1^4 y_2^5 y_3^{10} y_4^{11}+ 2016 y_1^5 
y_2^4 y_3^{10} y_4^{11} \\[.05cm] & + 4060 y_1^3 y_2^5 y_3^{10} y_4^{12}+ 5500 y_1^4 y_2^4 y_3^{10} 
y_4^{12}+ 3104 y_1^3 y_2^4 y_3^{10} y_4^{13}+ 360 y_1^2 y_2^4 y_3^{10} y_4^{14}+ 100 
y_1^5 y_2^6 y_3^{10} y_4^{10}+ 1154 y_1^4 y_2^6 y_3^{10} y_4^{11}+ 2292 y_1^5 y_2^5 
y_3^{10} y_4^{11}+ 7476 y_1^4 y_2^5 y_3^{10} y_4^{12} \\[.05cm] & + 1680 y_1^5 y_2^4 y_3^{10} y_4^{12}+ 
2884 y_1^3 y_2^5 y_3^{10} y_4^{13}+ 3106 y_1^4 y_2^4 y_3^{10} y_4^{13}+ 1144 y_1^3 
y_2^4 y_3^{10} y_4^{14}+ 80 y_1^2 y_2^4 y_3^{10} y_4^{15}+ 536 y_1^5 y_2^6 y_3^{10} 
y_4^{11}+ 2374 y_1^4 y_2^6 y_3^{10} y_4^{12}+ 2720 y_1^5 y_2^5 y_3^{10} y_4^{12} \\[.05cm] & + 5364 
y_1^4 y_2^5 y_3^{10} y_4^{13}+ 960 y_1^5 y_2^4 y_3^{10} y_4^{13}+ 1260 y_1^3 y_2^5 
y_3^{10} y_4^{14}+ 1144 y_1^4 y_2^4 y_3^{10} y_4^{14}+ 248 y_1^3 y_2^4 y_3^{10} y_4^{15}+ 8 
y_1^2 y_2^4 y_3^{10} y_4^{16}+ 70 y_1^4 y_2^7 y_3^{10} y_4^{12}+ 1192 y_1^5 y_2^6 y_3^{10} 
y_4^{12} \\[.05cm] & + 2554 y_1^4 y_2^6 y_3^{10} y_4^{13}+ 2068 y_1^5 y_2^5 y_3^{10} y_4^{13}+ 2372 
y_1^4 y_2^5 y_3^{10} y_4^{14}+ 360 y_1^5 y_2^4 y_3^{10} y_4^{14}+ 308 y_1^3 y_2^5 y_3^{10} 
y_4^{15}+ 248 y_1^4 y_2^4 y_3^{10} y_4^{15}+ 24 y_1^3 y_2^4 y_3^{10} y_4^{16}+ 88 y_1^5 
y_2^7 y_3^{10} y_4^{12} \\[.05cm] & + 220 y_1^4 y_2^7 y_3^{10} y_4^{13}+ 1408 y_1^5 y_2^6 y_3^{10} 
y_4^{13}+ 1508 y_1^4 y_2^6 y_3^{10} y_4^{14}+ 984 y_1^5 y_2^5 y_3^{10} y_4^{14}+ 592 
y_1^4 y_2^5 y_3^{10} y_4^{15}+ 80 y_1^5 y_2^4 y_3^{10} y_4^{15}+ 32 y_1^3 y_2^5 y_3^{10} 
y_4^{16}+ 24 y_1^4 y_2^4 y_3^{10} y_4^{16} \\[.05cm] & + 284 y_1^5 y_2^7 y_3^{10} y_4^{13}+ 248 y_1^4 
y_2^7 y_3^{10} y_4^{14}+ 932 y_1^5 y_2^6 y_3^{10} y_4^{14}+ 460 y_1^4 y_2^6 y_3^{10} 
y_4^{15}+ 268 y_1^5 y_2^5 y_3^{10} y_4^{15}+ 64 y_1^4 y_2^5 y_3^{10} y_4^{16}+ 8 y_1^5 
y_2^4 y_3^{10} y_4^{16}+ 336 y_1^5 y_2^7 y_3^{10} y_4^{14} \\[.05cm] & + 116 y_1^4 y_2^7 y_3^{10} 
y_4^{15}+ 328 y_1^5 y_2^6 y_3^{10} y_4^{15}+ 56 y_1^4 y_2^6 y_3^{10} y_4^{16}+ 32 y_1^5 
y_2^5 y_3^{10} y_4^{16}+ 28 y_1^5 y_2^8 y_3^{10} y_4^{14}+ 172 y_1^5 y_2^7 y_3^{10} 
y_4^{15}+ 18 y_1^4 y_2^7 y_3^{10} y_4^{16}+ 48 y_1^5 y_2^6 y_3^{10} y_4^{16} \\[.05cm] & + 32 y_1^5 
y_2^8 y_3^{10} y_4^{15}+ 32 y_1^5 y_2^7 y_3^{10} y_4^{16}+ 8 y_1^5 y_2^8 y_3^{10} y_4^{16}+ 
y_1^2 y_2^4 y_3^{11} y_4^6+ 4 y_1^3 y_2^4 y_3^{11} y_4^6+ 12 y_1^2 y_2^4 y_3^{11} 
y_4^7+ 6 y_1^4 y_2^4 y_3^{11} y_4^6+ 48 y_1^3 y_2^4 y_3^{11} y_4^7+ 66 y_1^2 y_2^4 
y_3^{11} y_4^8 \\[.05cm] & + 4 y_1^5 y_2^4 y_3^{11} y_4^6+ 72 y_1^4 y_2^4 y_3^{11} y_4^7+ 264 y_1^3 
y_2^4 y_3^{11} y_4^8+ 220 y_1^2 y_2^4 y_3^{11} y_4^9+ y_1^6 y_2^4 y_3^{11} y_4^6+ 48 
y_1^5 y_2^4 y_3^{11} y_4^7+ 28 y_1^3 y_2^5 y_3^{11} y_4^8+ 396 y_1^4 y_2^4 y_3^{11} 
y_4^8+ 880 y_1^3 y_2^4 y_3^{11} y_4^9 \\[.05cm] & + 495 y_1^2 y_2^4 y_3^{11} y_4^{10}+ 12 y_1^6 
y_2^4 y_3^{11} y_4^7+ 70 y_1^4 y_2^5 y_3^{11} y_4^8+ 264 y_1^5 y_2^4 y_3^{11} y_4^8+ 
256 y_1^3 y_2^5 y_3^{11} y_4^9+ 1320 y_1^4 y_2^4 y_3^{11} y_4^9+ 1980 y_1^3 y_2^4 
y_3^{11} y_4^{10}+ 792 y_1^2 y_2^4 y_3^{11} y_4^{11} \\[.05cm] & + 48 y_1^5 y_2^5 y_3^{11} y_4^8+ 66 
y_1^6 y_2^4 y_3^{11} y_4^8+ 656 y_1^4 y_2^5 y_3^{11} y_4^9+ 880 y_1^5 y_2^4 y_3^{11} 
y_4^9+ 1044 y_1^3 y_2^5 y_3^{11} y_4^{10}+ 2970 y_1^4 y_2^4 y_3^{11} y_4^{10}+ 3168 
y_1^3 y_2^4 y_3^{11} y_4^{11}+ 924 y_1^2 y_2^4 y_3^{11} y_4^{12} \\[.05cm] & + 6 y_1^6 y_2^5 y_3^{11} 
y_4^8+ 440 y_1^5 y_2^5 y_3^{11} y_4^9+ 220 y_1^6 y_2^4 y_3^{11} y_4^9+ 2714 y_1^4 
y_2^5 y_3^{11} y_4^{10}+ 1980 y_1^5 y_2^4 y_3^{11} y_4^{10}+ 2496 y_1^3 y_2^5 y_3^{11} 
y_4^{11}+ 4752 y_1^4 y_2^4 y_3^{11} y_4^{11}+ 3696 y_1^3 y_2^4 y_3^{11} y_4^{12} \\[.05cm] & + 792 
y_1^2 y_2^4 y_3^{11} y_4^{13}+ 60 y_1^6 y_2^5 y_3^{11} y_4^9+ 190 y_1^4 y_2^6 y_3^{11} 
y_4^{10}+ 1808 y_1^5 y_2^5 y_3^{11} y_4^{10}+ 495 y_1^6 y_2^4 y_3^{11} y_4^{10}+ 6532 
y_1^4 y_2^5 y_3^{11} y_4^{11}+ 3168 y_1^5 y_2^4 y_3^{11} y_4^{11}+ 3864 y_1^3 y_2^5 
y_3^{11} y_4^{12}
\end{array}
$} \nonumber
\eeq

\newpage

%=================================================================
\beq
\resizebox{.92\textwidth}{!}{$
\begin{array}{cl}
& + 5544 y_1^4 y_2^4 y_3^{11} y_4^{12}+ 3168 y_1^3 y_2^4 y_3^{11} y_4^{13}+ 
495 y_1^2 y_2^4 y_3^{11} y_4^{14}+ 160 y_1^5 y_2^6 y_3^{11} y_4^{10}+ 270 y_1^6 y_2^5 
y_3^{11} y_4^{10}+ 1236 y_1^4 y_2^6 y_3^{11} y_4^{11}+ 4384 y_1^5 y_2^5 y_3^{11} y_4^{11}+ 
792 y_1^6 y_2^4 y_3^{11} y_4^{11} \\[.05cm] & + 10128 y_1^4 y_2^5 y_3^{11} y_4^{12}+ 3696 y_1^5 
y_2^4 y_3^{11} y_4^{12}+ 4032 y_1^3 y_2^5 y_3^{11} y_4^{13}+ 4752 y_1^4 y_2^4 y_3^{11} 
y_4^{13}+ 1980 y_1^3 y_2^4 y_3^{11} y_4^{14}+ 220 y_1^2 y_2^4 y_3^{11} y_4^{15}+ 15 y_1^6 
y_2^6 y_3^{11} y_4^{10}+ 1084 y_1^5 y_2^6 y_3^{11} y_4^{11} \\[.05cm] & + 720 y_1^6 y_2^5 y_3^{11} 
y_4^{11}+ 3444 y_1^4 y_2^6 y_3^{11} y_4^{12}+ 6944 y_1^5 y_2^5 y_3^{11} y_4^{12}+ 924 
y_1^6 y_2^4 y_3^{11} y_4^{12}+ 10564 y_1^4 y_2^5 y_3^{11} y_4^{13}+ 3168 y_1^5 y_2^4 
y_3^{11} y_4^{13}+ 2856 y_1^3 y_2^5 y_3^{11} y_4^{14}+ 2970 y_1^4 y_2^4 y_3^{11} y_4^{14} \\[.05cm] & + 
880 y_1^3 y_2^4 y_3^{11} y_4^{15}+ 66 y_1^2 y_2^4 y_3^{11} y_4^{16}+ 120 y_1^6 y_2^6 
y_3^{11} y_4^{11}+ 70 y_1^4 y_2^7 y_3^{11} y_4^{12}+ 3172 y_1^5 y_2^6 y_3^{11} y_4^{12}+ 
1260 y_1^6 y_2^5 y_3^{11} y_4^{12}+ 5340 y_1^4 y_2^6 y_3^{11} y_4^{13}+ 7504 y_1^5 
y_2^5 y_3^{11} y_4^{13} \\[.05cm] & + 792 y_1^6 y_2^4 y_3^{11} y_4^{13}+ 7496 y_1^4 y_2^5 y_3^{11} 
y_4^{14}+ 1980 y_1^5 y_2^4 y_3^{11} y_4^{14}+ 1344 y_1^3 y_2^5 y_3^{11} y_4^{15}+ 1320 
y_1^4 y_2^4 y_3^{11} y_4^{15}+ 264 y_1^3 y_2^4 y_3^{11} y_4^{16}+ 12 y_1^2 y_2^4 y_3^{11} 
y_4^{17}+ 200 y_1^5 y_2^7 y_3^{11} y_4^{12} \\ % [.05cm]

& + 420 y_1^6 y_2^6 y_3^{11} y_4^{12}+ 320 y_1^4 
y_2^7 y_3^{11} y_4^{13}+ 5224 y_1^5 y_2^6 y_3^{11} y_4^{13}+ 1512 y_1^6 y_2^5 y_3^{11} 
y_4^{13}+ 5000 y_1^4 y_2^6 y_3^{11} y_4^{14}+ 5600 y_1^5 y_2^5 y_3^{11} y_4^{14}+ 495 
y_1^6 y_2^4 y_3^{11} y_4^{14}+ 3564 y_1^4 y_2^5 y_3^{11} y_4^{15} \\[.05cm] & + 880 y_1^5 y_2^4 
y_3^{11} y_4^{15}+ 396 y_1^3 y_2^5 y_3^{11} y_4^{16}+ 396 y_1^4 y_2^4 y_3^{11} y_4^{16}+ 48 
y_1^3 y_2^4 y_3^{11} y_4^{17}+ y_1^2 y_2^4 y_3^{11} y_4^{18}+ 20 y_1^6 y_2^7 y_3^{11} 
y_4^{12}+ 908 y_1^5 y_2^7 y_3^{11} y_4^{13}+ 840 y_1^6 y_2^6 y_3^{11} y_4^{13}+ 582 y_1^4 
y_2^7 y_3^{11} y_4^{14} \\[.05cm] & + 5280 y_1^5 y_2^6 y_3^{11} y_4^{14}+ 1260 y_1^6 y_2^5 y_3^{11} 
y_4^{14}+ 2860 y_1^4 y_2^6 y_3^{11} y_4^{15}+ 2848 y_1^5 y_2^5 y_3^{11} y_4^{15}+ 220 
y_1^6 y_2^4 y_3^{11} y_4^{15}+ 1082 y_1^4 y_2^5 y_3^{11} y_4^{16}+ 264 y_1^5 y_2^4 
y_3^{11} y_4^{16}+ 64 y_1^3 y_2^5 y_3^{11} y_4^{17} \\[.05cm] & + 72 y_1^4 y_2^4 y_3^{11} y_4^{17}+ 4 
y_1^3 y_2^4 y_3^{11} y_4^{18}+ 120 y_1^6 y_2^7 y_3^{11} y_4^{13}+ 1660 y_1^5 y_2^7 
y_3^{11} y_4^{14}+ 1050 y_1^6 y_2^6 y_3^{11} y_4^{14}+ 528 y_1^4 y_2^7 y_3^{11} y_4^{15}+ 
3340 y_1^5 y_2^6 y_3^{11} y_4^{15}+ 720 y_1^6 y_2^5 y_3^{11} y_4^{15} \\[.05cm] & + 956 y_1^4 y_2^6 
y_3^{11} y_4^{16}+ 944 y_1^5 y_2^5 y_3^{11} y_4^{16}+ 66 y_1^6 y_2^4 y_3^{11} y_4^{16}+ 188 
y_1^4 y_2^5 y_3^{11} y_4^{17}+ 48 y_1^5 y_2^4 y_3^{11} y_4^{17}+ 4 y_1^3 y_2^5 y_3^{11} 
y_4^{18}+ 6 y_1^4 y_2^4 y_3^{11} y_4^{18}+ 84 y_1^5 y_2^8 y_3^{11} y_4^{14}+ 300 y_1^6 
y_2^7 y_3^{11} y_4^{14} \\[.05cm] & + 1548 y_1^5 y_2^7 y_3^{11} y_4^{15}+ 840 y_1^6 y_2^6 y_3^{11} 
y_4^{15}+ 242 y_1^4 y_2^7 y_3^{11} y_4^{16}+ 1284 y_1^5 y_2^6 y_3^{11} y_4^{16}+ 270 
y_1^6 y_2^5 y_3^{11} y_4^{16}+ 164 y_1^4 y_2^6 y_3^{11} y_4^{17}+ 184 y_1^5 y_2^5 y_3^{11} 
y_4^{17}+ 12 y_1^6 y_2^4 y_3^{11} y_4^{17} \\[.05cm] & + 14 y_1^4 y_2^5 y_3^{11} y_4^{18}+ 4 y_1^5 
y_2^4 y_3^{11} y_4^{18}+ 15 y_1^6 y_2^8 y_3^{11} y_4^{14}+ 200 y_1^5 y_2^8 y_3^{11} 
y_4^{15}+ 400 y_1^6 y_2^7 y_3^{11} y_4^{15}+ 764 y_1^5 y_2^7 y_3^{11} y_4^{16}+ 420 y_1^6 
y_2^6 y_3^{11} y_4^{16}+ 48 y_1^4 y_2^7 y_3^{11} y_4^{17}+ 272 y_1^5 y_2^6 y_3^{11} 
y_4^{17} \\[.05cm] & + 60 y_1^6 y_2^5 y_3^{11} y_4^{17}+ 10 y_1^4 y_2^6 y_3^{11} y_4^{18}+ 16 y_1^5 
y_2^5 y_3^{11} y_4^{18}+ y_1^6 y_2^4 y_3^{11} y_4^{18}+ 60 y_1^6 y_2^8 y_3^{11} y_4^{15}+ 
160 y_1^5 y_2^8 y_3^{11} y_4^{16}+ 300 y_1^6 y_2^7 y_3^{11} y_4^{16}+ 184 y_1^5 y_2^7 
y_3^{11} y_4^{17}+ 120 y_1^6 y_2^6 y_3^{11} y_4^{17} \\[.05cm] & + 2 y_1^4 y_2^7 y_3^{11} y_4^{18}+ 24 
y_1^5 y_2^6 y_3^{11} y_4^{18}+ 6 y_1^6 y_2^5 y_3^{11} y_4^{18}+ 90 y_1^6 y_2^8 y_3^{11} 
y_4^{16}+ 48 y_1^5 y_2^8 y_3^{11} y_4^{17}+ 120 y_1^6 y_2^7 y_3^{11} y_4^{17}+ 16 y_1^5 
y_2^7 y_3^{11} y_4^{18}+ 15 y_1^6 y_2^6 y_3^{11} y_4^{18}+ 6 y_1^6 y_2^9 y_3^{11} y_4^{16} \\[.05cm] & + 
60 y_1^6 y_2^8 y_3^{11} y_4^{17}+ 4 y_1^5 y_2^8 y_3^{11} y_4^{18}+ 20 y_1^6 y_2^7 y_3^{11} 
y_4^{18}+ 12 y_1^6 y_2^9 y_3^{11} y_4^{17}+ 15 y_1^6 y_2^8 y_3^{11} y_4^{18}+ 6 y_1^6 
y_2^9 y_3^{11} y_4^{18}+ y_1^6 y_2^{10} y_3^{11} y_4^{18}+ 4 y_1^3 y_2^5 y_3^{12} y_4^8+ 14 
y_1^4 y_2^5 y_3^{12} y_4^8 \\[.05cm] & + 44 y_1^3 y_2^5 y_3^{12} y_4^9+ 16 y_1^5 y_2^5 y_3^{12} 
y_4^8+ 156 y_1^4 y_2^5 y_3^{12} y_4^9+ 220 y_1^3 y_2^5 y_3^{12} y_4^{10}+ 6 y_1^6 
y_2^5 y_3^{12} y_4^8+ 172 y_1^5 y_2^5 y_3^{12} y_4^9+ 782 y_1^4 y_2^5 y_3^{12} y_4^{10}+ 
660 y_1^3 y_2^5 y_3^{12} y_4^{11}+ 64 y_1^6 y_2^5 y_3^{12} y_4^9 \\[.05cm] & + 90 y_1^4 y_2^6 
y_3^{12} y_4^{10}+ 840 y_1^5 y_2^5 y_3^{12} y_4^{10}+ 2332 y_1^4 y_2^5 y_3^{12} y_4^{11}+ 
1320 y_1^3 y_2^5 y_3^{12} y_4^{12}+ 120 y_1^5 y_2^6 y_3^{12} y_4^{10}+ 310 y_1^6 y_2^5 
y_3^{12} y_4^{10}+ 726 y_1^4 y_2^6 y_3^{12} y_4^{11}+ 2460 y_1^5 y_2^5 y_3^{12} y_4^{11} \\[.05cm] & + 
4604 y_1^4 y_2^5 y_3^{12} y_4^{12}+ 1848 y_1^3 y_2^5 y_3^{12} y_4^{13}+ 33 y_1^6 y_2^6 
y_3^{12} y_4^{10}+ 988 y_1^5 y_2^6 y_3^{12} y_4^{11}+ 900 y_1^6 y_2^5 y_3^{12} y_4^{11}+ 
2580 y_1^4 y_2^6 y_3^{12} y_4^{12}+ 4800 y_1^5 y_2^5 y_3^{12} y_4^{12}+ 6328 y_1^4 
y_2^5 y_3^{12} y_4^{13} \\[.05cm] & + 1848 y_1^3 y_2^5 y_3^{12} y_4^{14}+ 286 y_1^6 y_2^6 y_3^{12} 
y_4^{11}+ 42 y_1^4 y_2^7 y_3^{12} y_4^{12}+ 3596 y_1^5 y_2^6 y_3^{12} y_4^{12}+ 1740 
y_1^6 y_2^5 y_3^{12} y_4^{12}+ 5292 y_1^4 y_2^6 y_3^{12} y_4^{13}+ 6552 y_1^5 y_2^5 
y_3^{12} y_4^{13}+ 6188 y_1^4 y_2^5 y_3^{12} y_4^{14} \\[.05cm] & + 1320 y_1^3 y_2^5 y_3^{12} y_4^{15}+ 
240 y_1^5 y_2^7 y_3^{12} y_4^{12}+ 1100 y_1^6 y_2^6 y_3^{12} y_4^{12}+ 260 y_1^4 y_2^7 
y_3^{12} y_4^{13}+ 7588 y_1^5 y_2^6 y_3^{12} y_4^{13}+ 2352 y_1^6 y_2^5 y_3^{12} y_4^{13}+ 
6888 y_1^4 y_2^6 y_3^{12} y_4^{14}+ 6384 y_1^5 y_2^5 y_3^{12} y_4^{14} \\[.05cm] & + 4312 y_1^4 
y_2^5 y_3^{12} y_4^{15}+ 660 y_1^3 y_2^5 y_3^{12} y_4^{16}+ 72 y_1^6 y_2^7 y_3^{12} 
y_4^{12}+ 1420 y_1^5 y_2^7 y_3^{12} y_4^{13}+ 2464 y_1^6 y_2^6 y_3^{12} y_4^{13}+ 678 
y_1^4 y_2^7 y_3^{12} y_4^{14}+ 10220 y_1^5 y_2^6 y_3^{12} y_4^{14}+ 2268 y_1^6 y_2^5 
y_3^{12} y_4^{14} \\[.05cm] & + 5880 y_1^4 y_2^6 y_3^{12} y_4^{15}+ 4440 y_1^5 y_2^5 y_3^{12} y_4^{15}+ 
2102 y_1^4 y_2^5 y_3^{12} y_4^{16}+ 220 y_1^3 y_2^5 y_3^{12} y_4^{17}+ 480 y_1^6 y_2^7 
y_3^{12} y_4^{13}+ 3548 y_1^5 y_2^7 y_3^{12} y_4^{14}+ 3542 y_1^6 y_2^6 y_3^{12} y_4^{14}+ 
960 y_1^4 y_2^7 y_3^{12} y_4^{15} \\[.05cm] & + 9100 y_1^5 y_2^6 y_3^{12} y_4^{15}+ 1560 y_1^6 y_2^5 
y_3^{12} y_4^{15}+ 3276 y_1^4 y_2^6 y_3^{12} y_4^{16}+ 2160 y_1^5 y_2^5 y_3^{12} y_4^{16}+ 
684 y_1^4 y_2^5 y_3^{12} y_4^{17}+ 44 y_1^3 y_2^5 y_3^{12} y_4^{18}+ 140 y_1^5 y_2^8 
y_3^{12} y_4^{14}+ 1368 y_1^6 y_2^7 y_3^{12} y_4^{14} \\[.05cm] & + 4836 y_1^5 y_2^7 y_3^{12} y_4^{15}+ 
3388 y_1^6 y_2^6 y_3^{12} y_4^{15}+ 790 y_1^4 y_2^7 y_3^{12} y_4^{16}+ 5348 y_1^5 y_2^6 
y_3^{12} y_4^{16}+ 750 y_1^6 y_2^5 y_3^{12} y_4^{16}+ 1140 y_1^4 y_2^6 y_3^{12} y_4^{17}+ 
700 y_1^5 y_2^5 y_3^{12} y_4^{17}+ 134 y_1^4 y_2^5 y_3^{12} y_4^{18} \\[.05cm] & + 4 y_1^3 y_2^5 
y_3^{12} y_4^{19}+ 78 y_1^6 y_2^8 y_3^{12} y_4^{14}+ 520 y_1^5 y_2^8 y_3^{12} y_4^{15}+ 
2160 y_1^6 y_2^7 y_3^{12} y_4^{15}+ 3864 y_1^5 y_2^7 y_3^{12} y_4^{16}+ 2156 y_1^6 
y_2^6 y_3^{12} y_4^{16}+ 372 y_1^4 y_2^7 y_3^{12} y_4^{17}+ 1996 y_1^5 y_2^6 y_3^{12} 
y_4^{17} \\[.05cm] & + 240 y_1^6 y_2^5 y_3^{12} y_4^{17}+ 222 y_1^4 y_2^6 y_3^{12} y_4^{18}+ 136 y_1^5 
y_2^5 y_3^{12} y_4^{18}+ 12 y_1^4 y_2^5 y_3^{12} y_4^{19}+ 364 y_1^6 y_2^8 y_3^{12} 
y_4^{15}+ 740 y_1^5 y_2^8 y_3^{12} y_4^{16}+ 2040 y_1^6 y_2^7 y_3^{12} y_4^{16}+ 1796 
y_1^5 y_2^7 y_3^{12} y_4^{17} \\[.05cm] & + 880 y_1^6 y_2^6 y_3^{12} y_4^{17}+ 90 y_1^4 y_2^7 y_3^{12} 
y_4^{18}+ 428 y_1^5 y_2^6 y_3^{12} y_4^{18}+ 46 y_1^6 y_2^5 y_3^{12} y_4^{18}+ 18 y_1^4 
y_2^6 y_3^{12} y_4^{19}+ 12 y_1^5 y_2^5 y_3^{12} y_4^{19}+ 676 y_1^6 y_2^8 y_3^{12} 
y_4^{16}+ 496 y_1^5 y_2^8 y_3^{12} y_4^{17} \\[.05cm] & + 1152 y_1^6 y_2^7 y_3^{12} y_4^{17}+ 444 
y_1^5 y_2^7 y_3^{12} y_4^{18}+ 209 y_1^6 y_2^6 y_3^{12} y_4^{18}+ 8 y_1^4 y_2^7 y_3^{12} 
y_4^{19}+ 40 y_1^5 y_2^6 y_3^{12} y_4^{19}+ 4 y_1^6 y_2^5 y_3^{12} y_4^{19}+ 42 y_1^6 
y_2^9 y_3^{12} y_4^{16}+ 624 y_1^6 y_2^8 y_3^{12} y_4^{17} \\[.05cm] & + 152 y_1^5 y_2^8 y_3^{12} 
y_4^{18}+ 360 y_1^6 y_2^7 y_3^{12} y_4^{18}+ 44 y_1^5 y_2^7 y_3^{12} y_4^{19}+ 22 y_1^6 
y_2^6 y_3^{12} y_4^{19}+ 112 y_1^6 y_2^9 y_3^{12} y_4^{17}+ 286 y_1^6 y_2^8 y_3^{12} 
y_4^{18}+ 16 y_1^5 y_2^8 y_3^{12} y_4^{19}+ 48 y_1^6 y_2^7 y_3^{12} y_4^{19} \\[.05cm] & + 98 y_1^6 
y_2^9 y_3^{12} y_4^{18}+ 52 y_1^6 y_2^8 y_3^{12} y_4^{19}+ 9 y_1^6 y_2^{10} y_3^{12} y_4^{18}+ 
28 y_1^6 y_2^9 y_3^{12} y_4^{19}+ 6 y_1^6 y_2^{10} y_3^{12} y_4^{19}+ 22 y_1^4 y_2^6 
y_3^{13} y_4^{10}+ 40 y_1^5 y_2^6 y_3^{13} y_4^{10}+ 212 y_1^4 y_2^6 y_3^{13} y_4^{11}+ 21 
y_1^6 y_2^6 y_3^{13} y_4^{10} \\[.05cm] & + 388 y_1^5 y_2^6 y_3^{13} y_4^{11}+ 918 y_1^4 y_2^6 y_3^{13} 
y_4^{12}+ 200 y_1^6 y_2^6 y_3^{13} y_4^{11}+ 14 y_1^4 y_2^7 y_3^{13} y_4^{12}+ 1692 y_1^5 
y_2^6 y_3^{13} y_4^{12}+ 2352 y_1^4 y_2^6 y_3^{13} y_4^{13}+ 160 y_1^5 y_2^7 y_3^{13} 
y_4^{12}+ 858 y_1^6 y_2^6 y_3^{13} y_4^{12} \\[.05cm] & + 112 y_1^4 y_2^7 y_3^{13} y_4^{13}+ 4368 
y_1^5 y_2^6 y_3^{13} y_4^{13}+ 3948 y_1^4 y_2^6 y_3^{13} y_4^{14}+ 96 y_1^6 y_2^7 y_3^{13} 
y_4^{12}+ 1172 y_1^5 y_2^7 y_3^{13} y_4^{13}+ 2184 y_1^6 y_2^6 y_3^{13} y_4^{13}+ 392 
y_1^4 y_2^7 y_3^{13} y_4^{14}+ 7392 y_1^5 y_2^6 y_3^{13} y_4^{14} \\[.05cm] & + 4536 y_1^4 y_2^6 
y_3^{13} y_4^{15}+ 720 y_1^6 y_2^7 y_3^{13} y_4^{13}+ 3740 y_1^5 y_2^7 y_3^{13} y_4^{14}+ 
3654 y_1^6 y_2^6 y_3^{13} y_4^{14}+ 784 y_1^4 y_2^7 y_3^{13} y_4^{15}+ 8568 y_1^5 y_2^6 
y_3^{13} y_4^{15}+ 3612 y_1^4 y_2^6 y_3^{13} y_4^{16}+ 140 y_1^5 y_2^8 y_3^{13} y_4^{14} \\[.05cm] & + 
2364 y_1^6 y_2^7 y_3^{13} y_4^{14}+ 6788 y_1^5 y_2^7 y_3^{13} y_4^{15}+ 4200 y_1^6 
y_2^6 y_3^{13} y_4^{15}+ 980 y_1^4 y_2^7 y_3^{13} y_4^{16}+ 6888 y_1^5 y_2^6 y_3^{13} 
y_4^{16}+ 1968 y_1^4 y_2^6 y_3^{13} y_4^{17}+ 165 y_1^6 y_2^8 y_3^{13} y_4^{14}+ 720 
y_1^5 y_2^8 y_3^{13} y_4^{15} \\[.05cm] & + 4440 y_1^6 y_2^7 y_3^{13} y_4^{15}+ 7660 y_1^5 y_2^7 
y_3^{13} y_4^{16}+ 3360 y_1^6 y_2^6 y_3^{13} y_4^{16}+ 784 y_1^4 y_2^7 y_3^{13} y_4^{17}+ 
3792 y_1^5 y_2^6 y_3^{13} y_4^{17}+ 702 y_1^4 y_2^6 y_3^{13} y_4^{18}+ 906 y_1^6 y_2^8 
y_3^{13} y_4^{15}+ 1520 y_1^5 y_2^8 y_3^{13} y_4^{16} \\[.05cm] & + 5220 y_1^6 y_2^7 y_3^{13} y_4^{16}+ 
5500 y_1^5 y_2^7 y_3^{13} y_4^{17}+ 1848 y_1^6 y_2^6 y_3^{13} y_4^{17}+ 392 y_1^4 y_2^7 
y_3^{13} y_4^{18}+ 1368 y_1^5 y_2^6 y_3^{13} y_4^{18}+ 148 y_1^4 y_2^6 y_3^{13} y_4^{19}+ 
2073 y_1^6 y_2^8 y_3^{13} y_4^{16}+ 1680 y_1^5 y_2^8 y_3^{13} y_4^{17} \\[.05cm] & + 3936 y_1^6 
y_2^7 y_3^{13} y_4^{17}+ 2452 y_1^5 y_2^7 y_3^{13} y_4^{18}+ 669 y_1^6 y_2^6 y_3^{13} 
y_4^{18}+ 112 y_1^4 y_2^7 y_3^{13} y_4^{19}+ 292 y_1^5 y_2^6 y_3^{13} y_4^{19}+ 14 y_1^4 
y_2^6 y_3^{13} y_4^{20}+ 126 y_1^6 y_2^9 y_3^{13} y_4^{16}+ 2532 y_1^6 y_2^8 y_3^{13} 
y_4^{17} \\[.05cm] & + 1020 y_1^5 y_2^8 y_3^{13} y_4^{18}+ 1860 y_1^6 y_2^7 y_3^{13} y_4^{18}+ 620 
y_1^5 y_2^7 y_3^{13} y_4^{19}+ 144 y_1^6 y_2^6 y_3^{13} y_4^{19}+ 14 y_1^4 y_2^7 y_3^{13} 
y_4^{20}+ 28 y_1^5 y_2^6 y_3^{13} y_4^{20}+ 440 y_1^6 y_2^9 y_3^{13} y_4^{17}+ 1743 y_1^6 
y_2^8 y_3^{13} y_4^{18} \\[.05cm] & + 320 y_1^5 y_2^8 y_3^{13} y_4^{19}+ 504 y_1^6 y_2^7 y_3^{13} 
y_4^{19}+ 68 y_1^5 y_2^7 y_3^{13} y_4^{20}+ 14 y_1^6 y_2^6 y_3^{13} y_4^{20}+ 576 y_1^6 
y_2^9 y_3^{13} y_4^{18}+ 642 y_1^6 y_2^8 y_3^{13} y_4^{19}+ 40 y_1^5 y_2^8 y_3^{13} 
y_4^{20}+ 60 y_1^6 y_2^7 y_3^{13} y_4^{20} \\[.05cm] & + 36 y_1^6 y_2^{10} y_3^{13} y_4^{18}+ 336 y_1^6 
y_2^9 y_3^{13} y_4^{19}+ 99 y_1^6 y_2^8 y_3^{13} y_4^{20}+ 54 y_1^6 y_2^{10} y_3^{13} 
y_4^{19}+ 74 y_1^6 y_2^9 y_3^{13} y_4^{20}+ 21 y_1^6 y_2^{10} y_3^{13} y_4^{20}+ 2 y_1^4 
y_2^6 y_3^{14} y_4^{10}+ 4 y_1^5 y_2^6 y_3^{14} y_4^{10}+ 22 y_1^4 y_2^6 y_3^{14} y_4^{11} \\[.05cm] & + 
3 y_1^6 y_2^6 y_3^{14} y_4^{10}+ 44 y_1^5 y_2^6 y_3^{14} y_4^{11}+ 110 y_1^4 y_2^6 
y_3^{14} y_4^{12}+ 30 y_1^6 y_2^6 y_3^{14} y_4^{11}+ 2 y_1^4 y_2^7 y_3^{14} y_4^{12}+ 220 
y_1^5 y_2^6 y_3^{14} y_4^{12}+ 330 y_1^4 y_2^6 y_3^{14} y_4^{13}+ 56 y_1^5 y_2^7 y_3^{14} 
y_4^{12}+ 138 y_1^6 y_2^6 y_3^{14} y_4^{12} \\[.05cm] & + 20 y_1^4 y_2^7 y_3^{14} y_4^{13}+ 660 y_1^5 
y_2^6 y_3^{14} y_4^{13}+ 660 y_1^4 y_2^6 y_3^{14} y_4^{14}+ 56 y_1^6 y_2^7 y_3^{14} 
y_4^{12}+ 488 y_1^5 y_2^7 y_3^{14} y_4^{13}+ 386 y_1^6 y_2^6 y_3^{14} y_4^{13}+ 90 y_1^4 
y_2^7 y_3^{14} y_4^{14}+ 1320 y_1^5 y_2^6 y_3^{14} y_4^{14} \\[.05cm] & + 924 y_1^4 y_2^6 y_3^{14} 
y_4^{15}+ 472 y_1^6 y_2^7 y_3^{14} y_4^{13}+ 1892 y_1^5 y_2^7 y_3^{14} y_4^{14}+ 730 
y_1^6 y_2^6 y_3^{14} y_4^{14}+ 240 y_1^4 y_2^7 y_3^{14} y_4^{15}+ 1848 y_1^5 y_2^6 
y_3^{14} y_4^{15}+ 924 y_1^4 y_2^6 y_3^{14} y_4^{16}+ 84 y_1^5 y_2^8 y_3^{14} y_4^{14} \\[.05cm] & + 
1778 y_1^6 y_2^7 y_3^{14} y_4^{14}+ 4288 y_1^5 y_2^7 y_3^{14} y_4^{15}+ 980 y_1^6 y_2^6 
y_3^{14} y_4^{15}+ 420 y_1^4 y_2^7 y_3^{14} y_4^{16}+ 1848 y_1^5 y_2^6 y_3^{14} y_4^{16}+ 
660 y_1^4 y_2^6 y_3^{14} y_4^{17}+ 180 y_1^6 y_2^8 y_3^{14} y_4^{14}+ 560 y_1^5 y_2^8 
y_3^{14} y_4^{15} \\[.05cm] & + 3932 y_1^6 y_2^7 y_3^{14} y_4^{15}+ 6272 y_1^5 y_2^7 y_3^{14} y_4^{16}+ 
952 y_1^6 y_2^6 y_3^{14} y_4^{16}+ 504 y_1^4 y_2^7 y_3^{14} y_4^{17}+ 1320 y_1^5 y_2^6 
y_3^{14} y_4^{17}+ 330 y_1^4 y_2^6 y_3^{14} y_4^{18}+ 1164 y_1^6 y_2^8 y_3^{14} y_4^{15}+ 
1600 y_1^5 y_2^8 y_3^{14} y_4^{16} \\[.05cm] & + 5632 y_1^6 y_2^7 y_3^{14} y_4^{16}+ 6160 y_1^5 
y_2^7 y_3^{14} y_4^{17}+ 668 y_1^6 y_2^6 y_3^{14} y_4^{17}+ 420 y_1^4 y_2^7 y_3^{14} 
y_4^{18}+ 660 y_1^5 y_2^6 y_3^{14} y_4^{18}+ 110 y_1^4 y_2^6 y_3^{14} y_4^{19}+ 3242 
y_1^6 y_2^8 y_3^{14} y_4^{16}+ 2544 y_1^5 y_2^8 y_3^{14} y_4^{17} \\[.05cm] & + 5428 y_1^6 y_2^7 
y_3^{14} y_4^{17}+ 4088 y_1^5 y_2^7 y_3^{14} y_4^{18}+ 331 y_1^6 y_2^6 y_3^{14} y_4^{18}+ 
240 y_1^4 y_2^7 y_3^{14} y_4^{19}+ 220 y_1^5 y_2^6 y_3^{14} y_4^{19}+ 22 y_1^4 y_2^6 
y_3^{14} y_4^{20}+ 210 y_1^6 y_2^9 y_3^{14} y_4^{16}+ 5050 y_1^6 y_2^8 y_3^{14} y_4^{17} \\[.05cm] & + 
2440 y_1^5 y_2^8 y_3^{14} y_4^{18}+ 3532 y_1^6 y_2^7 y_3^{14} y_4^{18}+ 1792 y_1^5 
y_2^7 y_3^{14} y_4^{19}+ 110 y_1^6 y_2^6 y_3^{14} y_4^{19}+ 90 y_1^4 y_2^7 y_3^{14} 
y_4^{20}+ 44 y_1^5 y_2^6 y_3^{14} y_4^{20}+ 2 y_1^4 y_2^6 y_3^{14} y_4^{21}+ 940 y_1^6 
y_2^9 y_3^{14} y_4^{17} \\[.05cm] & + 4766 y_1^6 y_2^8 y_3^{14} y_4^{18}+ 1424 y_1^5 y_2^8 y_3^{14} 
y_4^{19}+ 1508 y_1^6 y_2^7 y_3^{14} y_4^{19}+ 488 y_1^5 y_2^7 y_3^{14} y_4^{20}+ 22 y_1^6 
y_2^6 y_3^{14} y_4^{20}+ 20 y_1^4 y_2^7 y_3^{14} y_4^{21}+ 4 y_1^5 y_2^6 y_3^{14} y_4^{21}+ 
1694 y_1^6 y_2^9 y_3^{14} y_4^{18} \\[.05cm] &

+ 2744 y_1^6 y_2^8 y_3^{14} y_4^{19}+ 480 y_1^5 y_2^8 
y_3^{14} y_4^{20}+ 392 y_1^6 y_2^7 y_3^{14} y_4^{20}+ 72 y_1^5 y_2^7 y_3^{14} y_4^{21}+ 2 
y_1^6 y_2^6 y_3^{14} y_4^{21}+ 2 y_1^4 y_2^7 y_3^{14} y_4^{22}+ 84 y_1^6 y_2^{10} y_3^{14} 
y_4^{18}+ 1544 y_1^6 y_2^9 y_3^{14} y_4^{19}+ 910 y_1^6 y_2^8 y_3^{14} y_4^{20}\\[.05cm] & + 80 y_1^5 
y_2^8 y_3^{14} y_4^{21}+ 52 y_1^6 y_2^7 y_3^{14} y_4^{21}+ 4 y_1^5 y_2^7 y_3^{14} y_4^{22}+ 
210 y_1^6 y_2^{10} y_3^{14} y_4^{19}+ 722 y_1^6 y_2^9 y_3^{14} y_4^{20}+ 146 y_1^6 y_2^8 
y_3^{14} y_4^{21}+ 4 y_1^5 y_2^8 y_3^{14} y_4^{22}+ 2 y_1^6 y_2^7 y_3^{14} y_4^{22}+ 180 
y_1^6 y_2^{10} y_3^{14} y_4^{20}\\[.05cm] & + 148 y_1^6 y_2^9 y_3^{14} y_4^{21}+ 6 y_1^6 y_2^8 y_3^{14} 
y_4^{22}+ 56 y_1^6 y_2^{10} y_3^{14} y_4^{21}+ 6 y_1^6 y_2^9 y_3^{14} y_4^{22}+ 3 y_1^6 
y_2^{10} y_3^{14} y_4^{22}+ 8 y_1^5 y_2^7 y_3^{15} y_4^{12}+ 12 y_1^6 y_2^7 y_3^{15} y_4^{12}+ 
80 y_1^5 y_2^7 y_3^{15} y_4^{13}+ 112 y_1^6 y_2^7 y_3^{15} y_4^{13}\\[.05cm] & + 360 y_1^5 y_2^7 
y_3^{15} y_4^{14}+ 28 y_1^5 y_2^8 y_3^{15} y_4^{14}+ 472 y_1^6 y_2^7 y_3^{15} y_4^{14}+ 960 
y_1^5 y_2^7 y_3^{15} y_4^{15}+ 105 y_1^6 y_2^8 y_3^{15} y_4^{14}+ 232 y_1^5 y_2^8 y_3^{15} 
y_4^{15}+ 1184 y_1^6 y_2^7 y_3^{15} y_4^{15}+ 1680 y_1^5 y_2^7 y_3^{15} y_4^{16} \\[.05cm] & + 796 
y_1^6 y_2^8 y_3^{15} y_4^{15}+ 848 y_1^5 y_2^8 y_3^{15} y_4^{16}+ 1960 y_1^6 y_2^7 
y_3^{15} y_4^{16}+ 2016 y_1^5 y_2^7 y_3^{15} y_4^{17}+ 2656 y_1^6 y_2^8 y_3^{15} y_4^{16}+ 
1792 y_1^5 y_2^8 y_3^{15} y_4^{17}+ 2240 y_1^6 y_2^7 y_3^{15} y_4^{17}+ 1680 y_1^5 
y_2^7 y_3^{15} y_4^{18} \\[.05cm] & + 210 y_1^6 y_2^9 y_3^{15} y_4^{16}+ 5108 y_1^6 y_2^8 y_3^{15} 
y_4^{17}+ 2408 y_1^5 y_2^8 y_3^{15} y_4^{18}+ 1792 y_1^6 y_2^7 y_3^{15} y_4^{18}+ 960 
y_1^5 y_2^7 y_3^{15} y_4^{19}+ 1180 y_1^6 y_2^9 y_3^{15} y_4^{17}+ 6218 y_1^6 y_2^8 
y_3^{15} y_4^{18}+ 2128 y_1^5 y_2^8 y_3^{15} y_4^{19} \\[.05cm] & + 992 y_1^6 y_2^7 y_3^{15} y_4^{19}+ 
360 y_1^5 y_2^7 y_3^{15} y_4^{20}+ 2786 y_1^6 y_2^9 y_3^{15} y_4^{18}+ 4940 y_1^6 y_2^8 
y_3^{15} y_4^{19}+ 1232 y_1^5 y_2^8 y_3^{15} y_4^{20}+ 364 y_1^6 y_2^7 y_3^{15} y_4^{20}+ 
80 y_1^5 y_2^7 y_3^{15} y_4^{21}+ 126 y_1^6 y_2^{10} y_3^{15} y_4^{18} \\[.05cm] & + 3560 y_1^6 y_2^9 
y_3^{15} y_4^{19}+ 2536 y_1^6 y_2^8 y_3^{15} y_4^{20}+ 448 y_1^5 y_2^8 y_3^{15} y_4^{21}+ 
80 y_1^6 y_2^7 y_3^{15} y_4^{21}+ 8 y_1^5 y_2^7 y_3^{15} y_4^{22}+ 462 y_1^6 y_2^{10} 
y_3^{15} y_4^{19}+ 2630 y_1^6 y_2^9 y_3^{15} y_4^{20}+ 796 y_1^6 y_2^8 y_3^{15} y_4^{21} \\[.05cm] & + 
92 y_1^5 y_2^8 y_3^{15} y_4^{22}+ 8 y_1^6 y_2^7 y_3^{15} y_4^{22}+ 651 y_1^6 y_2^{10} 
y_3^{15} y_4^{20}+ 1100 y_1^6 y_2^9 y_3^{15} y_4^{21}+ 133 y_1^6 y_2^8 y_3^{15} y_4^{22}+ 8 
y_1^5 y_2^8 y_3^{15} y_4^{23}+ 432 y_1^6 y_2^{10} y_3^{15} y_4^{21}+ 230 y_1^6 y_2^9 
y_3^{15} y_4^{22}+ 8 y_1^6 y_2^8 y_3^{15} y_4^{23} \\[.05cm] & + 129 y_1^6 y_2^{10} y_3^{15} y_4^{22}+ 16 
y_1^6 y_2^9 y_3^{15} y_4^{23}+ 12 y_1^6 y_2^{10} y_3^{15} y_4^{23}+ 4 y_1^5 y_2^8 y_3^{16} 
y_4^{14}+ 30 y_1^6 y_2^8 y_3^{16} y_4^{14}+ 40 y_1^5 y_2^8 y_3^{16} y_4^{15}+ 264 y_1^6 
y_2^8 y_3^{16} y_4^{15}+ 180 y_1^5 y_2^8 y_3^{16} y_4^{16}+ 1036 y_1^6 y_2^8 y_3^{16} 
y_4^{16} \\[.05cm] & + 480 y_1^5 y_2^8 y_3^{16} y_4^{17}+ 126 y_1^6 y_2^9 y_3^{16} y_4^{16}+ 2384 
y_1^6 y_2^8 y_3^{16} y_4^{17}+ 840 y_1^5 y_2^8 y_3^{16} y_4^{18}+ 872 y_1^6 y_2^9 y_3^{16} 
y_4^{17}+ 3556 y_1^6 y_2^8 y_3^{16} y_4^{18}+ 1008 y_1^5 y_2^8 y_3^{16} y_4^{19}+ 2610 
y_1^6 y_2^9 y_3^{16} y_4^{18} 
\end{array}
$} \nonumber
\eeq

%=================================================================

\beq
\resizebox{\textwidth}{!}{$
\begin{array}{cl}
& + 3584 y_1^6 y_2^8 y_3^{16} y_4^{19}+ 840 y_1^5 y_2^8 
y_3^{16} y_4^{20}+ 126 y_1^6 y_2^{10} y_3^{16} y_4^{18}+ 4404 y_1^6 y_2^9 y_3^{16} y_4^{19}+ 
2464 y_1^6 y_2^8 y_3^{16} y_4^{20}+ 480 y_1^5 y_2^8 y_3^{16} y_4^{21}+ 630 y_1^6 y_2^{10} 
y_3^{16} y_4^{19}+ 4570 y_1^6 y_2^9 y_3^{16} y_4^{20} \\[.05cm] & + 1136 y_1^6 y_2^8 y_3^{16} y_4^{21}+ 
180 y_1^5 y_2^8 y_3^{16} y_4^{22}+ 1290 y_1^6 y_2^{10} y_3^{16} y_4^{20}+ 2976 y_1^6 
y_2^9 y_3^{16} y_4^{21}+ 334 y_1^6 y_2^8 y_3^{16} y_4^{22}+ 40 y_1^5 y_2^8 y_3^{16} 
y_4^{23}+ 1380 y_1^6 y_2^{10} y_3^{16} y_4^{21}+ 1182 y_1^6 y_2^9 y_3^{16} y_4^{22} \\[.05cm] & + 56 
y_1^6 y_2^8 y_3^{16} y_4^{23}+ 4 y_1^5 y_2^8 y_3^{16} y_4^{24}+ 810 y_1^6 y_2^{10} y_3^{16} 
y_4^{22}+ 260 y_1^6 y_2^9 y_3^{16} y_4^{23}+ 4 y_1^6 y_2^8 y_3^{16} y_4^{24}+ 246 y_1^6 
y_2^{10} y_3^{16} y_4^{23}+ 24 y_1^6 y_2^9 y_3^{16} y_4^{24}+ 30 y_1^6 y_2^{10} y_3^{16} 
y_4^{24}+ 3 y_1^6 y_2^8 y_3^{17} y_4^{14} \\[.05cm] & + 30 y_1^6 y_2^8 y_3^{17} y_4^{15}+ 135 y_1^6 
y_2^8 y_3^{17} y_4^{16}+ 42 y_1^6 y_2^9 y_3^{17} y_4^{16}+ 360 y_1^6 y_2^8 y_3^{17} 
y_4^{17}+ 352 y_1^6 y_2^9 y_3^{17} y_4^{17}+ 630 y_1^6 y_2^8 y_3^{17} y_4^{18}+ 1304 
y_1^6 y_2^9 y_3^{17} y_4^{18}+ 756 y_1^6 y_2^8 y_3^{17} y_4^{19} \\[.05cm] & + 84 y_1^6 y_2^{10} y_3^{17} 
y_4^{18}+ 2800 y_1^6 y_2^9 y_3^{17} y_4^{19}+ 630 y_1^6 y_2^8 y_3^{17} y_4^{20}+ 546 
y_1^6 y_2^{10} y_3^{17} y_4^{19}+ 3836 y_1^6 y_2^9 y_3^{17} y_4^{20}+ 360 y_1^6 y_2^8 
y_3^{17} y_4^{21}+ 1515 y_1^6 y_2^{10} y_3^{17} y_4^{20}+ 3472 y_1^6 y_2^9 y_3^{17} y_4^{21} \\[.05cm] & + 
135 y_1^6 y_2^8 y_3^{17} y_4^{22}+ 2328 y_1^6 y_2^{10} y_3^{17} y_4^{21}+ 2072 y_1^6 
y_2^9 y_3^{17} y_4^{22}+ 30 y_1^6 y_2^8 y_3^{17} y_4^{23}+ 2145 y_1^6 y_2^{10} y_3^{17} 
y_4^{22}+ 784 y_1^6 y_2^9 y_3^{17} y_4^{23}+ 3 y_1^6 y_2^8 y_3^{17} y_4^{24}+ 1194 y_1^6 
y_2^{10} y_3^{17} y_4^{23}  \\[.05cm] 
& + 170 y_1^6 y_2^9 y_3^{17} y_4^{24}+ 381 y_1^6 y_2^{10} y_3^{17} 
y_4^{24}+ 16 y_1^6 y_2^9 y_3^{17} y_4^{25}+ 60 y_1^6 y_2^{10} y_3^{17} y_4^{25}+ 3 y_1^6 
y_2^{10} y_3^{17} y_4^{26}+ 6 y_1^6 y_2^9 y_3^{18} y_4^{16}+ 60 y_1^6 y_2^9 y_3^{18} y_4^{17}+ 
270 y_1^6 y_2^9 y_3^{18} y_4^{18}+ 36 y_1^6 y_2^{10} y_3^{18} y_4^{18} \\[.05cm] & + 720 y_1^6 y_2^9 
y_3^{18} y_4^{19}+ 294 y_1^6 y_2^{10} y_3^{18} y_4^{19}+ 1260 y_1^6 y_2^9 y_3^{18} y_4^{20}+ 
1056 y_1^6 y_2^{10} y_3^{18} y_4^{20}+ 1512 y_1^6 y_2^9 y_3^{18} y_4^{21}+ 2184 y_1^6 
y_2^{10} y_3^{18} y_4^{21}+ 1260 y_1^6 y_2^9 y_3^{18} y_4^{22}+ 2856 y_1^6 y_2^{10} y_3^{18} 
y_4^{22} \\[.05cm] & + 720 y_1^6 y_2^9 y_3^{18} y_4^{23}+ 2436 y_1^6 y_2^{10} y_3^{18} y_4^{23}+ 270 
y_1^6 y_2^9 y_3^{18} y_4^{24}+ 1344 y_1^6 y_2^{10} y_3^{18} y_4^{24}+ 60 y_1^6 y_2^9 
y_3^{18} y_4^{25}+ 456 y_1^6 y_2^{10} y_3^{18} y_4^{25}+ 6 y_1^6 y_2^9 y_3^{18} y_4^{26}+ 84 
y_1^6 y_2^{10} y_3^{18} y_4^{26} \\[.05cm] & + 6 y_1^6 y_2^{10} y_3^{18} y_4^{27}+ 9 y_1^6 y_2^{10} y_3^{19} 
y_4^{18}+ 90 y_1^6 y_2^{10} y_3^{19} y_4^{19}+ 405 y_1^6 y_2^{10} y_3^{19} y_4^{20}+ 1080 
y_1^6 y_2^{10} y_3^{19} y_4^{21}+ 1890 y_1^6 y_2^{10} y_3^{19} y_4^{22}+ 2268 y_1^6 y_2^{10} 
y_3^{19} y_4^{23}+ 1890 y_1^6 y_2^{10} y_3^{19} y_4^{24} \\[.05cm] & + 1080 y_1^6 y_2^{10} y_3^{19} 
y_4^{25}+ 405 y_1^6 y_2^{10} y_3^{19} y_4^{26}+ 90 y_1^6 y_2^{10} y_3^{19} y_4^{27}+ 9 y_1^6 
y_2^{10} y_3^{19} y_4^{28}+ y_1^6 y_2^{10} y_3^{20} y_4^{18}+ 12 y_1^6 y_2^{10} y_3^{20} y_4^{19}+ 
66 y_1^6 y_2^{10} y_3^{20} y_4^{20}+ 220 y_1^6 y_2^{10} y_3^{20} y_4^{21} \\[.05cm] & + 495 y_1^6 y_2^{10} 
y_3^{20} y_4^{22}+ 792 y_1^6 y_2^{10} y_3^{20} y_4^{23}+ 924 y_1^6 y_2^{10} y_3^{20} y_4^{24}+ 
792 y_1^6 y_2^{10} y_3^{20} y_4^{25}+ 495 y_1^6 y_2^{10} y_3^{20} y_4^{26}+ 220 y_1^6 
y_2^{10} y_3^{20} y_4^{27}+ 66 y_1^6 y_2^{10} y_3^{20} y_4^{28}+ 12 y_1^6 y_2^{10} y_3^{20} 
y_4^{29} \\[.05cm] & + y_1^6 y_2^{10} y_3^{20} y_4^{30}
\end{array}
$}
\eeq

We have explicitly computed the partition functions up to $n=8$. We do not include $Z_7$ and $Z_8$ here due to their length. To give a flavor of it, the numbers of terms in the $Z_6$, $Z_7$ and $Z_8$  are 1,191, 5,767 and 24,920, respectively.

While these partition functions encode a vast amount of detailed information about the crystals, for example the number of melting configurations with given numbers of atoms of each type, they are not particularly illuminating. There are some obvious features, like the linear term encoding the top atoms and the highest order term summarizing all the atoms in the crystal, but going beyond them is challenging. In Section \sref{section_stable_variables}, we will introduce new variables that reveal additional underlying patterns. 

\bigskip

%=================================================================
\subsubsection*{Unrefining Partition Functions}
%=================================================================

The partition functions become more manageable if we set all fugacities to the same value, namely $y_i=y$ for $i=1,\ldots,4$. In this case, the term of order $n$ counts the number of melting configurations with $n$ atoms removed. The unrefined partition functions are given below.

\beq
\resizebox{.8\textwidth}{!}{$
\begin{array}{cl}
%=================================================================
Z_0 & =1 \\[.3cm]
%=================================================================
Z_1 & =1 + y \\[.3cm]
 %=================================================================
Z_2 & =1 + y + 2 y^2 + y^3 \\[.3cm]
 %=================================================================
Z_3 & =1 + 2 y + y^{2} + y^{3} + 2 y^{4} + 5 y^{5} + 6 y^{6} + 6 y^{7} + 4 y^{8} + y^{9} \\[.3cm]
 %=================================================================
Z_4 & = 1+ 2 y + 5 y^{2} + 7 y^{3} + 10 y^{4} + 10 y^{5} + 7 y^{6} + 9 y^{7} + 16 y^{8} + 
 21 y^{9} + 20 y^{10} + 18 y^{11} + 20 y^{12} \\ & + 27 y^{13} + 32 y^{14} + 31 y^{15} + 
 24 y^{16} + 15 y^{17} + 6 y^{18} + y^{19} 
 %=================================================================
  \end{array}
 $} \nonumber
 \eeq
 
\beq
\resizebox{\textwidth}{!}{$
\begin{array}{cl}
 %=================================================================
 Z_6 & =1 + 3 y+ 9 y^{2} + 18 y^{3} + 34 y^{4} + 52 y^{5} + 71 y^{6} + 92 y^{7} + 
 116 y^{8} + 162 y^{9} + 240 y^{10} + 331 y^{11} \\
 & + 416 y^{12} + 528 y^{13} + 
 724 y^{14} + 1029 y^{15} + 1408 y^{16} + 1844 y^{17} + 2404 y^{18} + 
 3222 y^{19} + 4400 y^{20} \\
 & + 5994 y^{21} + 8050 y^{22} + 10657 y^{23} + 
 13950 y^{24} + 18068 y^{25} + 23020 y^{26} + 28616 y^{27} + 34472 y^{28} \\
 & + 
 40134 y^{29} + 45152 y^{30} + 49174 y^{31} + 51916 y^{32} + 53287 y^{33} + 
 53451 y^{34} + 52709 y^{35} + 51375 y^{36} \\ 
 &+ 49688 y^{37} + 47670 y^{38} + 
 45162 y^{39} + 42138 y^{40} + 38856 y^{41} + 35531 y^{42} + 32118 y^{43} + 
 28617 y^{44} \\ & + 25325 y^{45} + 22551 y^{46} + 20226 y^{47} + 18042 y^{48} + 
 15885 y^{49} + 13916 y^{50} + 12254 y^{51} + 10811 y^{52} \\ & + 9452 y^{53} + 
 8175 y^{54} + 7063 y^{55} + 6116 y^{56} + 5213 y^{57} + 4227 y^{58} + 
 3147 y^{59} + 2088 y^{60} + 1203 y^{61} \\ & + 585 y^{62} + 229 y^{63} + 66 y^{64} + 
 12 y^{65} + y^{66} \\[.3cm]
 %=================================================================
  Z_7 & = 1 + 4 y + 6 y^{2} + 7 y^{3} + 13 y^{4} + 32 y^{5} + 65 y^{6} + 110 y^{7} + 
 182 y^{8} + 318 y^{9} + 575 y^{10} + 1018 y^{11} \\
 & + 1716 y^{12} + 2778 y^{13} + 
 4392 y^{14} + 6822 y^{15} + 10368 y^{16} + 15290 y^{17} + 21729 y^{18} + 
 29693 y^{19} + 39248 y^{20} \\
 & + 50452 y^{21} + 63242 y^{22} + 77544 y^{23} + 
 93502 y^{24} + 111530 y^{25} + 132246 y^{26} + 156258 y^{27} + 184012 y^{28} \\
 & + 
 215977 y^{29} + 253036 y^{30} + 296402 y^{31} + 346998 y^{32} + 
 405560 y^{33} + 473868 y^{34} + 555164 y^{35} + 652568 y^{36} \\
 & + 
 768009 y^{37} + 903768 y^{38} + 1064486 y^{39} + 1256054 y^{40} + 
 1483651 y^{41} + 1753196 y^{42} + 2074660 y^{43} + 2461782 y^{44} \\
 &+ 
 2928759 y^{45} + 3489375 y^{46} + 4160046 y^{47} + 4961596 y^{48} + 
 5917103 y^{49} + 7049035 y^{50} + 8378921 y^{51} + 9926283 y^{52} \\
 &+ 
 11702652 y^{53} + 13701651 y^{54} + 15889945 y^{55} + 18202072 y^{56} + 
 20541291 y^{57} + 22787499 y^{58} + 24811876 y^{59} + 26494993 y^{60} \\
 &+ 
 27743965 y^{61} + 28503274 y^{62} + 28757193 y^{63} + 28524899 y^{64} + 
 27851479 y^{65} + 26799572 y^{66} + 25442924 y^{67} + 23859830 y^{68} \\
 &+ 
 22125737 y^{69} + 20307551 y^{70} + 18461281 y^{71} + 16632048 y^{72} + 
 14855069 y^{73} + 13157481 y^{74} + 11560244 y^{75} + 10079359 y^{76} \\
 &+ 
 8726270 y^{77} + 7507453 y^{78} + 6423179 y^{79} + 5466975 y^{80} + 
 4628255 y^{81} + 3897161 y^{82} + 3266741 y^{83} + 2730229 y^{84} \\
 &+ 
 2277676 y^{85} + 1896757 y^{86} + 1576259 y^{87} + 1307575 y^{88} + 
 1083418 y^{89} + 896715 y^{90} + 741067 y^{91} + 611292 y^{92} \\
 &+ 
 503028 y^{93} + 412328 y^{94} + 335988 y^{95} + 271790 y^{96} + 
 217948 y^{97} + 172403 y^{98} + 132963 y^{99} + 98192 y^{100} \\
 &+ 
 67999 y^{101} + 43256 y^{102} + 24785 y^{103} + 12538 y^{104} + 5468 y^{105} + 
 1988 y^{106} + 572 y^{107} + 120 y^{108} + 16 y^{109} + y^{110} \\[.3cm]
  %=================================================================
  Z_8 & = 1 + 4 y + 14 y^{2} + 35 y^{3} + 79 y^{4} + 151 y^{5} + 261 y^{6} + 414 y^{7} + 
 618 y^{8} + 907 y^{9} + 1328 y^{10} \\
 & + 1965 y^{11} + 2911 y^{12} + 4219 y^{13} + 
 5973 y^{14} + 8504 y^{15} + 12382 y^{16} + 18184 y^{17} + 26405 y^{18} \\ & + 
 37927 y^{19} + 54646 y^{20} + 79615 y^{21} + 116650 y^{22} + 170479 y^{23} + 
 247936 y^{24} + 359587 y^{25} + 520827 y^{26} 
 \\ & + 752659 y^{27} + 
 1082908 y^{28} + 1548976 y^{29} + 2202201 y^{30} + 3113344 y^{31} + 
 4377660 y^{32} + 6119308 y^{33} + 8494965 y^{34} 
 \\ & + 11697464 y^{35} + 
 15958510 y^{36} + 21546783 y^{37} + 28758311 y^{38} + 37899931 y^{39} + 
 49267872 y^{40} + 63124516 y^{41} + 79677932 y^{42} 
 \\ & + 99069317 y^{43} + 
 121370666 y^{44} + 146595958 y^{45} + 174726334 y^{46} + 205743294 y^{47} + 
 239657802 y^{48} + 276524261 y^{49} + 316440446 y^{50} 
 \\ & + 359546184 y^{51} + 
 406022098 y^{52} + 456080991 y^{53} + 509959193 y^{54} + 567927730 y^{55} + 
 630320582 y^{56} + 697549486 y^{57} + 770090836 y^{58} 
 \\ & + 848457986 y^{59} + 
 933169609 y^{60} + 1024717268 y^{61} + 1123554436 y^{62} + 
 1230117840 y^{63} + 1344854765 y^{64} + 1468243943 y^{65} 
 \\ & + 
 1600854894 y^{66} + 1743450645 y^{67} + 1897038521 y^{68} + 
 2062813754 y^{69} + 2242112307 y^{70} + 2436487227 y^{71} + 
 2647811321 y^{72} \\ & 
 + 2878253576 y^{73} + 3130218613 y^{74} + 
 3406448335 y^{75} + 3710222154 y^{76} + 4045396262 y^{77} + 
 4416269826 y^{78} + 4827550346 y^{79} \\ & 
 + 5284527034 y^{80} + 
 5793236201 y^{81} + 6360468118 y^{82} + 6993772677 y^{83} + 
 7701626172 y^{84} + 8493616986 y^{85} + 9380397681 y^{86} 
 \\ & + 
 10373380644 y^{87} + 11484339117 y^{88} + 12724969803 y^{89} + 
 14106282891 y^{90} + 15637661836 y^{91} + 17325551501 y^{92} + 
 19171817360 y^{93} 
 \\ & + 21171803004 y^{94} + 23312097235 y^{95} + 
 25568150651 y^{96} + 27902159772 y^{97} + 30261878202 y^{98} + 
 32581021756 y^{99} + 34781708325 y^{100} 
 \\ & + 36779047955 y^{101} + 
 38487605919 y^{102} + 39828955870 y^{103} + 40739082969 y^{104} + 
 41174315578 y^{105} + 41114854887 y^{106} + 40565575967 y^{107} 
 \\ & + 
 39554300750 y^{108} + 38128087253 y^{109} + 36348256083 y^{110} + 
 34284921022 y^{111} + 32011680521 y^{112} + 29600938430 y^{113} + 
 27120176235 y^{114} 
 \\ & + 24629366897 y^{115} + 22179540290 y^{116} + 
 19812338096 y^{117} + 17560288898 y^{118} + 15447504612 y^{119} + 
 13490540649 y^{120} + 11699269933 y^{121} 
 \\ & + 10077746871 y^{122} + 
 8625091461 y^{123} + 7336395546 y^{124} + 6203622621 y^{125} + 
 5216469677 y^{126} + 4363154060 y^{127} + 3631089257 y^{128} 
 \\ & + 
 3007443714 y^{129} + 2479604438 y^{130} + 2035550113 y^{131} + 
 1664111013 y^{132} + 1355110724 y^{133} + 1099416270 y^{134} + 
 888921099 y^{135} 
 \\ & + 716477520 y^{136} + 575818016 y^{137} + 
 461510088 y^{138} + 368935464 y^{139} + 294233950 y^{140} + 
 234186117 y^{141} + 186083671 y^{142} 
 \\ & + 147646061 y^{143} + 
 116981441 y^{144} + 92549519 y^{145} + 73106186 y^{146} + 57645602 y^{147} + 
 45356208 y^{148} + 35587698 y^{149} 
 \\ & + 27821190 y^{150} + 21643375 y^{151} + 
 16727386 y^{152} + 12817172 y^{153} + 9711261 y^{154} + 7247977 y^{155} + 
 5297234 y^{156} 
 \\ & + 3758917 y^{157} + 2561559 y^{158} + 1655566 y^{159} + 
 1001542 y^{160} + 559540 y^{161} + 284674 y^{162} + 129860 y^{163} + 
 52128 y^{164} \\ & + 17964 y^{165} + 5133 y^{166} + 1156 y^{167} + 190 y^{168} + 
 20 y^{169} + y^{170}
 \end{array}
 $}
 \label{unrefined_Zs_in_y_variables}
 \eeq

Let us compare the sizes of these expression to those of the refined partition functions discussed in the previous sections. After unrefinement, the number of terms in $Z_6$, $Z_7$ and $Z_8$  are 67, 111 and 171, respectively.
 
\begin{figure}[H]
	\centering
	\includegraphics[width=9cm]{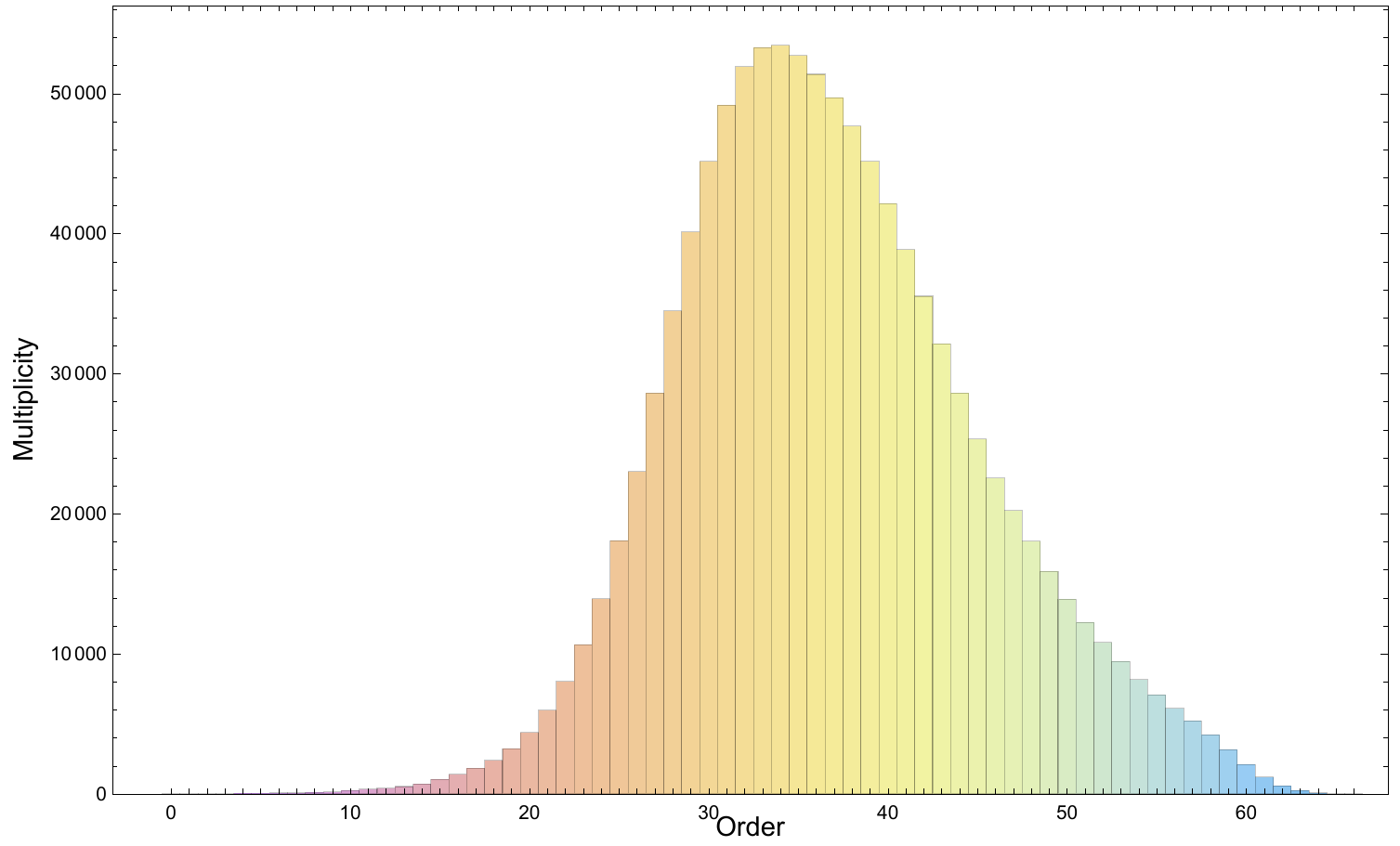}
\caption{Profile of the partition function $Z_6$, namely the number of melting configurations as a function of the order.}
	\label{Histogram_Z6}
\end{figure}

It is interesting to visualize how the number of melting configurations depends on the number of atoms in them. We will reflect to such a plot as the {\it profile of the partition function}. This is shown for $Z_6$, whose crystal is large enough for the distribution to become smooth, in \fref{Histogram_Z6}. 

It is natural to ask whether these distributions (after some appropriate rescaling) approach some limit shape as the step in the cascade goes to infinity and how their shape depends on the underlying geometry and flavor configuration. We attempted fitting this profile to some qualitatively similar distributions, such as $\Gamma$ (which includes $\chi^2$), Lognormal, Generalized $\Gamma$ and Poisson-like. None of these functions lead to a good fit. We will not try to determine the functional form of these profiles any further in this paper.
 
%===================================================================
%===================================================================

\begin{figure}[H]
	\centering
	\includegraphics[width=9cm]{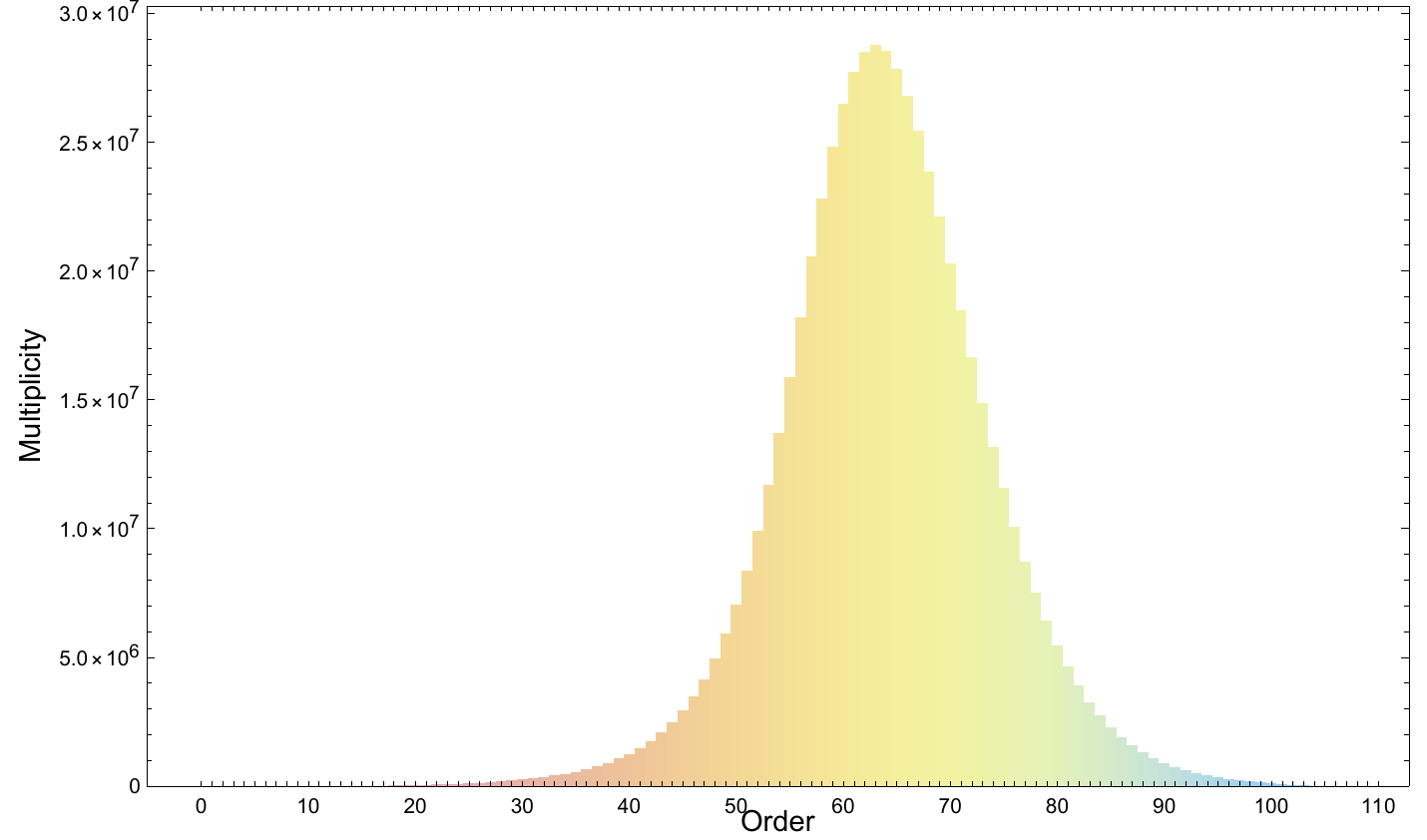}
\caption{Profile of the partition function $Z_7$.}
	\label{Histogram_Z7}
\end{figure}

\begin{figure}[H]
	\centering
	\includegraphics[width=9cm]{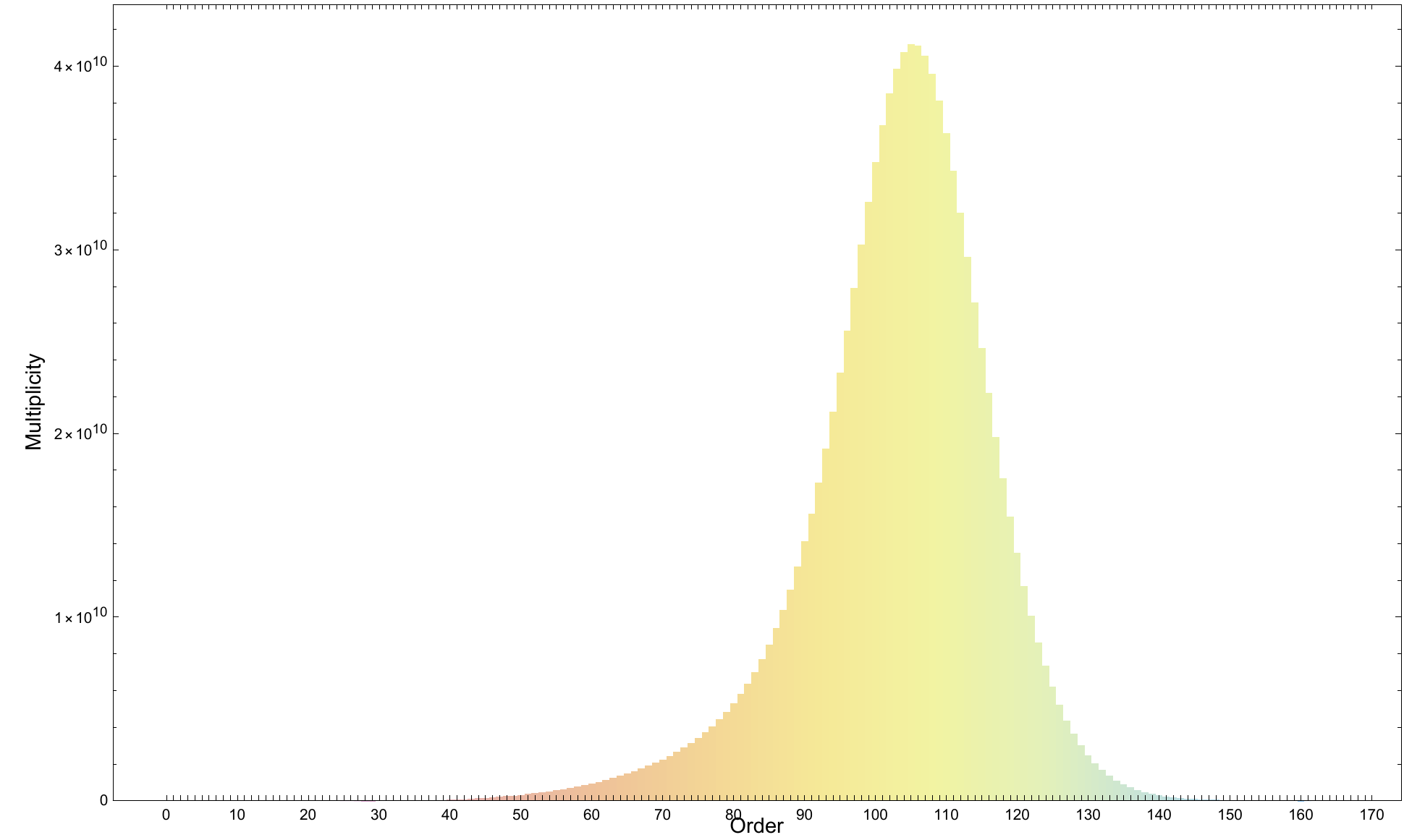}
\caption{Profile of the partition function $Z_8$.}
	\label{Histogram_Z8}
\end{figure}

In Figures \ref{Histogram_Z7} and \ref{Histogram_Z8}, we show the profiles for $Z_7$ and $Z_8$.
 
 %===================================================================
%===================================================================

 %===================================================================
%===================================================================

%=================================================================
\subsubsection*{Counting Melting Configurations}
%=================================================================

The number of melting configurations $N_{melt}$ of a crystal is another piece of highly non-trivial information, which can potentially guide the discovery of recursive approaches or, more ambitiously, a generalization of cluster algebras along the lines discussed in Section \sref{section_generalization_cluster_algebras}. This number can be obtained by fully unrefining the corresponding partition function, i.e. by setting all $y_i$ fugacities equal to 1. We used this approach to determine these multiplicities for the first eight crystals in the cascade:
\beq
\begin{array}{|c|c|}
\hline
\ \mbox{Step} \ & N_{melt} \\ \hline
1 & 2 \\
2 & 5 \\
3 & 29 \\
4 & 282 \\
5 & 10,044 \\
6 & 1,063,496 \\
7 & 616,911,740 \\
8 & \ 1,103,801,218,198 \ \\
\hline
\end{array}
\label{multiplicities_melting_configurations}
\eeq
Reproducing these numbers would constitute a highly non-trivial test of any generalization of cluster algebras. It would be analogous to studying cluster algebras without coefficients and would therefore serve as a natural intermediate step before tackling the full refined partition functions.

\begin{figure}[H]
	\centering
	\includegraphics[width=8cm]{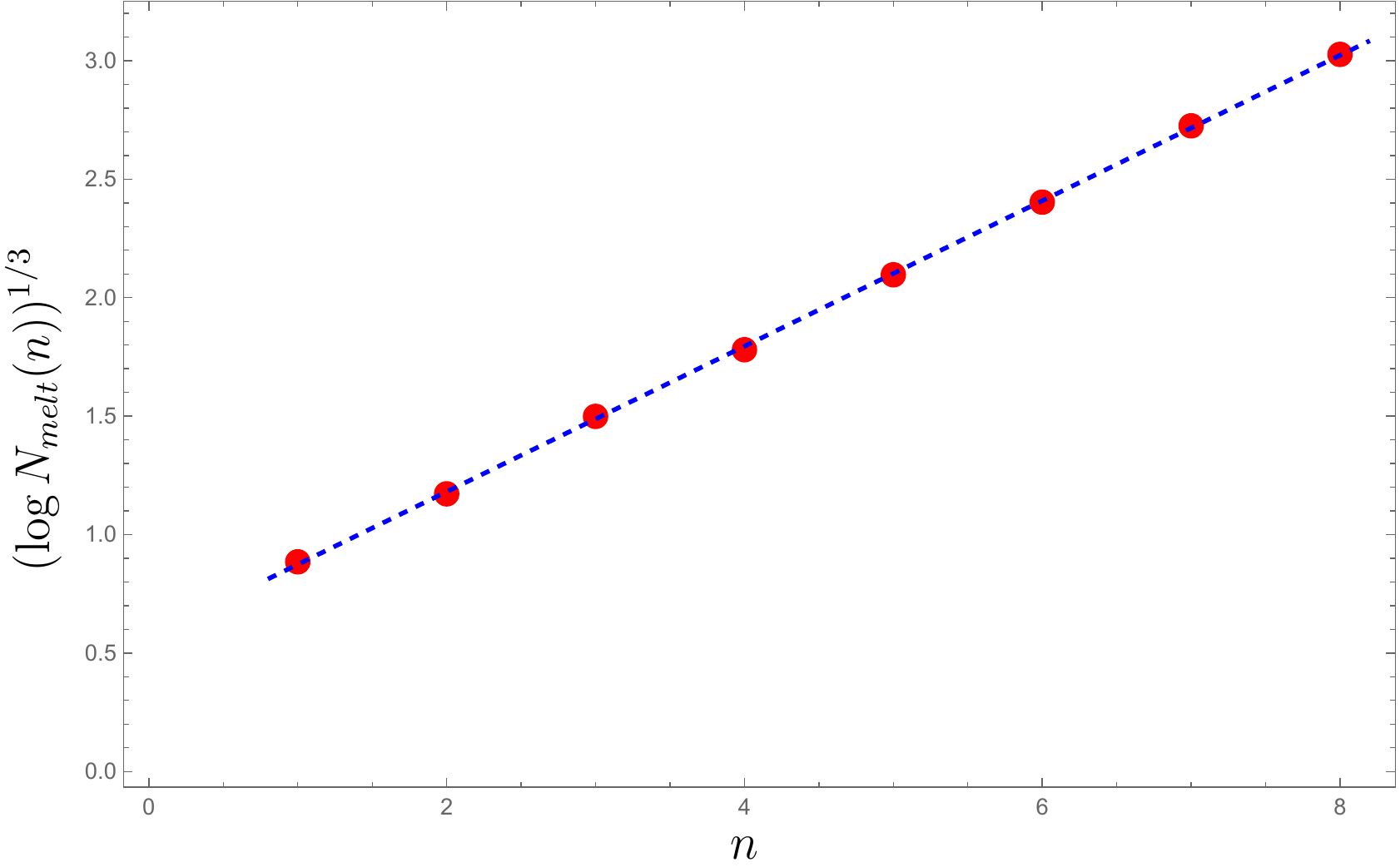}
\caption{A fit to the growth of $N_{melt}$ with the step in the cascade.}
	\label{fit_multiplicities}
\end{figure}

\fref{fit_multiplicities} shows that $(\log N_{melt}(n))^{1/3}$ is very well, albeit not exactly, described by a linear fit. While this is simply an educated guess, it suggests the number of melting configurations grows as $e^{\# n^3}$ for large $n$.

 %===================================================================
%===================================================================

%=================================================================
\paragraph{Melting Configurations and Brick Matchings}
%=================================================================

In \cite{Franco:2023tly}, it was explained how melting configurations for infinite crystals are in one-to-one correspondence with brick matchings, the brane brick model analogues of perfect matchings for brane tilings, on the universal cover of the corresponding brane brick model. We expect that, similarly to the case of $3d$ crystals for toric CY $3$-folds, the melting configurations of finite crystals are also in one-to-one correspondence with brick matchings on a certain finite region of the brane brick model, subject to appropriate boundary conditions. See e.g. \cite{Eager:2011ns} for CY$_3$ examples. This interesting direction is beyond the scope of this paper and we leave it for future work.

%=================================================================
\section{Stable Variables}
%=================================================================

\label{section_stable_variables}

So far, we have written partition functions in terms of the variables $y_i$, which can be regarded as fugacities for each gauge node at a given step in the cascade. As noted in Section \sref{section_partition_functions}, the resulting expressions are quite obscure. In this section we propose a different set of variables, which are defined iteratively as follows. Given a triality transformation on node $k$
\beq
y_j'=\left\{
\begin{array}{ll}
y_k^{-1}, & \text{if } j = k,\\[7pt]
y_j \displaystyle\prod_{\chi_{j\to k}} y_k, \ \ \ & \text{if } j \neq k.
\end{array}
\right.
\label{transfomation_stable_variables}
\eeq
where $y_i$ and $y_i'$, $i=1,\ldots, G$, are the variables before and after triality, respectively. This transformation is motivated by the transformation of brane charges in the case of theories engineered via branes at singularities, which has been used to derive triality \cite{Franco:2016qxh}, quadrality of $0d$ $\mathcal{N}=1$ theories, as well as the order $(m+1)$ dualities for $m$-graded quivers associated to CY $(m+2)$-folds \cite{Franco:2016tcm,Franco:2017lpa}. The transformations in \eqref{transfomation_stable_variables} correspond to variables that are exponentially related to brane charges. Therefore, it represents a physically motivated generalization of the transformation of coefficients in ordinary cluster algebras given by equation \eqref{cluster_algebra_transformation_coefficients} to $2d$ (0,2) quivers and triality.\footnote{While this paper focuses on $2d$ (0,2) quivers and triality, it is worth noting that \eqref{transfomation_stable_variables} would indeed be a natural generalization of \eqref{cluster_algebra_transformation_coefficients} to order $(m+1)$ dualities for $m$-graded quivers.}

\begin{figure}[ht!]
	\centering
	\includegraphics[width=\textwidth]{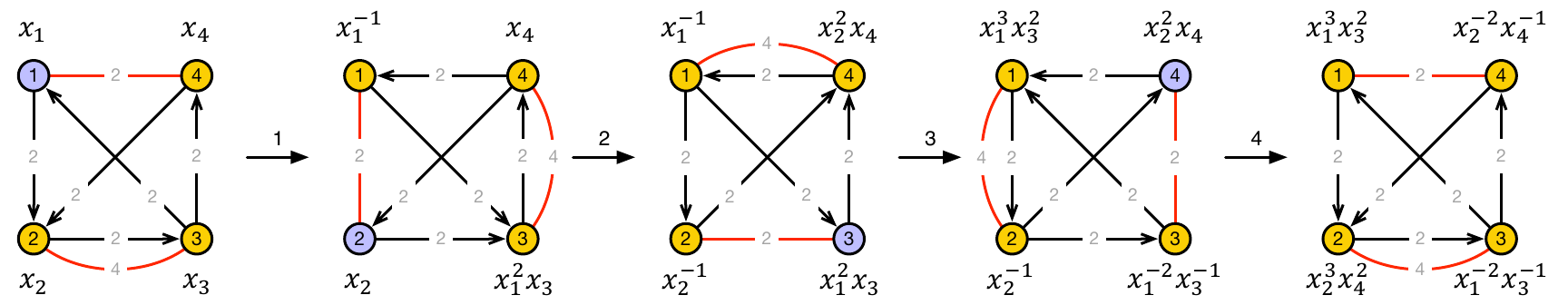}
\caption{Evolution of stable variables along the cascade.}
	\label{cascade_Q111_stable_variables}
\end{figure}

\fref{cascade_Q111_stable_variables} illustrates this transformation of variables along the $Q^{1,1,1}$ cascade under consideration. The starting point is a quiver where the variable associated with each node $i$, $i=1,\ldots,4$, is simply set to $x_i$. Then, we follow the effect of the transformation of variables in \eqref{cascade_Q111_stable_variables} over the four triality transformations that comprise one period of the cascade. 

%===================================================================
%===================================================================

\begin{figure}[ht!]
	\centering
	\includegraphics[width=10cm]{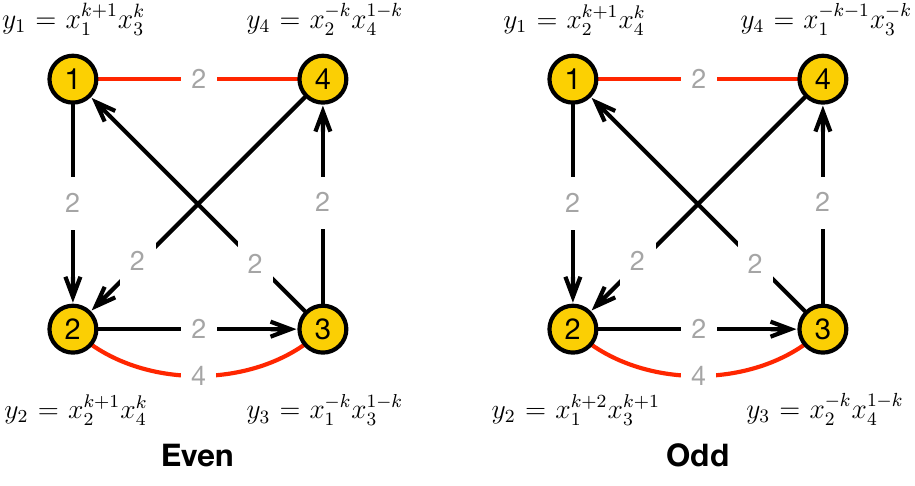}
\caption{Stable variables at arbitrary even ($n=2k$, $k=0,1,2,\ldots$) and odd ($n=2k+1$, $k=0,1,2,\ldots$) steps in the cascade.}
	\label{stable_variables_even_odd}
\end{figure}

Extrapolating from \fref{cascade_Q111_stable_variables}, it is straightforward to determine what the variables at every node are for a general step $n$ in the cascade when expressed in terms of the initial basis of $x_i$ variables. The final result is summarized in \fref{stable_variables_even_odd}, where, for convenience, we have distinguished between even and odd steps in the cascade. We have also relabeled nodes when necessary, so as to rotate the quiver to the canonical form in \fref{quiver_Q111}. 

%===================================================================
%===================================================================

In summary, we will consider the following changes of variables:

%=================================================================
\paragraph{Even Steps.}
%=================================================================

Step $n=2k$, $k=0,1,2,\ldots$:
\beq
\begin{array}{ccl}
y_1 & = & x_1^{k+1} x_3^k \\[.1cm]
y_2 & = & x_2^{k+1} x_4^k \\[.1cm]
y_3 & = & x_1^{-k} x_3^{1-k} \\[.1cm]
y_4 & = & x_2^{-k} x_4^{1-k}
\end{array}
\label{stable_variables_even}
\eeq

%=================================================================
\paragraph{Odd Steps}
%=================================================================

Step $n=2k+1$, $k=0,1,2,\ldots$:
\beq
\begin{array}{ccl}
y_1 & = & x_2^{k+1} x_4^{k} \\[.1cm]
y_2 & = & x_1^{k+2} x_3^{k+1} \\[.1cm]
y_3 & = & x_2^{-k} x_4^{1-k} \\[.1cm]
y_4 & = & x_1^{-k-1} x_3^{-k}
\end{array}
\label{stable_variables_odd}
\eeq

In \eqref{partition_functions_stable_variables}, we express the partition functions $Z_1$ to $Z_4$ from \eqref{Z1_to_Z5} in terms of $x_i$ variables, after performing the changes of variables defined in \eqref{stable_variables_even} and \eqref{stable_variables_odd}.\footnote{More precisely, to simplify the final expressions, we have replaced all $x_i$ by $x_i^{-1}$.} We have organized the terms vertically to facilitate the comparison of the partition functions at different steps. The resulting expressions exhibit a remarkable stabilization property: once a term appears in a partition function $Z_n$, it appears unchanged in all higher steps of the cascade. 
\beq
\begin{array}{|c|c|c|c|c|}
\hline
\ \ \ \ \ \ Z_0 \ \ \ \ \ \ & \ \ \ \ \ \ Z_1 \ \ \ \ \ \ & \ \ \ \ \ \ Z_2 \ \ \ \ \ \ & \ \ \ \ \ \ Z_3 \ \ \ \ \ \ & \ \ \ \ \ \ Z_4 \ \ \ \ \ \ \\ \hline
1 & 1 & 1 & 1 & 1 \\
 & x_1 & x_1 &  x_1 &  x_1  \\
 & & 2 x_1 x_2 & 2 x_1 x_2 & 2 x_1 x_2  \\
 & & x_1 x_2^2 & x_1 x_2^2 & x_1 x_2^2 \\
 & & & 2 x_1^2 x_3 & 2 x_1^2 x_3 \\
 & & &                      & 4 x_1 x_14 x_2^2 \\
& & & 4 x_1^3 x_2 x_3 & 4 x_1^3 x_2 x_3 \\
& & &                      & 4 x_1 x_14 x_2^3 \\
& & & x_1^4 x_3^2 & x_1^4 x_3^2 \\
& & & 4 x_1^3 x_2^2 x_3 & 4 x_1^3 x_2^2 x_3 \\
& & &                      & 4 x_1^2 x_14 x_2^2 x_3 \\
& & &                      & 6 x_1 x_14^2 x_2^4 \\
& & & 2 x_1^5 x_2 x_3^2 & 2 x_1^5 x_2 x_3^2 \\
& & &                      & 2 x_1 x_14^2 x_2^5 \\
& & &                      & 12 x_1^3 x_14 x_2^3 x_3 \\
& & & 6 x_1^5 x_2^2 x_3^2 & 6 x_1^5 x_2^2 x_3^2 \\
& & &                      & 2 x_1^2 x_14^2 x_2^4 x_3 \\
& & &                      & 4 x_1^4 x_14 x_2^2 x_3^2 \\
& & &                      & 4 x_1^3 x_14 x_2^4 x_3 \\
& & &                      & 4 x_1 x_14^3 x_2^6 \\
& & &                      & 12 x_1^3 x_14^2 x_2^5 x_3 \\
& & &                      & 8 x_1^5 x_14 x_2^3 x_3^2 \\
& & & 4 x_1^7 x_2^2 x_3^3 & 4 x_1^7 x_2^2 x_3^3 \\
& & &                           & 6 x_1^4 x_14^2 x_2^4 x_3^2 \\
& & &                           & 14 x_1^5 x_14 x_2^4 x_3^2 \\
& & &                           & x_1 x_14^4 x_2^8 \\
& & &                           & 4 x_1^3 x_14^3 x_2^7 x_3 \\
& & &                           & 12 x_1^5 x_14^2 x_2^5 x_3^2 \\
& & & x_1^9 x_2^2 x_3^4 & x_1^9 x_2^2 x_3^4 \\
& & &                           & 4 x_1^4 x_14^3 x_2^6 x_3^2 \\
& & &                           & \vdots \\ \hline
\end{array}
\label{partition_functions_stable_variables}
\eeq
We have confirmed the stabilization of the partition functions at higher steps in the cascade, but the results are too long to present them here. We conjecture that, in the $x_i$ variables, partition functions converge as formal power series expansions. Analogous variables were first introduced for crystals associated to toric CY 3-folds in \cite{Eager:2011ns}, were it was also observed that they lead to the stabilization of the partition functions. Following \cite{Eager:2011ns}, we will refer to the $x_i$ variables, which are connected to the $y_i$'s by the changes of variables in \eqref{stable_variables_even} and \eqref{stable_variables_odd}, as {\it stable variables}. Although we illustrated the concept of stable variables using the specific example of $Q^{1,1,1}$, the construction extends straightforwardly to arbitrary CY $4$-folds.

The stabilization for CY $3$-folds, originally conjectured in \cite{Eager:2011ns}, was later rigorously proved for the conifold in \cite{Zhang:2018uur}, where partial results were also presented for $F_0$. Additional results were presented in \cite{Gupta2018}, including some for $dP_1$. In the mathematical language of cluster algebras used in that paper, the use of stable variables for partition functions is regarded as a change of basis leading to the stabilization of $F$-polynomials.

It interesting to observe that, in \fref{stable_variables_even_odd}, the products the of the variables for the two pairs of diagonally opposite nodes are constant along the cascade and equal to $x_1x_3$ and $x_2x_4$. It is natural to expect that this property may lead to interesting simplifications of the partition functions when considering something like the {\it node condensation} operation introduced in \cite{Eager:2011ns} for CY 3-folds.

The stabilization property exhibited by the crystal partition functions suggests that they play a role analogous to that of $F$-polynomials for CY 4-folds, something that is currently unknown. This observation further supports our proposal that these partition functions are natural physical quantities to study in the search for a generalization of cluster algebras.

%=================================================================
\subsubsection*{Unrefined Partition Functions in Stable Variables}
%=================================================================

Switching to the basis of stable variables does not lead to a recombination of terms in a fully refined partition function but reshuffles them in interesting ways and even modifies their orders. To investigate these effects, it is interesting to unrefine the partition function in stable variables. For concreteness, let  us consider $Z_6$ and set $x_i=x$ for $i=1,\ldots,4$. We obtain:
\beq
\resizebox{\textwidth}{!}{$
\begin{array}{cl}
 %=================================================================
 Z_6 & =1 + x + 2 x^{2} + 3 x^{3} + 4 x^{4} + 11 x^{5} + 14 x^{6} + 18 x^{7} + 32 x^{8} + 
 36 x^{9} + 65 x^{10} + 83 x^{11} + 92 x^{12} + 154 x^{13} + 168 x^{14} \\ 
 & + 
 243 x^{15} + 320 x^{16} + 348 x^{17} + 508 x^{18} + 574 x^{19} + 734 x^{20} + 
 950 x^{21} + 1070 x^{22} + 1362 x^{23} + 1590 x^{24} + 1939 x^{25} \\
 &  + 
 2367 x^{26} + 2749 x^{27} + 3216 x^{28} + 3764 x^{29} + 4495 x^{30} + 5188 x^{31} + 5988 x^{32} + 6765 x^{33} + 7726 x^{34} + 8996 x^{35} \\ 
 & + 
 10019 x^{36} + 11171 x^{37} + 12341 x^{38} + 13768 x^{39} + 15298 x^{40} + 
 16778 x^{41} + 17918 x^{42} + 19129 x^{43} + 21162 x^{44} \\ &  
 + 22085 x^{45} + 
 24084 x^{46} + 24791 x^{47} + 25126 x^{48} + 27982 x^{49} + 27288 x^{50} + 29616 x^{51} + 29692 x^{52} + 28195 x^{53} \\ &
  + 31824 x^{54} + 29299 x^{55} + 
 31394 x^{56} + 30908 x^{57} + 27356 x^{58} + 31176 x^{59} + 27912 x^{60} + 
 28887 x^{61} + 28074 x^{62} \\ 
 &  + 23290 x^{63} + 26340 x^{64} + 24230 x^{65} + 
 23129 x^{66} + 22286 x^{67} + 17783 x^{68} + 19254 x^{69} + 19690 x^{70} + 
 16015 x^{71} \\ & + 15424 x^{72} + 12687 x^{73} + 12282 x^{74} + 15185 x^{75} + 
 9427 x^{76} + 9234 x^{77} + 9089 x^{78} + 6944 x^{79} + 11016 x^{80} \\ & + 
 4568 x^{81} + 4724 x^{82} + 7051 x^{83} + 3532 x^{84} + 7322 x^{85} + 
 1734 x^{86} + 2033 x^{87} + 5956 x^{88} + 1604 x^{89} + 4322 x^{90} \\ & + 
 482 x^{91} + 714 x^{92} + 5127 x^{93} + 624 x^{94} + 2192 x^{95} + 88 x^{96} + 
 191 x^{97} + 4203 x^{98} + 192 x^{99} + 912 x^{100} + 8 x^{101} \\ & + 34 x^{102} + 
 3144 x^{103}  + 40 x^{104} + 288 x^{105} + 3 x^{107} + 2088 x^{108} + 4 x^{109} + 
 60 x^{110} + 1203 x^{113} + 6 x^{115} + 585 x^{118} \\ & + 229 x^{123} + 66 x^{128} + 
 12 x^{133} + x^{138}
 %=================================================================
 \end{array}
 $}
 \label{unrefined_Z6_stable_variables}
 \eeq

%=================================================================
\subsubsection*{Emergence of an Intriguing Profile}
%=================================================================
 
\begin{figure}[H]
	\centering
	\includegraphics[width=10cm]{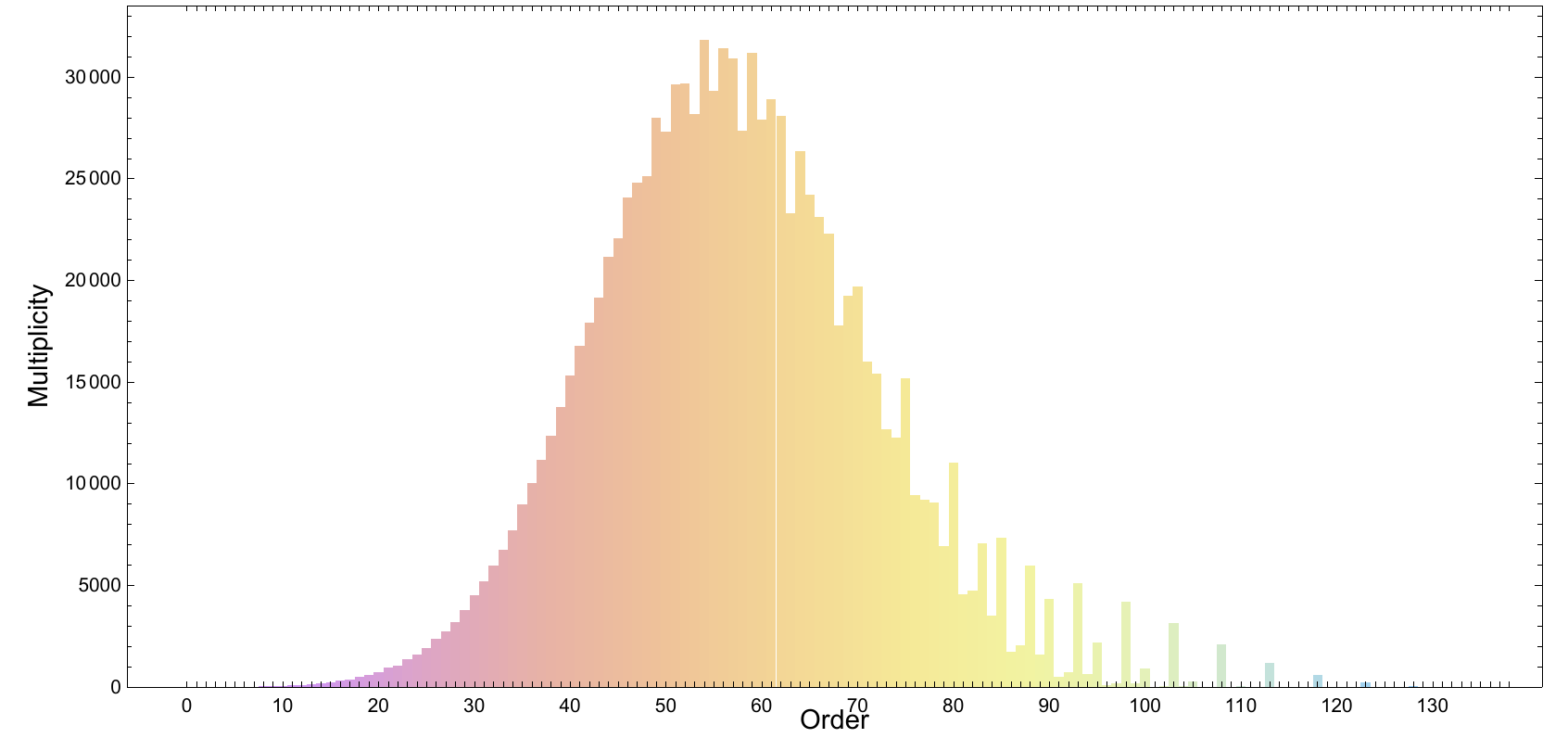}
\caption{Profile of the partition function $Z_6$ in stable variables.}
	\label{Histogram_Z6_stable_variables}
\end{figure}

This expression should be compared to $Z_6$ in $y$ variables, which was given in \eqref{unrefined_Zs_in_y_variables}. We can trace how individual terms map between the fully refined partition functions in original and stable variables. Therefore, we cal also follow the transformation of terms between the unrefined expressions \eqref{unrefined_Zs_in_y_variables} and \eqref{unrefined_Z6_stable_variables}. We immediately note that the relative orders of pairs of terms can be flipped by the change of variables. Similarly, the order of the partition function changes from 66 to 138. 
Moreover, even though terms do not mix when passing from $y_i$ to $x_i$ variables, they combine in different ways when the partition functions are unrefined. In other words, the sets of terms that have the same order in $y_i$ variables, and hence combine into single terms when unrefining, are, generically, different from the sets of terms that have the same order in $x_i$ variables, which combine under unrefinement.  For example, the unrefined version of $Z_6$ has 67 non-vanishing coefficients in original variables, while it has 117 non-vanishing coefficients in stable variables. This raises the question of whether all such rearrangements of terms has any special effect. \fref{Histogram_Z6_stable_variables} shows the resulting profile of the partition function. 

%===================================================================
%===================================================================

\begin{figure}[htbp]
  \centering
  \begin{subfigure}{0.49\textwidth}
    \centering
    \includegraphics[width=\linewidth]{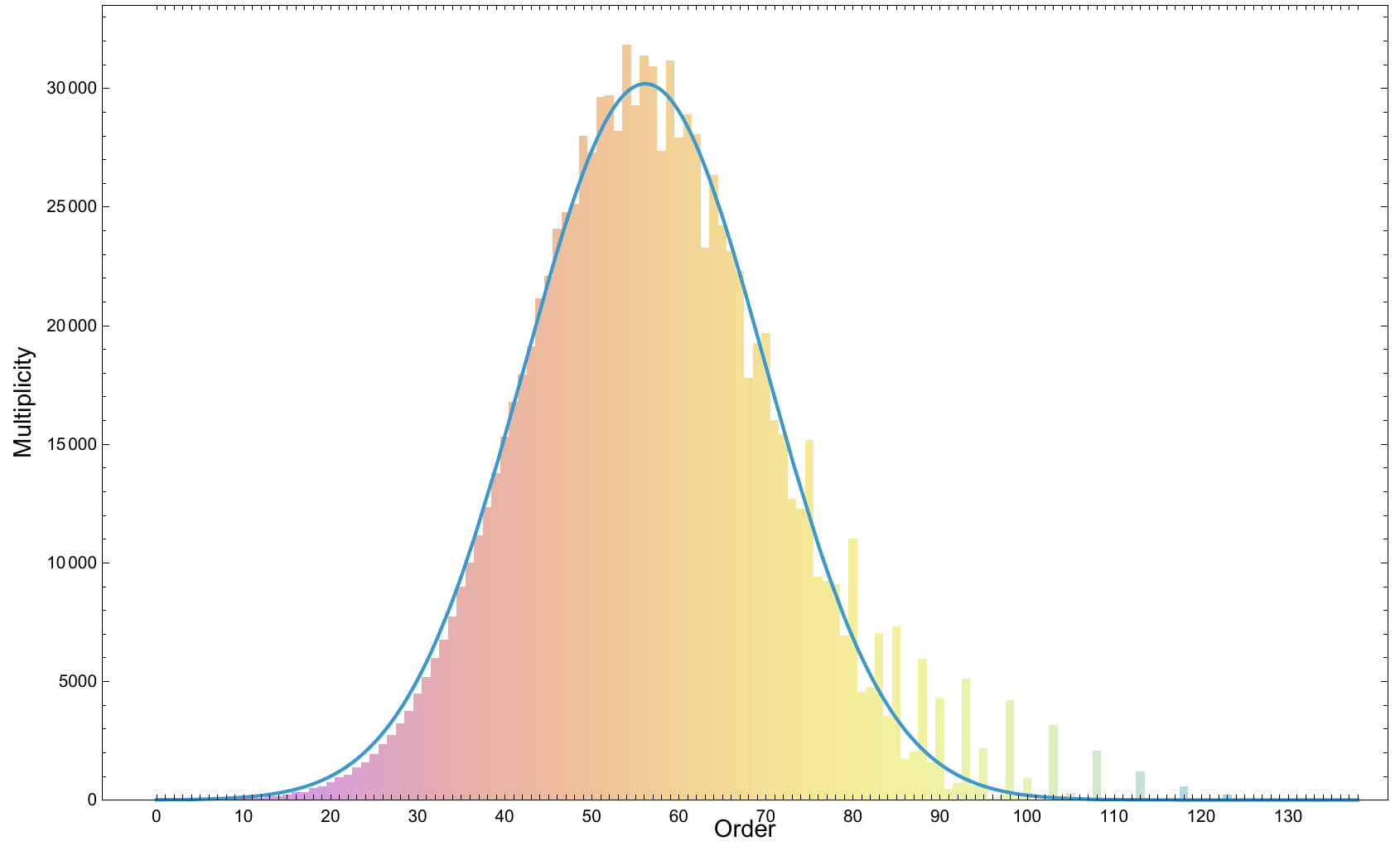}
    \caption{}
    \label{fig:img1}
  \end{subfigure}\hfill
  \begin{subfigure}{0.49\textwidth}
    \centering
    \includegraphics[width=\linewidth]{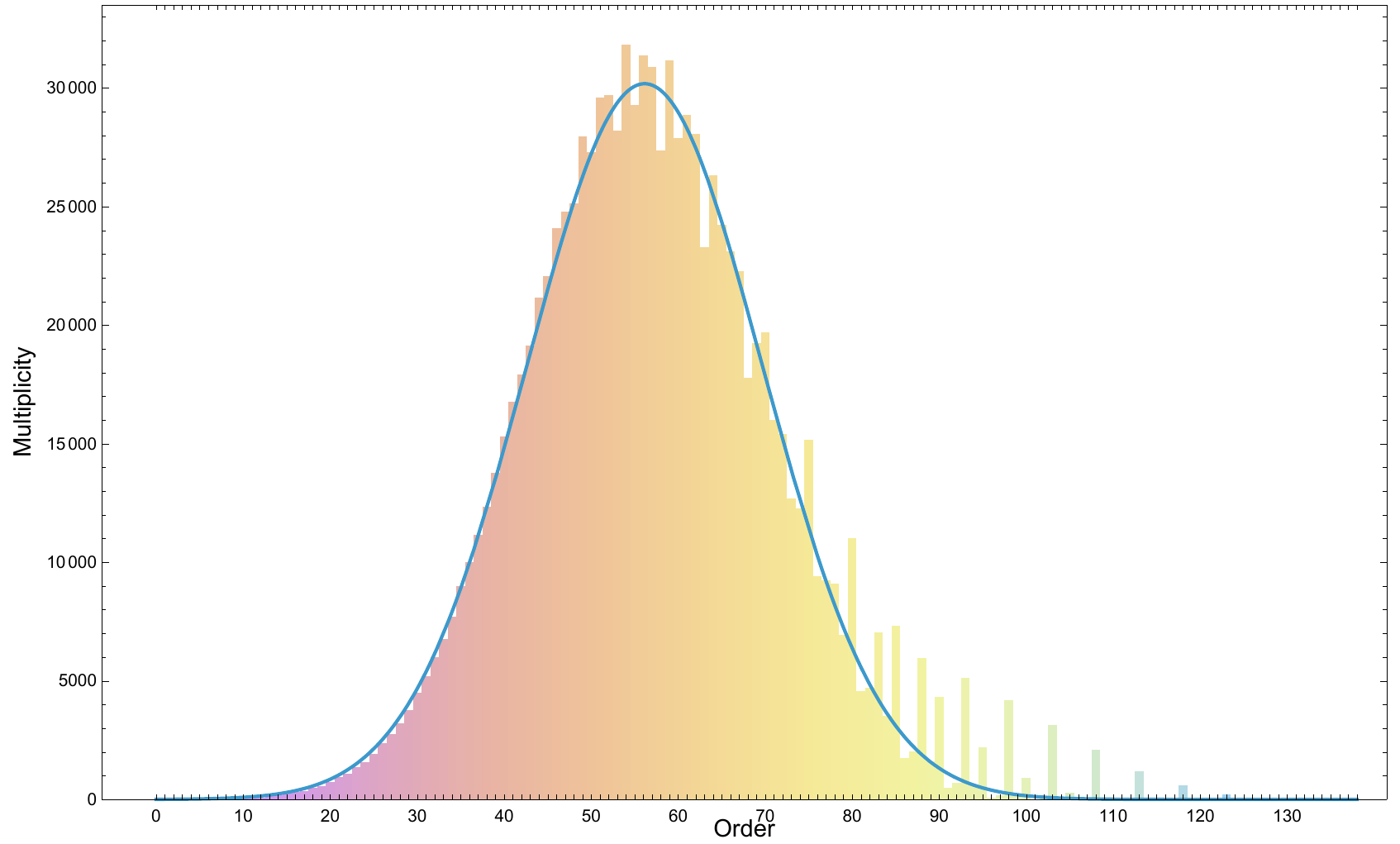}
    \caption{}
    \label{fig:img2}
  \end{subfigure}
  \caption{a) Gaussian fit to the profile of $Z_6$ in stable variables, which has $R^2=0.991435$. b) A Gaussian fit to the data after excluding several spikes on the right side of the distribution.}
  \label{Histogram_Z6_stable_variables_Gaussian_fit}
\end{figure}

Surprisingly, switching to stable variables transforms the profile of the partition function into something that resembles a Gaussian! To test whether this is indeed the case, we performed a Gaussian fit to the profile, which is shown in \fref{fig:img1}. Apart from the the spikes on the right-hand side of the bell, the agreement is already quite good, although the fitted curve appears slightly broader than the data. By explicitly excluding these spikes from the fit, we obtain the result shown in \fref{fig:img2}, which exhibits excellent agreement with the data, particularly on the rising side of the bell.

%===================================================================
%===================================================================

\beq
\resizebox{\textwidth}{!}{$
\begin{array}{cl}
 %=================================================================
Z_7 & = 1 + x + 2 x^{2} + 3 x^{3} + 4 x^{4} + 11 x^{5} + 14 x^{6} + 22 x^{7} + 30 x^{8} + 
 48 x^{9} + 77 x^{10} + 95 x^{11} \\ & + 146 x^{12} + 206 x^{13} + 284 x^{14} + 
 402 x^{15} + 524 x^{16} + 750 x^{17} + 942 x^{18} + 1293 x^{19} + 1662 x^{20} + 
 2106 x^{21} \\ & + 2886 x^{22} + 3538 x^{23} + 4704 x^{24} + 5674 x^{25} + 
 7181 x^{26} + 9117 x^{27} + 10763 x^{28} + 13899 x^{29} + 16625 x^{30} + 
 20401 x^{31} \\ &  + 24512 x^{32} + 29199 x^{33} + 35394 x^{34} + 41340 x^{35} + 
 49969 x^{36} + 58485 x^{37} + 69063 x^{38} + 80367 x^{39} + 94058 x^{40} \\ &  + 
 108708 x^{41} + 124758 x^{42} + 146517 x^{43} + 165768 x^{44} + 
 193087 x^{45} + 217572 x^{46} + 249876 x^{47} + 284778 x^{48} \\ &  + 
 314492 x^{49} + 367284 x^{50} + 404674 x^{51} + 459654 x^{52} + 
 519653 x^{53} + 570449 x^{54} + 655200 x^{55} + 704832 x^{56} \\ & + 
 801929 x^{57} + 893794 x^{58} + 962802 x^{59} + 1114976 x^{60} + 
 1175544 x^{61} + 1333348 x^{62} + 1453601 x^{63} \\ & + 1557538 x^{64} + 
 1803532 x^{65} + 1837087 x^{66} + 2139593 x^{67} + 2250165 x^{68} + 
 2419600 x^{69} + 2760482 x^{70} \\ & + 2756378 x^{71} + 3277932 x^{72} + 
 3257995 x^{73} + 3660668 x^{74} + 3984069 x^{75} + 3971315 x^{76} + 
 4743295 x^{77} \\ & + 4503043 x^{78} + 5308448 x^{79} + 5339116 x^{80} + 
 5600623 x^{81} + 6402082 x^{82} + 5952064 x^{83} + 7259834 x^{84} + 
 6766136 x^{85} \\ &  + 7641217 x^{86} + 7929474 x^{87} + 7698010 x^{88} + 
 9183965 x^{89} + 8138340 x^{90} + 9837219 x^{91} + 9212860 x^{92} + 
 9783132 x^{93} \\ & + 10505450 x^{94} + 9520736 x^{95} + 11655437 x^{96} + 
 10073731 x^{97} + 11803227 x^{98} + 11198278 x^{99} + 11143638 x^{100} \\ & + 
 12320560 x^{101} + 10582946 x^{102} + 13047098 x^{103} + 11181347 x^{104} + 
 12571150 x^{105} + 12045957 x^{106} + 11280202 x^{107} \\ & + 12758876 x^{108} + 
 10560938 x^{109} + 12887416 x^{110} + 11036141 x^{111} + 11917814 x^{112} + 
 11439629 x^{113} + 10133786 x^{114} \\ & + 11660065 x^{115} + 9458761 x^{116} + 
 11265259 x^{117} + 9623905 x^{118} + 10076746 x^{119} + 9612054 x^{120} + 
 8077174 x^{121} \\ & + 9396734 x^{122} + 7605879 x^{123} + 8759031 x^{124} + 
 7373798 x^{125} + 7611973 x^{126} + 7185485 x^{127} + 5708054 x^{128} \\ & + 
 6679427 x^{129} + 5495336 x^{130} + 6087234 x^{131} + 4947645 x^{132} + 
 5154271 x^{133} + 4807598 x^{134} + 3567272 x^{135} \\ & + 4210368 x^{136} + 
 3565526 x^{137} + 3786811 x^{138} + 2924384 x^{139} + 3139746 x^{140} + 
 2888046 x^{141} + 1970006 x^{142} \\ & + 2383379 x^{143} + 2073011 x^{144} + 
 2106297 x^{145} + 1560596 x^{146} + 1717880 x^{147} + 1560278 x^{148} + 
 972472 x^{149} \\ & + 1229583 x^{150} + 1087043 x^{151} + 1046621 x^{152} + 
 785712 x^{153} + 838190 x^{154} + 761148 x^{155} + 447678 x^{156} \\ & + 
 583656 x^{157} + 532046 x^{158} + 462900 x^{159} + 389421 x^{160} + 
 366798 x^{161} + 335763 x^{162} + 212608 x^{163} \\ & + 256204 x^{164} + 
 257930 x^{165} + 179908 x^{166} + 191411 x^{167} + 154828 x^{168} + 
 131920 x^{169} + 117570 x^{170} \\ & + 104222 x^{171} + 127464 x^{172} + 
 61758 x^{173} + 89311 x^{174} + 77228 x^{175} + 44696 x^{176} + 
 72804 x^{177} \\ & + 38721 x^{178} + 61735 x^{179} + 22590 x^{180} + 
 36424 x^{181} + 51877 x^{182} + 12721 x^{183} + 41704 x^{184} + 
 12565 x^{185} \\ & + 27568 x^{186} + 14139 x^{187} + 11756 x^{188} + 
 38158 x^{189} + 3032 x^{190} + 18904 x^{191} + 3336 x^{192} + 10783 x^{193} \\ & + 
 13190 x^{194} + 2692 x^{195} + 24304 x^{196} + 594 x^{197} + 6164 x^{198} + 
 670 x^{199} + 3464 x^{200} + 11458 x^{201} + 378 x^{202} \\ & + 12030 x^{203} + 
 84 x^{204} + 1323 x^{205} + 90 x^{206} + 827 x^{207} + 8008 x^{208} + 
 24 x^{209} + 4368 x^{210} + 6 x^{211} + 162 x^{212} \\ & + 6 x^{213} + 126 x^{214} + 
 4368 x^{215} + 1092 x^{217} + 8 x^{219} + 9 x^{221} + 1820 x^{222} + 
 168 x^{224} + 560 x^{229} + 12 x^{231} \\ & + 120 x^{236} + 16 x^{243} + x^{250}
 %=================================================================
 \end{array}
 $}
 \label{unrefined_Z7_stable_variables}
 \eeq

\begin{figure}[H]
	\centering
	\includegraphics[width=10cm]{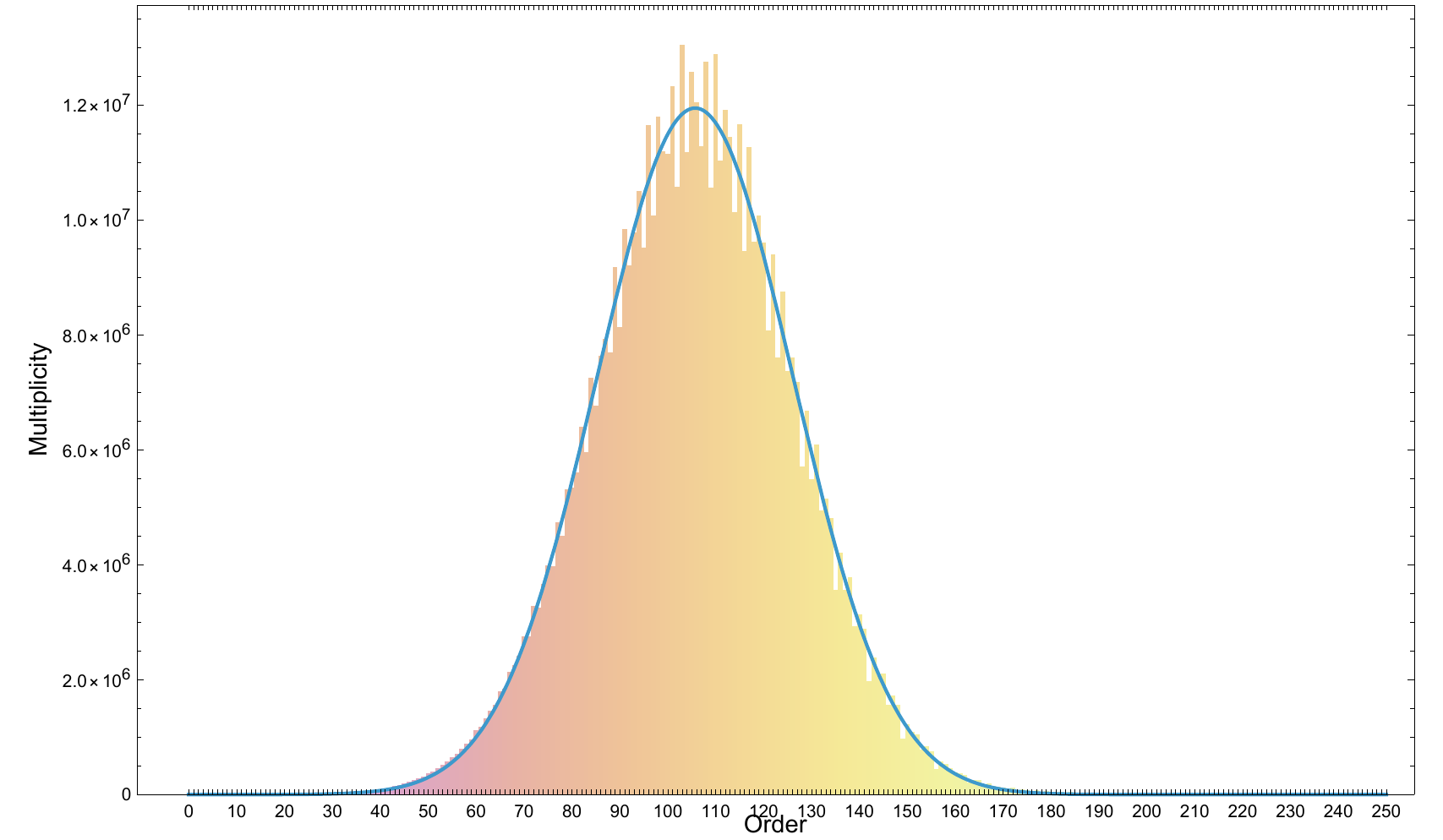}
\caption{Profile of the partition function $Z_7$ in stable variables. We also show a Gaussian fit to it, which has $R^2=0.995214$.}
	\label{Histogram_Z7_stable_variables_Gaussian_fit}
\end{figure}

\fref{Histogram_Z7_stable_variables_Gaussian_fit} displays the profile of $Z_7$ in stable variables together with a Gaussian fit. The agreement is excellent and represents a clear improvement over the $Z_6$ case. Despite this apparent agreement, one should be cautious: in principle, the data should not be exactly Gaussian, since the orders of terms in the partition function are constrained to be positive semidefinite. However, this might be possible due to the discrete nature of the data or emerge in some appropriate infinite limit, as the center of the Gaussian moves to higher values. 

%===================================================================
%===================================================================

\begin{figure}[ht!]
	\centering
	\includegraphics[width=10cm]{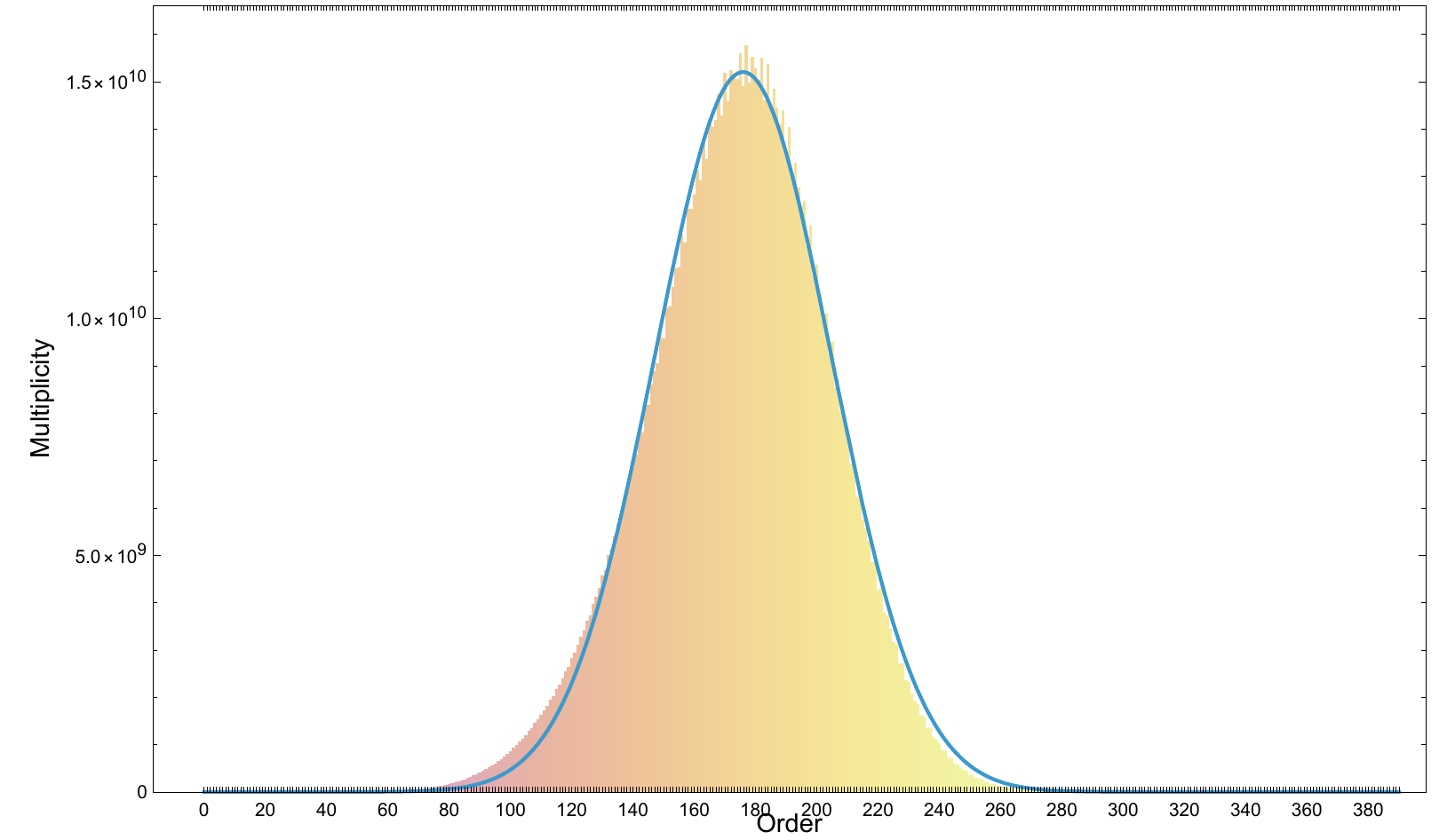}
\caption{Profile of the partition function $Z_8$ in stable variables.  We also show a Gaussian fit to it, which has $R^2=0.998216$.}
	\label{Histogram_Z8_stable_variables_Gaussian_fit}
\end{figure}

It would be remarkable if our observations were evidence indicating that stable variables reorganize the terms in the partition function so that, at large steps in the cascade, the profile approaches a universal form. With this motivation, we constructed $Z_8$ in stable variables and fitted a Gaussian to it. The partition function is given below and \fref{Histogram_Z8_stable_variables_Gaussian_fit} shows the corresponding profile plot.

\beq
\resizebox{\textwidth}{!}{$
\begin{array}{cl}
 %=================================================================
 Z_8 & = 1 + x + 2 x^{2} + 3 x^{3} + 4 x^{4} + 11 x^{5} + 14 x^{6} + 22 x^{7} + 38 x^{8} + 
 52 x^{9} + 85 x^{10} + 111 x^{11} + 178 x^{12} + 254 x^{13} 
 \\ & + 332 x^{14} + 
 510 x^{15} + 672 x^{16} + 970 x^{17} + 1282 x^{18} + 1761 x^{19} + 2448 x^{20} + 
 3130 x^{21} + 4380 x^{22} + 5678 x^{23} \\ &
   + 7652 x^{24} + 9998 x^{25} + 
 12987 x^{26} + 17329 x^{27} + 21919 x^{28} + 29044 x^{29} + 37011 x^{30} + 
 47718 x^{31} + 61254 x^{32} 
 \\ & 
 + 77116 x^{33} + 99370 x^{34} + 123965 x^{35} + 
 157667 x^{36} + 197771 x^{37} + 246107 x^{38} + 309232 x^{39}  + 380174 x^{40}  \\ & 
+ 474177 x^{41} + 582182 x^{42} + 715343 x^{43} + 
 879958 x^{44} + 1065634 x^{45} + 1303724 x^{46} + 1572091 x^{47}
 \\ &  + 
 1900754 x^{48} + 2290994 x^{49} + 2737414 x^{50} + 3288045 x^{51} + 
 3899014 x^{52} + 4638940 x^{53} + 5491924 x^{54} \\ &
  + 6464788 x^{55} + 
 7630184 x^{56} + 8918519 x^{57} + 10449772 x^{58} + 12168011 x^{59} + 
 14120958 x^{60} + 16405102 x^{61} \\ &
  + 18895126 x^{62} + 21832619 x^{63} + 
 25046761 x^{64} + 28703904 x^{65} + 32839663 x^{66} + 37338263 x^{67} + 
 42581987 x^{68}
 \\ &  + 48149390 x^{69} + 54598150 x^{70} + 61545552 x^{71} + 
 69275252 x^{72} + 77910581 x^{73} + 87084818 x^{74} + 97707120 x^{75} \\ &
  + 
 108612310 x^{76} + 121327965 x^{77} + 134466633 x^{78} + 149172426 x^{79} + 
 165229610 x^{80} + 181841820 x^{81} + 201433714 x^{82} \\ &
  + 220254889 x^{83} + 
 243202007 x^{84} + 265466331 x^{85} + 290708361 x^{86} + 318221901 x^{87} + 
 344954470 x^{88}  + 378582072 x^{89}
 \\ &
 + 407775328 x^{90} + 445743880 x^{91} + 
 480907863 x^{92} + 519802409 x^{93} + 564565414 x^{94} + 603239937 x^{95} + 
 657150437 x^{96} 
 \\ & + 699629621 x^{97} + 756647619 x^{98} + 810764690 x^{99} + 
 864540644 x^{100} + 934449894 x^{101} + 986632398 x^{102} 
  \end{array}
 $} \nonumber
 \eeq
 \beq
\resizebox{\textwidth}{!}{$
\begin{array}{cl}
 %=================================================================
 & + 
 1066350969 x^{103} + 1128026885 x^{104} + 1204852089 x^{105} + 
 1287993785 x^{106} + 1355999567 x^{107} + 1459488340 x^{108}
 \\ &  + 
 1529907150 x^{109} + 1636553550 x^{110} + 1729949508 x^{111} + 
 1822697974 x^{112} + 1948487655 x^{113} + 2031782489 x^{114} 
 \\ & + 
 2173165674 x^{115} + 2275281961 x^{116} + 2402016304 x^{117} + 
 2547601471 x^{118} + 2650364801 x^{119} + 2830747167 x^{120} 
 \\ & + 
 2939229915 x^{121} + 3112143295 x^{122} + 3272023773 x^{123} + 
 3403940905 x^{124} + 3625291171 x^{125} + 3737472219 x^{126}  \\ &
 + 
 3972925082 x^{127} + 4132002434 x^{128} + 4314618156 x^{129} + 
 4567975427 x^{130} + 4687204509 x^{131} + 4999195190 x^{132}  \\ & + 
 5133808349 x^{133} + 5403176028 x^{134} + 5653609775 x^{135} + 
 5809181930 x^{136} + 6189430869 x^{137} + 6282908304 x^{138} \\ & + 
 6677390751 x^{139} + 6863360393 x^{140} + 7119327128 x^{141} + 
 7506328361 x^{142} + 7589293541 x^{143} + 8110588137 x^{144} 
 \\ & + 
 8174512123 x^{145} + 8611483678 x^{146} + 8884819890 x^{147} + 
 9058942208 x^{148} + 9613297195 x^{149} + 9577503033 x^{150} 
 \\ & + 
 10221642511 x^{151} 
 + 10257818135 x^{152} + 10668969576 x^{153} + 
 11054585761 x^{154} + 11071325292 x^{155} + 11791678020 x^{156}  \\ & + 
 11596881305 x^{157} + 12312618961 x^{158} + 12317827394 x^{159} + 
 12623804929 x^{160} + 13116951425 x^{161} 
 + 12912536923 x^{162} 
  \\ & + 
 13758701464 x^{163} + 13374641850 x^{164} + 14091731379 x^{165} + 
 14047519586 x^{166} + 14183321525 x^{167} + 14741396322 x^{168} \\ & +
 14282262300 x^{169} + 15183486466 x^{170} + 14594233563 x^{171}  + 
 15246074987 x^{172} + 15117720145 x^{173} + 15055987407 x^{174} \\ &  + 
 15606219978 x^{175} + 14902227748 x^{176} + 15775939517 x^{177} + 
 14987022583 x^{178} + 15530512723 x^{179} + 15278746613 x^{180}
 \\ &   + 
 15040638120 x^{181} + 15500737393 x^{182} + 14608914517 x^{183} + 
 15380759373 x^{184} + 14427381599 x^{185} + 14850297593 x^{186}
 \\ &  + 
 14451046708 x^{187} + 14099243345 x^{188} + 14402190677 x^{189} + 
 13417793632 x^{190} + 14035468662 x^{191} + 12985538341 x^{192} 
 \\ & + 
 13299640660 x^{193} + 12761204416 x^{194} + 12377575985 x^{195} + 
 12490953370 x^{196} + 11526005387 x^{197} + 11964867339 x^{198} 
 \\ & + 
 10910451848 x^{199} + 11137057466 x^{200} + 10504673325 x^{201} + 
 10162546271 x^{202} + 10095945049 x^{203} + 9251011524 x^{204}  
 \\ & + 
 9513595485 x^{205} + 8550300291 x^{206} + 8708230445 x^{207} + 
 8052589574 x^{208} + 7796041278 x^{209} + 7594861797 x^{210}  
 \\ & + 
 6934450216 x^{211} + 7046301528 x^{212} + 6248243178 x^{213} + 
 6350026788 x^{214} + 5745020828 x^{215} + 5583226091 x^{216}  
 \\ & + 
 5312017292 x^{217} + 4853492750 x^{218} + 4855946429 x^{219} + 
 4257992451 x^{220} + 4313014398 x^{221} + 3813548777 x^{222} \\ & + 
 3729561188 x^{223} + 3451738764 x^{224} + 3171195478 x^{225} + 
 3111334529 x^{226} + 2706607854 x^{227} + 2725527014 x^{228} \\ &
  + 
 2355325989 x^{229} + 2321568386 x^{230} + 2083307897 x^{231} + 
 1933411306 x^{232} + 1853209643 x^{233} + 1605147119 x^{234} 
 \\ & + 
 1601031253 x^{235} + 1353945820 x^{236} + 1345438618 x^{237} + 
 1168891821 x^{238} + 1098941380 x^{239} + 1027218452 x^{240}  
 \\ & + 
 888153535 x^{241} + 874014328 x^{242} + 725028230 x^{243} + 
 725540216 x^{244} + 611590843 x^{245} + 581469517 x^{246} + 
 531449535 x^{247}  
 \\ & + 458376480 x^{248} + 443861340 x^{249} + 
 362410668 x^{250} + 364152626 x^{251} + 300733645 x^{252} + 
 285775695 x^{253} + 258142884 x^{254}  
 \\ & + 220513405 x^{255} + 
 210381481 x^{256} + 169976379 x^{257} + 170373927 x^{258} + 
 141242718 x^{259} + 130078216 x^{260} + 118818025 x^{261} 
 \\ & + 
 98786475 x^{262} + 93710052 x^{263} + 75828661 x^{264} + 74526817 x^{265} + 
 65182712 x^{266} + 54646621 x^{267} + 52444189 x^{268}  
 \\ & + 41157207 x^{269} + 
 39689139 x^{270} + 33288091 x^{271} + 30587696 x^{272} + 30693253 x^{273} + 
 21111033 x^{274} + 22445837 x^{275} 
 \\ & + 15917984 x^{276} + 16241393 x^{277} + 
 15409677 x^{278} + 11799257 x^{279} + 15178558 x^{280} + 7472666 x^{281} + 
 9356986 x^{282}  
 \\ & + 5697813 x^{283} + 6520884 x^{284} + 8204959 x^{285} + 
 4262514 x^{286} + 7855136 x^{287} + 2413742 x^{288} + 3770885 x^{289} \\ & + 
 1877506 x^{290}
 + 2581815 x^{291} + 5146010 x^{292} + 1424338 x^{293} + 
 4103283 x^{294} + 706659 x^{295} + 1439432 x^{296}  
 \\ & + 564326 x^{297} + 
 995178 x^{298} + 3541130 x^{299} + 429999 x^{300} + 2065562 x^{301} + 
 185003 x^{302} + 505079 x^{303} + 152345 x^{304} 
 \\ & + 362563 x^{305} + 
 2449918 x^{306} + 113022 x^{307} + 962210 x^{308} + 42243 x^{309} + 
 157194 x^{310} + 35971 x^{311} + 119950 x^{312} \\ & 
  + 1616296 x^{313} + 
 24500 x^{314} + 401136 x^{315} + 8060 x^{316} + 41650 x^{317} + 7104 x^{318} + 
 34360 x^{319} + 991414 x^{320} \\ &  + 4042 x^{321} + 144993 x^{322} + 
 1194 x^{323} + 8926 x^{324} + 1090 x^{325} + 8030 x^{326} + 557686 x^{327} + 
 444 x^{328} + 43812 x^{329} \\ &  + 120 x^{330} + 1437 x^{331} + 114 x^{332} + 
 1409 x^{333} + 284458 x^{334} + 24 x^{335} + 10536 x^{336} + 6 x^{337} + 
 153 x^{338} + 6 x^{339}\\ &  + 162 x^{340} + 129848 x^{341} + 1872 x^{343} + 
 8 x^{345} + 9 x^{347} + 52128 x^{348} + 216 x^{350} + 17964 x^{355} + 
 12 x^{357} \\ & + 5133 x^{362} + 1156 x^{369} + 190 x^{376} + 20 x^{383} + x^{390}
 %=================================================================
 \end{array}
 $}
 \label{unrefined_Z8_stable_variables}
 \eeq

%===================================================================
%===================================================================

While the fit in this case is even better than for $Z_7$ according to the higher $R^2$, the distribution appears to be asymmetric. It would be desirable to compute and analyze $Z_9$ to determine whether the apparent trend persists, but we have been unable to do so due to computational limitations.

We prefer to remain agnostic about whether the profiles converge to a particular distribution, such as a Gaussian, or not. Perhaps, the appearance of a Gaussian profile is just an accidental coincidence for the partition functions we considered. Nevertheless, we find the observations intriguing, worth reporting, and deserving of further investigation. More broadly, our results in Sections \sref{section_crystals_Q111} and \sref{section_stable_variables} suggest that the profiles of partition functions, both in original and stable variables, are interesting objects in their own right and merit deeper study.

If the profiles in stable variables indeed converge to a Gaussian distribution, it may be an indication of a universal behavior, since it is unlikely that this is an exclusive feature of $Q^{1,1,1}$.

%=================================================================
\section{Conclusions}
%=================================================================

\label{section_conclusions}

In this paper, we extended the study of the crystal melting models associated to toric CY 4-folds introduced in \cite{Franco:2023tly} in various directions.

First, we explored in further detail how the models work for general toric CY 4-folds, focusing on the explicit example of $Q^{1,1,1}$. To do so, we introduced an efficient algorithm for the construction of crystals based on periodic quivers. 
One of our primary goals was to investigate how crystals and their partition functions transform under triality. Using $Q^{1,1,1}$ as a representative example, we identified a periodic triality cascade and explicitly constructed the associated crystals at several stages along the cascade. We generated detailed data for several of these models, including their Hasse diagrams, the number of atoms of each type, the corresponding partition functions, and the multiplicities of melting configurations. Although our analysis focused on finite crystals, the methods we developed extend naturally to the infinite case, as well as to other flavor configurations and geometries.

We also introduced the concept of stable variables and showed how they lead to the stabilization of partition functions.

In addition, we introduced and initiated a study of the profiles of the partition functions, both in original and in stable variables. We observed that, when expressed in terms of stable variables, the profiles seem to suggest the emergence of a universal behavior at large steps of the cascade. More generally, our findings indicate that these profiles merit further investigation, both to determine whether and how they encode information about the underlying geometry, and to clarify whether they exhibit universal features.

Another central motivation for our work is the search for a physically motivated generalization of cluster algebras based on $2d$ (0,2) quiver theories and triality. We have argued that toric CY$_4$ crystals, together with their evolution under triality, provide a natural candidate class of systems governed by such novel mathematical structures. Let us briefly summarize the current status of this proposed generalization. Clearly, $2d$ (0,2) quivers and triality play the role of ordinary quivers and their mutations in the definition of cluster algebras \eqref{quiver_mutation_ordinary_cluster}. Moreover, our results suggest that the stable variables and their transformation rules are the counterparts of cluster coefficients and their transformations \eqref{cluster_algebra_transformation_coefficients}. Moreover, the stabilization exhibited by crystal partition functions suggests that they are the analogues of $F$-polynomials for CY 4-folds. The appropriate generalization of the cluster transformations \eqref{cluster_transformation} themselves, however, remains an open problem. We are optimistic that the explicit results obtained in this work---such as partition functions and multiplicities of melting configurations---as well as further data accessible through the techniques we have developed, will provide valuable empirical input to guide the formulation of these transformations. Indeed, reproducing the multiplicities of melting configurations, such as those in \eqref{multiplicities_melting_configurations}, may provide a natural initial target. This is analogous to studying cluster algebras without coefficients and, therefore, more accessible than deriving the fully refined partition functions. We plan to revisit this question in the future.

Although the idea of the profiles of partition functions arose while studying CY$_4$ crystals, it would be interesting to investigate them in the context of CY$_3$ crystals as well. This case should be considerably more tractable. For CY$_3$’s, it is well understood how to compute partition functions iteratively via cluster transformations. Therefore, it would be easier to investigate questions at large steps in cascades, such as whether profiles in stable variables converge to a universal distribution. We plan to explore this question in future work.

 %======================================================================  
\acknowledgments
%======================================================================  

We would like to thank Gregg Musiker for enjoyable discussions. This work is supported by the U.S. National Science Foundation grants PHY-2112729 and PHY-2412479.

%======================================================================
\bibliographystyle{JHEP}
\bibliography{mybib}
%======================================================================

\end{document}